\documentclass[aps, prd, onecolumn, tightenlines, notitlepage, superscriptaddress, nofootinbib, preprintnumbers, floatfix,showkeys,11pt,altaffilletter]{revtex4-1}

\usepackage{csquotes}
\usepackage[official]{eurosym}
\usepackage[normalem]{ulem}
\usepackage{amstext}
\usepackage{graphicx}
\usepackage{dirtytalk}
\graphicspath{{figures/}}
\usepackage{url}
\usepackage{color}
\usepackage{ulem}
\usepackage[version=4]{mhchem}
\usepackage[utf8]{inputenc}
\usepackage{fontawesome}
\usepackage{yfonts}
\usepackage[normalem]{ulem}
\usepackage{slashed}
\usepackage{makecell}

\usepackage{epsfig,amsfonts,amsmath,amssymb,graphicx,color,slashed,multirow}
\usepackage{amsmath,latexsym,amssymb,graphicx,slashed,color,enumerate,url,cancel,gensymb}
\usepackage{textcomp}

\usepackage[x11names]{xcolor}
\usepackage[colorlinks,pdfstartview=FitV,breaklinks=true]{hyperref}
\usepackage{booktabs}
\usepackage{adjustbox}
\usepackage{textgreek} % To add greek letters to the text
\usepackage{caption} % Better figure customization
\usepackage{ragged2e}

\makeatletter
\long\def\@makecaption#1#2{%
  \vskip\abovecaptionskip
  \sbox\@tempboxa{\textbf{#1.} #2}%
  \ifdim \wd\@tempboxa >\hsize
    \textbf{#1.} \justifying #2\par
  \else
    \global \@minipagefalse
    \hb@xt@\hsize{\hfil\box\@tempboxa\hfil}%
  \fi
  \vskip\belowcaptionskip}
\makeatother
\usepackage{booktabs} % Cool tables
\usepackage{float}
\usepackage{orcidlink}
\usepackage{charter}

\usepackage{appendix}

\hypersetup{colorlinks,citecolor= nicered,linkcolor= blue}
\definecolor{nicered}{rgb}{0.7,0.1,0.1}
\definecolor{nicegreen}{rgb}{0.1,0.5,0.1}
\makeatletter
    \newcommand{\colorboxed}[3][white]{\fcolorbox{#2}{#1}{\m@th$\displaystyle#3$}}
\makeatother

\def\cevns{CE\textnu NS}
\def\d{\mathrm{d}}

\usepackage[T1]{fontenc}
\graphicspath{{Figures/}}
\usepackage{appendix}

\def\d{\mathrm{d}}

\definecolor{blue(ncs)}{rgb}{0.0, 0.53, 0.74}
\definecolor{chromeyellow}{rgb}{1.0, 0.56, 0.0}
\definecolor{amber(sae/ece)}{rgb}{1.0, 0.49, 0.0}
\definecolor{regalia}{rgb}{0.32, 0.18, 0.5}
\definecolor{royalblue(web)}{rgb}{0.25, 0.41, 0.88}
\definecolor{vdrgreen}{rgb}{0.0, 0.6, 0.0}
\definecolor{darkraspberry}{rgb}{0.53, 0.15, 0.34}

\newcommand{\AddrAHEP}{%
Instituto de F\'{i}sica Corpuscular (IFIC), CSIC‐Universitat de Val\'encia, E-46980 Valencia, Spain}

\AtBeginDocument{\hypersetup{citecolor=darkraspberry,linkcolor=darkraspberry,urlcolor=darkraspberry}}
\usepackage{appendix}

\begin{document}

\title{\Large Sensitivity to sub-GeV dark matter in forthcoming\\ 
spallation-source neutrino experiments}%

\author{D. Aristizabal Sierra~\orcidlink{}}%
\email{daristizabal@uliege.be}\affiliation{Universidad T\'ecnica Federico Santa Mar\'{i}a - Departamento de F\'{i}sica Casilla 110-V, Avda. Espa\~na 1680, Valpara\'{i}so, Chile}%

\author{V. De Romeri~\orcidlink{0000-0003-3585-7437}}%
\email{deromeri@ific.uv.es}%
\affiliation{\AddrAHEP}%

\author{D. K. Papoulias~\orcidlink{0000-0003-0453-8492}}%
\email{dimitrios.papoulias@uni-hamburg.de}
\affiliation{Institute of Experimental Physics, University of Hamburg, 22761, Hamburg, Germany}

\author{G. Sanchez Garcia~\orcidlink{0000-0003-1830-2325}}%
\email{g.sanchez@ciencias.unam.mx}%
\affiliation{Departamento de F\'{i}sica, Facultad de Ciencias, Universidad Nacional Aut\'onoma de M\'exico,
Apartado Postal 70-542, Ciudad de M\'exico 04510, M\'exico}%


\begin{abstract}

Sub-GeV thermal dark matter weakly interacting with the Standard Model through vector-portal mediators provides a well-motivated and predictive framework that remains challenging to probe with conventional direct detection experiments. Motivated by the rapid development of neutrino facilities based on spallation neutron sources, we study the sensitivity of future coherent elastic neutrino–nucleus scattering experiments to light dark matter produced in neutral pion decays. We consider scalar dark matter interactions mediated by two different vector portals, a generic dark photon and a baryophilic vector mediator. The neutral pion yield is calculated through a \texttt{GEANT4} simulation and the results are compared with those obtained with the \textit{Sandford-Wang} parametrization. We show that predictions based on either approach do not produce significant differences. Our results demonstrate that upcoming low-threshold neutrino detectors at the European Spallation Source (ESS), the Japan Proton Accelerator Research Complex (J-PARC) and the China Spallation Neutron Source (CSNS) may test regions in parameter space not yet explored, or be competitive with existing bounds. We point out that these facilities will strengthen the global experimental program searching for secluded sectors.

%----------------------------------------
\end{abstract}
\maketitle
%----------------------------------------

\section{Introduction}
The nature of a large fraction of the Universe’s matter content remains unknown and is commonly referred to as dark matter (DM)~\cite{Cirelli:2024ssz}. Despite extensive experimental efforts, no DM signals other than its gravitational evidence have yet been observed~\cite{XENON:2024hup,XENON:2025vwd,Zhang:2025ajc,PandaX:2024qfu,LZ:2024zvo,DarkSide-50:2022qzh}. From a theoretical perspective, weakly interacting massive particles (WIMPs) have long been regarded as well-motivated candidates~\cite{RevModPhys.90.045002,Cirelli:2024ssz}, inspiring decades of searches through direct and indirect detection techniques as well as their possible production at colliders. These traditional search strategies, particularly direct detection, achieve their highest sensitivity for WIMPs in the GeV-TeV mass range~\cite{Schumann:2019eaa,Billard:2021uyg}, while their limitations in the recoil-energy threshold significantly reduce their sensitivity to DM in the sub-GeV regime. Current  multi-ton liquid-xenon detectors operate with $\sim 1\,\text{keV}_\text{nr}$ thresholds and are therefore blind to potential nuclear recoils induced by sub-GeV DM (see, e.g., Refs. \cite{PandaX:2024muv,XENON:2024ijk}). In this regard, fixed-target experiments offer a complementary approach by probing the sub-GeV mass range~\cite{Batell:2009di,deNiverville:2012ij,Izaguirre:2013uxa,Izaguirre:2014bca,Berlin:2020uwy,NA64:2023wbi}. In particular, these experiments are  well suited for searches of new particles feebly interacting with the SM degrees of freedom through light mediators~\cite{Boehm:2003hm}, thereby providing access to potential dark sectors. 

Accelerator-based neutrino experiments provide an ideal environment for DM searches \cite{deNiverville:2011it,deNiverville:2015mwa,Ge:2017mcq,Dutta:2019nbn,Dutta:2020vop,DeRomeri:2019kic,Batell:2022xau,Buonocore:2019esg,deNiverville:2016rqh,Jordan:2018gcd}. In particular, high-intensity proton beams colliding against a fixed dense target open different production channels including meson decays, proton Bremsstrahlung, and Drell–Yan processes (see, e.g., Ref.~\cite{Ovchynnikov:2023cry}). Hence, spallation neutron sources offer a remarkably clean setting to search for sub-GeV DM. Because of their relatively low proton beam energies, secluded sector states are predominantly produced via neutral pion decays, resulting in a single dominant production channel.
Combined with the large luminosities, low-threshold detectors optimized for nuclear recoil measurements, and powerful background rejection enabled by timing information, these facilities are exceptionally well suited for MeV-scale dark sector searches. In fact, the first dedicated searches by the COHERENT and Coherent CAPTAIN-Mills (CCM) collaborations have already delivered  compelling results~\cite{COHERENT:2019kwz,COHERENT:2021pvd,CCM:2021leg,CCM:2021yzc,COHERENT:2022pli}. With several detectors moving toward multi-ton target masses as well as with continued improvements in background discrimination, the projected discovery reach is expected to get boosted~\cite{COHERENT:2023sol}.

Building on the experience of the COHERENT collaboration, several European and Asian experimental initiatives are planning to deploy detectors aiming at a strong coherent elastic neutrino-nucleus scattering (\cevns)~\cite{Abdullah:2022zue} program. Many of these projects are already in an advanced development stage~\cite{Baxter:2019mcx,Collar:2025sle,Su:2023klh}. Therefore, assessing their physics potential for dark sector searches is both timely and well motivated. In this paper, we explore this opportunity by focusing on two representative classes of models. First, we study a minimal vector-portal model based on a simple extended gauge symmetry, $G_\text{SM} \otimes U(1)^\prime$ \cite{Pospelov:2007mp,Arkani-Hamed:2008hhe}, with $G_\text{SM}$ the Standard Model (SM) gauge group. In this framework, kinetic mixing between $U(1)_Y$  and $U(1)^\prime$ enables
interactions between the DM and the SM through a dark photon that feebly couples to the SM electromagnetic current. This coupling is instrumental, since it allows production of the new gauge boson in neutral pion decays, followed by DM detection through scattering processes. As a second benchmark, we consider the case in which the additional Abelian gauge factor is associated with baryon number (see, e.g., Refs.~\cite{Batell:2014yra,Coloma:2015pih,Dror:2017ehi,Berlin:2018bsc,CCM:2021yzc} and references therein). The new gauge boson then couples predominantly to quarks, resulting in a 
\enquote{baryophilic} scenario that leverages the capabilities of the detectors considered in this paper, which are optimized to measure energy deposition induced by nuclear recoils. Together, these two frameworks capture a wide class of theoretically motivated light DM scenarios and illustrate the discovery potential of next-generation \cevns~dedicated detectors  to probe dark sector interactions.

We consider detectors to be deployed at three spallation neutron source facilities: the European Spallation Source (ESS)~\cite{Baxter:2019mcx}, currently under construction in Lund, Sweden; the Japan Proton Accelerator Research Complex (J-PARC)~\cite{Collar:2025sle}; and the China Spallation Neutron Source (CSNS)\cite{Su:2023klh}. In each of these  installations, a high-intensity proton beam impinges on a dense fixed target, typically mercury or tungsten. The resulting collisions induce intranuclear cascade processes that produce neutrons (the primary purpose of spallation sources) and other secondary particles. Among these, a large number  of $\pi^+$ mesons decay at rest ($\pi^-$ mesons are absorbed by nuclei), yielding a well-characterized flux of  electron and muon (anti-)neutrinos. As expected from nuclear isospin invariance, neutral pions are also abundantly produced. With short lifetimes even in the laboratory frame,  the $\pi^0$ mesons decay promptly into photon pairs.
However, this electromagnetic component is extremely suppressed since most of the photons are absorbed by the target material nuclei given their short mean free path in highly dense environments.

This picture is qualitatively modified in the presence of a secluded vector portal in which $\pi^0$ can also decay into dark vector bosons. In contrast to photons, the produced dark vector bosons can escape the target material without significant attenuation and further decay into pairs of DM particles. As a result, a flux of DM particles reaches the same detectors designed for \cevns~measurements \cite{deNiverville:2015mwa,Ge:2017mcq}. 
The dominant DM detection channel is nuclear recoils, mirroring the \cevns~signal.
All in all, as already stressed, the combination of high-intensity beams, well-understood neutrino backgrounds, and low-threshold detectors makes spallation-based experiments uniquely sensitive to MeV-scale DM searches.

To estimate the DM production rate, the neutral pion flux can be determined either through a dedicated simulation with existing toolkits like \texttt{GEANT4}~\cite{GEANT4:2002zbu} or, at the proton energies relevant in these installations, through the semi-empirical Sandford and Wang parametrization~\cite{SanfordWang1967}. We compare the results from both approaches and then adopt the \texttt{GEANT4}-based fluxes for the full numerical analysis. 
Our simulations are validated against publicly available results from the COHERENT Collaboration at the Spallation Neutron Source (SNS)~\cite{COHERENT:2021yvp}.
For the ESS and J-PARC facilities, we consider medium-scale germanium, gaseous xenon, and CsI detectors~\cite{Baxter:2019mcx,Collar:2025sle}, while for CSNS we assume a large-scale CsI detector~\cite{CSNS}. Under realistic assumptions on detector performance and backgrounds, we show that these setups can be competitive with other experiments in testing regions of parameter space close or beyond the relic-density target, including areas that remain challenging for other existing or proposed experimental searches.

The remainder of this manuscript is organized as follows. In Sec.~\ref{sec:theory}, we describe the theoretical frameworks for the two vector-portal DM scenarios. In Sec.~\ref{sec:DMprod}, we discuss neutral pion production at spallation neutron source experiments, considering two approaches: analytical and numerical. The details of our statistical analysis are presented in Sec.~\ref{sec:analysis}. Projected sensitivities for scalar DM with a dark photon mediator using \enquote{standard} experimental configurations are shown in Sec.~\ref{sec:results}. We present our conclusions in Sec.~\ref{sec:concl}. Results for other experimental configurations along with results for the baryophilic scenario are displayed in App. \ref{app:baryophilic_results}.

%----------------------------------------
\section{Theoretical framework}
\label{sec:theory}
%----------------------------------------
DM searches have long been guided by the WIMP paradigm: a weakly interacting particle with a mass at the electroweak scale whose relic abundance is set by thermal freeze-out. In its simplest realization, achieving the observed relic density requires heavy mediators and $\mathcal{O}(1)$ couplings. However, for sub-GeV DM, annihilation through heavy mediators becomes inefficient, typically leading to overclosure of the Universe~\cite{Lee:1977ua}. This constraint can be relaxed if the mediator is light, since the reduced DM mass is compensated by an enhanced annihilation rate mediated by a lighter dark-visible \enquote{messenger}~\cite{Boehm:2003hm,Fayet:2004bw}. Such scenarios open well-motivated regions of parameter space that are hard to explore with conventional direct detection experiments. While sub-GeV DM models are abundant, many constructions can be described within local $U(1)^\prime$ extensions of the SM gauge group, i.e.,  $G_\text{SM}\otimes U(1)^\prime$. These scenarios introduce a new Abelian gauge boson, $A^\prime$, and additional stable states that can act as DM candidates. The exact phenomenology depends on how the SM fields are charged under $U(1)^\prime$, determining whether the new mediator couples preferentially to leptons or quarks, and thus shaping the expected detector signatures. 

Motivated by this, we consider two representative frameworks. First, a generic scenario in which $U(1)^\prime$ has no relation to SM quantum numbers, leading to universal couplings of $A^\prime$ to electrically charged fermions. Second, a baryophilic realization where $U(1)^\prime$ is identified with gauged baryon number, $U(1)_B$, enhancing interactions with nuclei and hence maximizing the sensitivity of \cevns-oriented detectors.

%----------------------------------------
\subsection{Model 1: Generic dark photon scenario}
\label{sec:dark_photon_scenario}
%----------------------------------------

Scenarios of this type were first proposed in the early 1980s by Okun and Holdom~\cite{Okun:1982xi,Holdom:1985ag}. In general, an additional local Abelian factor $U(1)^\prime$ kinetically mixes with the SM hypercharge $U(1)_Y$, resulting in an interesting and rich phenomenology (see, e.g., Ref.~\cite{Berlin:2018bsc} and references therein). The new gauge boson is typically assumed to acquire a mass through spontaneous symmetry breaking at a higher scale of a larger symmetry group. For MeV phenomenology though, it can can be effectively described as a massive Proca field. 

The relevant terms in the Lagrangian can be written as
\begin{equation}
    \label{eq:U1_Y_U_prime_Lag}
    \mathcal{L}_{A^\prime}^\epsilon \supset
    -\frac{1}{4}F_{\mu\nu}^\prime F^{\prime\mu\nu} 
    - \frac{1}{4}B_{\mu\nu}B^{\mu\nu}
    -\frac{\epsilon^\prime}{2}F_{\mu\nu}^\prime B^{\mu\nu}
    - g^\prime j_\mu^Y B^\mu 
    - g_D j_\mu^\prime A^{\prime\mu}
    + \frac{1}{2} m_{A^\prime}^2 A_\mu^\prime A^{\prime\mu}\, ,
\end{equation}
where $F_{\mu\nu}^\prime$ and $B_{\mu\nu}$ denote the field-strength tensors of the dark and hypercharge gauge fields, respectively. Moreover, $g_D$ denotes the $U(1)'$ gauge coupling constant, and $m_{A^\prime}$ the mass of the new gauge boson. The currents $j_\mu^Y$ and $j_\mu^\prime$ correspond to the SM hypercharge and dark-sector currents. 
The latter involves the DM particle,  assumed to be a complex scalar with mass $m_\phi$, so that the dark current reads
\begin{equation}
    \label{eq:currents}
    j^\prime_\mu = \phi^*\partial_\mu \phi
    - \phi\partial_\mu\phi^* \equiv \phi^*\overleftrightarrow{\partial}_\mu\phi\ .
\end{equation}
The mixed field tensor term in Eq.~\eqref{eq:U1_Y_U_prime_Lag} implies non-canonical kinetic terms, through the kinetic mixing $\epsilon '$. These can be diagonalized through appropriate field redefinitions in an $\epsilon^\prime$-less basis. A further rotation to the mass eigenstate basis leads to the usual SM couplings of the photon and the $Z
$ to the electromagnetic (EM) and weak currents, as well as of $A^\prime$ to the DM current. In addition, kinetic mixing generates two phenomenologically relevant interactions: a suppressed coupling of $A^\prime$ to the EM current (millicharge coupling), and a coupling of the $Z$ to the DM current. The resulting renormalizable interactions can be written as
\begin{equation}
    \label{eq:new_couplings}
    \mathcal{L}_{A^\prime}^\text{M} \supset
    -\cos\theta_W\;\epsilon^\prime\;e\,j^\mu_\text{EM}\,A^\prime_\mu
    + g_Dj^\prime_\mu\,A^{\prime\mu}
    + \sin\theta_W\;\epsilon^\prime\,g_D\, j^{\prime}_\mu\,Z^\mu\ .
\end{equation}
Here, $\theta_W$ denotes the weak mixing angle and we define $\epsilon\equiv\epsilon^\prime\,\cos\theta_W$. Note that we are using the same notation for the new gauge boson in both interaction and mass bases. In the MeV regime, the first two terms in Eq.~\eqref{eq:new_couplings} are the most relevant. The first allows for $A^\prime$ production through neutral pion decays and mediates DM detection via $t$-channel scattering processes with either nuclei or electrons. The second term governs the decay of $A^\prime$ into DM pairs and thus accounts for the DM flux.
Since $A^\prime$ couples to the EM current, in principle we expect signals both in the electron and nuclear channels. However, since the coupling to nuclei is enhanced compared to that of electrons by a factor $Z$ (atomic number), nuclear-recoil scattering signatures are expected to be dominant in heavy-target materials.  This enhancement, combined with the optimization of the detection technique for nuclear recoils, makes DM-electron scattering subdominant in this context.

In this framework, the DM relic abundance can be achieved through two distinct annihilation mechanisms. If $m_\phi > m_{A^\prime}$, DM predominantly annihilates via the $t$-channel process $\phi + \phi^* \to A^\prime + A^\prime$. In this case, the annihilation cross section is typically large, leading to an underabundant relic density unless additional mechanisms are invoked. Conversely, when $m_\phi < m_{A^\prime}$, annihilation proceeds mainly through the $s$-channel process $\phi + \phi^* \to \text{SM} + \text{SM}$, mediated by an off-shell $A^\prime$. For suitable choices of masses and couplings, this channel can reproduce the observed DM relic density~\cite{Izaguirre:2013uxa}.
On dimensional grounds, the $s$-channel annihilation cross section for scalar DM can be cast as
\begin{equation}
    \label{eq:cross_section}
    \sigma_\text{s-channel}\simeq 
    \frac{16\pi^2\,\alpha}{m_\phi^2}
    \,v^2\,\epsilon^2\alpha_D
    \left(\frac{m_\phi}{m_{A^\prime}}\right)^4
    = \frac{16\pi^2\alpha}{m_\phi^2} v^2\,Y \ .
\end{equation}
Here, $\alpha_D=g_D^2/4 \pi$ and $\alpha$ are the new-sector and EM fine-structure constants, and $v$ is the DM relative velocity. We have introduced the dimensionless parameter
$
Y \equiv \epsilon^2 \alpha_D \left(m_\phi/m_{A^\prime}\right)^4,
$
which conveniently captures the combination of couplings and mass ratios relevant for thermal freeze-out. 
Requiring the relic abundance to match the observed DM density, fixes the annihilation cross section and therefore selects a target value of $Y$ for each $m_\phi$. Following standard practice, we will present our results in terms of this parameter to facilitate direct comparison with existing constraints and projected sensitivities from other facilities.

The velocity dependence in Eq.~\eqref{eq:cross_section} arises from the derivative coupling of the scalar field to the vector mediator, which enforces $p$-wave annihilation. As a consequence, the annihilation cross section is proportional to $v^2$, a crucial feature for the cosmological viability of this scenario. MeV-scale DM candidates are subject to stringent cosmological and astrophysical constraints, in particular from late-time energy injection during recombination. If DM annihilation produces electromagnetically interacting particles at late times, it modifies the ionization history of the Universe and leaves imprints on cosmic microwave background (CMB) anisotropies, which are tightly constrained by observations~\cite{Planck:2018vyg}. The simplest scalar DM model evades this constraint because of the velocity-dependent cross section. At late times, the DM species is highly non-relativistic and so the thermally averaged cross section is small.

More generally, viable scenarios are those in which late-time annihilation is suppressed. This condition is naturally fulfilled when $p$-wave processes control freeze-out, as in the scalar DM case. By contrast, in the simplest fermionic DM realization with vector couplings, the relic density is typically set by $s$-wave $s$-channel annihilation, which remains efficient at late times and is therefore severely constrained by CMB data. Nonetheless, consistent fermionic constructions also exist. For example, pseudo-Dirac DM models, where the relic abundance is governed by co-annihilation between nearly degenerate states, avoid these bounds because the heavier state becomes Boltzmann suppressed at late times, and so the co-annihilation process~\cite{CarrilloGonzalez:2021lxm,Brahma:2023psr,Garcia:2024uwf}. Majorana fermion DM provides another viable option: its axial coupling to the vector mediator leads again to $p$-wave–suppressed annihilation~\cite{Berlin:2018bsc}. Finally, asymmetric fermionic DM scenarios can also evade recombination limits, since an early depletion of the anti-DM population renders late-time annihilation and the associated energy injection ineffective~\cite{Berlin:2018bsc}. These realizations seem also to be more favorable in light of recent direct-detection data, see, e.g.,~\cite{Krnjaic:2025noj,Cheek:2025nul,Han:2026ozp}.
%----------------------------------------
\subsection{Model 2: Gauged baryon number scenario}
\label{sec:baryophilic}
\noindent
%----------------------------------------

In this class of models the $U(1)^\prime$ is identified with local baryon number: we refer to them as \textit{baryophilic}. In scenarios where global baryon number is promoted to a local symmetry, classical symmetries are broken at the quantum level thus demanding the presence of extra fermions. Since new heavy SM chiral fermions have been ruled out by LHC data~\cite{CMS:2012wdt}, viable options are fermions that are vector-like in the SM sector and chiral under the new gauged quantum number. These models are subject to multiple constraints, but scenarios compatible with theoretical and phenomenological considerations are viable and have been thoroughly studied in the literature (see, e.g., Refs. \cite{Aranda:1998fr,FileviezPerez:2010gw,Gondolo:2011eq,Duerr:2013dza,Dobrescu:2014fca,Batell:2014yra,Coloma:2015pih,Butterworth:2024eyr,FileviezPerez:2024fzc}). Interestingly, depending on the $G_\text{SM}\otimes U(1)_B$ quantum numbers some of these realizations naturally involve a heavy, electrically neutral and stable fermion that can be of either Dirac or Majorana nature (see, e.g.,~\cite{Butterworth:2024eyr,FileviezPerez:2024fzc}). In these constructions, the fermionic DM particle mass, deriving from anomaly cancellation, is typically above 100 GeV.

With the new extra Abelian local factor identified with local baryon number, the MeV-scale phenomenological implications are \enquote{decoupled} from kinetic mixing. By construction, the new gauge boson $A^\prime$ couples directly not only to the \enquote{baryonic DM} but also to the SM baryon (quark) current. Thus, even in the limit $\epsilon^\prime=0$ production of $A^\prime$ through $\pi^0$ decays and its further decay into DM pairs is possible\footnote{Note that even if the kinetic mixing parameter vanishes at tree level it will be generated at one-loop order via quark loops. Its value, however, will be suppressed by loop factors: $\epsilon^\prime_{\text{1-loop}}\simeq e\times g_D/16/\pi^2$.}. Detection will be dominated by nuclear recoils, with a cross section enhanced by the total number of nucleons. In contrast to the vector portal discussed in Sec. \ref{sec:dark_photon_scenario}, the vector boson couples to both protons and neutrons.

This kind of scenarios is minimal in the sense that the new-sector dynamics is entirely controlled by the new sector coupling $g_B$, namely 
\begin{equation}
    \label{eq:baryon_current_couplings}
    \mathcal{L}_{A^\prime}^B \supset 
    \frac{g_B}{3}\sum_q \overline{q}\gamma_\mu\,q\;A^{\prime\mu}
    + g_B Q_B j_\mu^B\,A^{\prime\mu}\ .
\end{equation}
Here, $g_B$ is the $U(1)_B$ gauge coupling constant, $Q_B$ the baryonic charge of the DM, while the sum runs over the SM quarks. Particularly relevant in our case are the couplings to first generation quarks. The structure of the baryon current $j_\mu^B$ is determined by Eq. \eqref{eq:currents}. Some comments regarding Eq. \eqref{eq:baryon_current_couplings} are in order. One can understand the couplings in that equation as the low-energy remnant of a $U(1)_B$ ultra-violet (UV) realization. If this is the case, it might be that the UV-complete model already involves a heavy DM candidate. In that sense, the MeV-scale DM particle defines along with the high-scale DM a multi-component DM scenario. Here, however, we assume that the scalar DM is responsible for 100\% of the DM relic density.

A self-contained scenario for fixed-target neutrino experiments involves only $g_B$. Thus, results in this case will be, instead, presented in the $\alpha_B-m_{A^\prime}$ plane ($\alpha_B=g_B^2Q_B^2/ 4\pi$ being the baryon-number structure constant). We remark, however, that for the MeV-scale DM particle to produce the right DM relic density a certain amount of kinetic mixing seems mandatory \cite{Batell:2021snh}. Without kinetic mixing the only possible annihilation mode would be $\text{DM}+\text{DM}\to 4\gamma$, enabled by off-shell $\pi^0$'s and thus phase-space suppressed. Radiatively-induced kinetic mixing will open the lepton channel, which, although suppressed by loop factors, might still suffice to produce the right DM relic abundance. A self-contained framework with $\epsilon^\prime\neq 0$ at the tree-level, but loop-induced, is likely viable. For the phenomenological analysis presented in Sec.~\ref{sec:numerical_approach} we adopt this assumption.
% -----------
% Section
% -----------
\section{Production and detection of DM: Sanford and Wang parametrization }
\label{sec:DMprod}
At spallation neutron sources, a pulsed proton beam (typically 1–3 GeV) strikes a dense target material  such as mercury or tungsten. Unlike most accelerator-based neutrino experiments, which use thin targets to suppress hadronic re-interactions, 
spallation facilities exploit intranuclear cascades to maximize neutron production. The hadronic cascades from collisions against thick targets in these experiments produce mainly charged and neutral pions, with subleading contributions from other  mesons like $\eta^0$ at higher energies\footnote{This, of course, is valid only at this energy regime. In accelerator-based neutrino experiments at Fermilab and in accelerator-based DM facilities (e.g. SHIP) a whole spectrum of mesons is abundantly produced.}.

The positively charged pions decay mainly at rest, producing low-energy neutrinos, while negatively charged pions are absorbed by nuclei via Coulomb interactions (almost entirely at 1 GeV). Hence, their contribution to the neutrino flux is subdominant, although increasing with proton beam energy~\cite{COHERENT:2021yvp}. 
On the other hand, neutral pions decay almost exclusively into two photons, which are absorbed within the target material given their short mean free path in highly dense environments\footnote{For both mercury and tungsten the mean free path for a 50 MeV photons is about 1 cm \cite{NIST-XCOM}.}. The mechanism by which photons are absorbed is mainly pair creation in the electromagnetic field of nuclei, $\gamma+Z\to e^+ + e^- +Z$. Most photons create pairs that are then further reprocessed by the target material either through electron-positron pair annihilation or atomic electron capture. Hence, electromagnetic activity is strongly suppressed in a thick target. This picture, however, is modified in the presence of a dark photon. If $\pi^0 \to \gamma A'$ is allowed through kinetic mixing, the produced dark photon can then decay into DM pairs. Since the dark photon and the DM feebly interact with matter, no process can induce their absorption. Then, the DM particle can escape the target material, and could be detected through nuclear recoil signals as those induced by neutrinos through \cevns. 

The dual capability of spallation neutron sources to measure  \cevns~and search for dark-sector candidates has been demonstrated by COHERENT and CCM~\cite{COHERENT:2021pvd,CCM:2021leg}. Moreover, complementary European and Asian spallation source facilities are now under development, and a dedicated analysis of the reach of forthcoming projects in DM searches seems mandatory. Motivated by this, from now we will focus our analysis on the ESS, J-PARC, and the CSNS proposed experiments.

As we are considering the production of dark photons from neutral pion decays, a key ingredient for DM searches in our study is an accurate determination of the $\pi^0$ flux originated within the target material.
Unlike charged pions, neutral pion production is poorly constrained experimentally, and  systematic uncertainties in hidden sector searches are dominated by the precision with which the $\pi^0$ flux can be predicted. Hence, we adopt two different methods to determine the $\pi^0$ flux and compare the results. First, we rely on \texttt{GEANT4} simulations, and second, we use semi-empirical parametrizations of charged-pion  double differential cross sections, and adapt them to $\pi^0$. 
Regardless of the adopted approach, the calculation pipeline of DM searches follow two main stages that are discussed below:

\begin{enumerate}
\item\underline{DM distributions at production:}\\
A sample of $\pi^0$ four-momenta is generated in the laboratory frame. Then, dark photons are produced in the $\pi^0$ rest frame, boosted back to the lab frame, and subsequently used to generate DM particles. The resulting lab-frame normalized DM energy and angular distributions serve as probability distribution functions (PDFs). These PDFs are then weighted by the branching ratio\footnote{Note that stringent constraints on the $m_{A^\prime}-\epsilon$ parameter space follow from a variety of laboratory searches and astrophysical and cosmological considerations (see e.g. \cite{Caputo:2025avc,Bauer:2018onh}.)} $\text{Br}(\pi^0\to\gamma A')$, and by the $\pi^0$ yield per proton on target (POT). 

\item\underline{DM distributions at detection:}\\
Accounting for detector geometry and fiducial volume, the DM energy distribution at detection is convolved with the DM–nucleus differential cross section to obtain event-rate predictions.
Because of the $\pi^0$ mass and the incoming proton kinetic energies involved,
one expects the $\pi^0$ production to be forward-peaked along the proton beam\footnote{Since $\pi^0$ is a spin zero pseudo-scalar the azimuthal distribution is isotropic.}. As a consequence, the DM angular distribution is also largely collinear. Detectors placed on or near the beam axis therefore maximize acceptance. 
\end{enumerate}

%----------------------------------------
\subsection{Neutral pion PDF's: Analytical versus simulation-based results}
\label{sec:analytical_assessment}
%----------------------------------------
In general, the calculation of $\pi^0$ four-momenta can be carried out in two  different ways. The first one relies on Monte Carlo (MC) transport simulations of the proton–target interaction, for instance with \texttt{GEANT4}~\cite{GEANT4:2002zbu}, a standard toolkit extensively benchmarked by the neutrino and DM experimental communities. Such simulations include not only inclusive $\pi^0$ production in primary $p$–$\mathcal{N}$ collisions ($\mathcal{N}$ denoting the target nucleus), but also secondary production from intranuclear cascades. In this way, subleading processes beyond the simplified picture of pion production in a single hard collision are  incorporated. Although the results from these simulations depend on nuclear models, and therefore carry systematic uncertainties, they provide robust predictions for sensitivity estimates (see, e.g.,~\cite{Baxter:2019mcx}).

An alternative approach relies on phenomenological parametrizations of charged-pion double differential cross sections in the relevant energy range. Because of the expertise required to run \texttt{GEANT4} simulations, parametrizations may offer a more accessible alternative to approximate the $\pi^0$ distributions. Properly adapted, these parametrizations allow the construction of $\pi^0$ kinetic-energy and polar-angle PDFs. While they only capture leading production effects, and cannot account for transport phenomena, they offer a fast and convenient approach for phenomenological analyses, provided the induced uncertainties remain under control. This strategy is well established in neutrino physics~\cite{K2K:2006yov,MiniBooNE:2008hfu}, where rapid estimates of charged-pion (and kaon) yields are needed for flux predictions, and has been used also for sub-GeV DM studies~\cite{deNiverville:2016rqh,Chu:2020ysb}.
Note that different energy regimes require different parametrizations, and that all current parametrizations were originally developed to model neutrino production from charged-pion decays, describing inclusive processes of the type $p+\mathcal{N}\to\pi^\pm+X$. Nevertheless, assuming isospin symmetry, they can be used to infer the $\pi^0$~ PDFs. Up to 1 GeV, the Burman–Smith parametrization~\cite{Burman:1989ds} provides a good description of the results reported by COHERENT~\cite{COHERENT:2021yvp}. Between roughly 1 and 12 GeV, the range relevant for spallation neutron sources, the Sanford and Wang (S \& W)~\cite{SanfordWang1967} parametrization is instead applicable. Given the proton energies involved at the ESS, J-PARC and CSNS (see below), in our present analysis we adopt the latter parametrization.

We now quantify the precision with which the $\pi^0$ kinetic-energy and polar-angle PDFs can be predicted using the S \& W parametrization. We first construct the corresponding PDF and then generate a sample of $5\times10^5$ $(p_i^{\pi^0}, T_{\pi^0})$ pairs via rejection sampling. Taking the \texttt{GEANT4} output as a proxy for the pion phase-space distribution, we compare the parametrized results with those obtained from the full simulation. Details of the simulation setup and benchmarking are discussed in Sec.~\ref{sec:numerical_approach}.

\renewcommand{\arraystretch}{1.7}
\setlength{\tabcolsep}{8pt}
\begin{table}[t]
    \centering
    \begin{tabular}{|c||c|c|c|c|c|c|c|c|c|}\hline
         Case & $c_1$ & $c_2$ &
         $c_3$ & $c_4$ & $c_5$ & $c_6$ &
         $c_7$ & $c_8$ & $c_9\,[\text{GeV}]$ \\\hline\hline
         $\pi^+$ & 220.7 & 1.080 & 1.000  & 1.978 & 1.320 & 5.572
        & 0.0868 & 9.686 & 1.0\\\hline
        $\pi^-$ & 213.7 & 0.9379 & 5.454 & 1.210 & 1.284 & 4.781
        & 0.07338 & 8.329 & 1.0\\\hline
    \end{tabular}
    \caption{Parameters for the S \& W parametrization derived by MiniBooNE \cite{MiniBooNE:2008hfu}. Values in the first (second) raw define the double differential cross section in momentum and solid angle for $\pi^+$ ($\pi^-$). See Eq. \eqref{eq:SW_double_diff_cross_section}.}
    \label{tab:S_W_parametrization_parameter_values}
\end{table}

The S \& W double differential cross section in three-momentum and solid angle is given by
\begin{equation}
    \label{eq:SW_double_diff_cross_section}
    \frac{\d^2\sigma}{\d p\,\d\Omega}(p,\theta) = c_1\,p^{c_2}\,
    \left(1 - \frac{p}{p_B-c_9}\right)\;e^{-\rho(c_a)}\quad
    \left[\frac{\text{mb}}{\text{GeV}\times\text{sr}}\right]\ ,
\end{equation}
where $p_B$ is the beam momentum in GeV. The values for the parameters $c_{1,2,9}$ and those in the argument of the exponential ($a=3,\cdots, 8$) follow from MiniBooNE fits \cite{MiniBooNE:2008hfu}, and are shown in Tab. \ref{tab:S_W_parametrization_parameter_values}. The function $\rho(c_i)$ in Eq. \eqref{eq:SW_double_diff_cross_section} reads
\begin{equation}
    \label{eq:argument_S_W}
    \rho(c_i) = c_3\,\frac{p^{c_4}}{p_B^{c_5}} 
    + c_6\,\theta\,\left(p - c_7\,p_B\,\cos^{c_8}\theta\right)\ .
\end{equation}
As can be seen from Eq. \eqref{eq:SW_double_diff_cross_section}, the $\pi^\pm$ differential cross sections have no dependence on the azimuthal angle, and the $\pi^0$ PDF constructed from these expressions will then inherit this feature. 
The cross sections in~Eq.~\eqref{eq:SW_double_diff_cross_section} only account for charged pion inclusive production. Thus, isospin symmetry of the strong interactions dictates
\begin{equation}
    \label{eq:pi0_DDCS}
    \frac{\d^2\sigma}{\d p\,\d\Omega}=
    \frac{1}{2}
    \left(
    \frac{\d^2\sigma_{\pi^+}}{\d p\,\d\Omega}
    +
    \frac{\d^2\sigma_{\pi^-}}{\d p\,\d\Omega}
    \right)\ .
\end{equation}
Normalization of the individual $\pi^{\pm}$ terms in Eq. \eqref{eq:pi0_DDCS} to their integrated values defines the normalized $\pi^0$ two-dimensional PDF in momentum and solid angle space. However, a more convenient set of variables are the kinetic energy, $T_{\pi^0}$, and $\cos\theta_{\pi^0}$, with $\theta_{\pi^0}$ being the polar angle of the $\pi^0$ measured along the proton beam-line in the laboratory frame. Hence, we can write
\begin{equation}
    \label{eq:pi0_DDCS_T_costh}
    \mathcal{F}_{\pi^0}(T_{\pi^0},\cos\theta_{\pi^0})\equiv
    \frac{\d^2\overline{\sigma}}{\d T_{\pi^0}\,\d\cos\theta_{\pi^0}}
    = 2\,\pi\,\frac{\d^2\overline{\sigma}}{\d p\,\d\Omega}
    \;\frac{T_{\pi^0} + m_{\pi^0}}{\sqrt{T_{\pi^0}^2 + 2\,m_{\pi^0}\,T_{\pi^0}}}\ ,
\end{equation}
where the last factor in Eq. \eqref{eq:pi0_DDCS_T_costh} follows from the Jacobian $\d p_{\pi^0}/\d T_{\pi^0}$, while $\overline{\sigma}_{\pi^0}$ is dimensionless and refers to Eq.~\eqref{eq:pi0_DDCS} after normalization. The next step involves a random sampling of $\mathcal{F}_{\pi^0}$ over the relevant kinetic variables $\{T_{\pi^0}, \theta_{\pi^0}, \varphi_{\pi^0}\}$. The azimuthal angle, $\varphi_{\pi^0}$, is not constrained by the two-dimensional S \& W PDF but enables the construction of the four momentum sample.
\begin{figure}[t]
    \centering
    \includegraphics[scale=0.6]{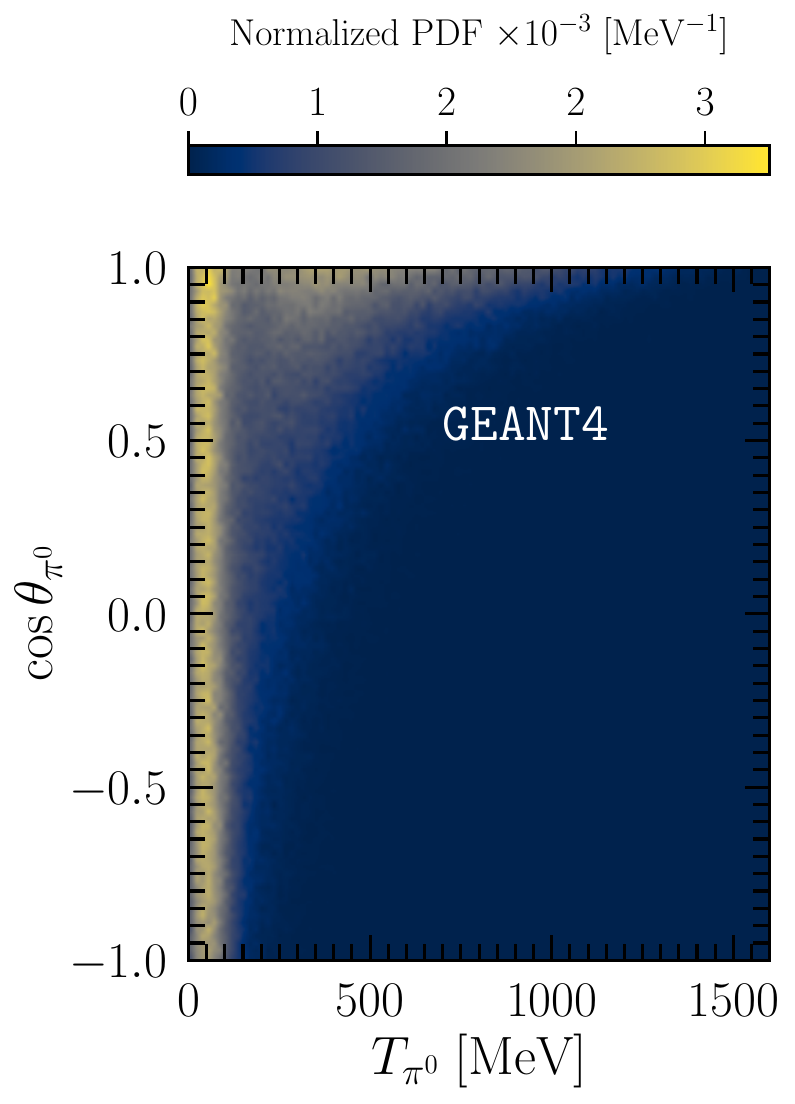}
    \includegraphics[scale=0.6]{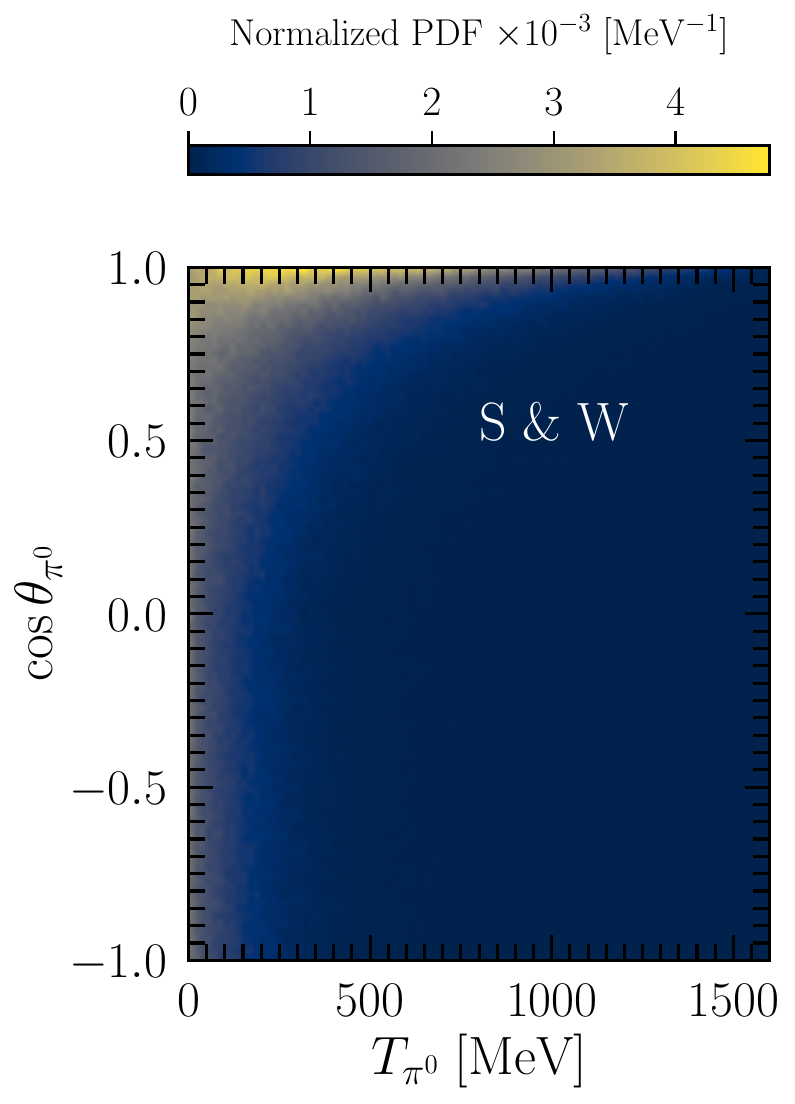}
    \caption{\textbf{Left graph}: Two-dimensional normalized neutral pion PDF in the $T_{\pi^0}-\cos\theta_{\pi^0}$ plane derived from the $\pi^0$ four momentum sample obtained through a \texttt{GEANT4} simulation. The sample accounts for about $3\times 10^5$ pion events and applies for the ESS forecasted experimental configuration (see Tab. \ref{tab:dim_Geant4}) \cite{Baxter:2019mcx}. \textbf{Right graph}: Two-dimensional normalized neutral pion PDF in the $T_{\pi^0}-\cos\theta_{\pi^0}$ plane obtained from the S \& W parametrization. The result has been derived by constructing a four-momentum sample of $5\times 10^5$ events by rejection sampling based on Eq.~\eqref{eq:pi0_DDCS_T_costh}. The difference in the samples size does not affect the conclusion. It has been chosen that way only to improve resolution.}
    \label{fig:2_dim_PDFs_Geant4_analytical_result}
\end{figure}
 Note that, when sampling from Eq.~\eqref{eq:pi0_DDCS_T_costh}, the material target plays no role, only the proton beam energy and the range over which $T_{\pi^0}$ run are relevant. Also, the S \& W parametrization is well defined for $0\leq\theta_{\pi^0}\leq\pi/2$. For values above $\pi/2$ we adopt the following convention \cite{deniverville_PhD}
\begin{equation}
    \label{eq:S_W_for_theta_larger_90}
    \mathcal{F}_{\pi^0}(T_{\pi^0},\cos\theta_{\pi^0}<0) =
    \mathcal{F}_{\pi^0}(T_{\pi^0},\cos\theta_{\pi^0}=0)\ .
\end{equation}

 For the sake of the illustration  we now consider the ESS experimental setup, where the proton beam energy is 2 GeV.  In Fig.~\ref{fig:2_dim_PDFs_Geant4_analytical_result} we show the obtained results for the normalized two-dimensional PDF in the $T_{\pi^0}-\cos\theta_{\pi^0}$ kinematic plane\footnote{Since the proton beam kinetic energies for J-PARC and CSNS differ little from 2 GeV, results for these two cases are  similar and are not displayed.}. The left panel in the figure is obtained through the \texttt{GEANT4} output with $3\times10^5$ events, while the right panel through the S \& W PDF in Eq. \eqref{eq:pi0_DDCS_T_costh} by generating $5\times10^5$ four-momenta via rejection sampling.

Overall, we can see that the S \& W PDF successfully  reproduces the main features of the full simulation: an approximately isotropic component at low kinetic energies, and a forward-peaked component along the beam axis. Nevertheless, two differences are  evident. First, in \texttt{GEANT4} the isotropic population extends to slightly higher kinetic energies. Second, once normalized, the S \& W PDF yields a somewhat larger pion rate.
These discrepancies clearly reflect subleading effects absent in the parametrization. For instance, low-energy (“soft”) pions are likely generated in secondary interactions, such as $\pi^+ + n \to p + \pi^0$ and $\pi^- + p \to n + \pi^0$, which populate the low-$T_{\pi^0}$ region. Such processes, as well as the absorption of charged pions in the thick target, are included in the simulation but not in the analytical parametrization.

\begin{figure}[t!]
    \centering
    \includegraphics[scale=0.6]{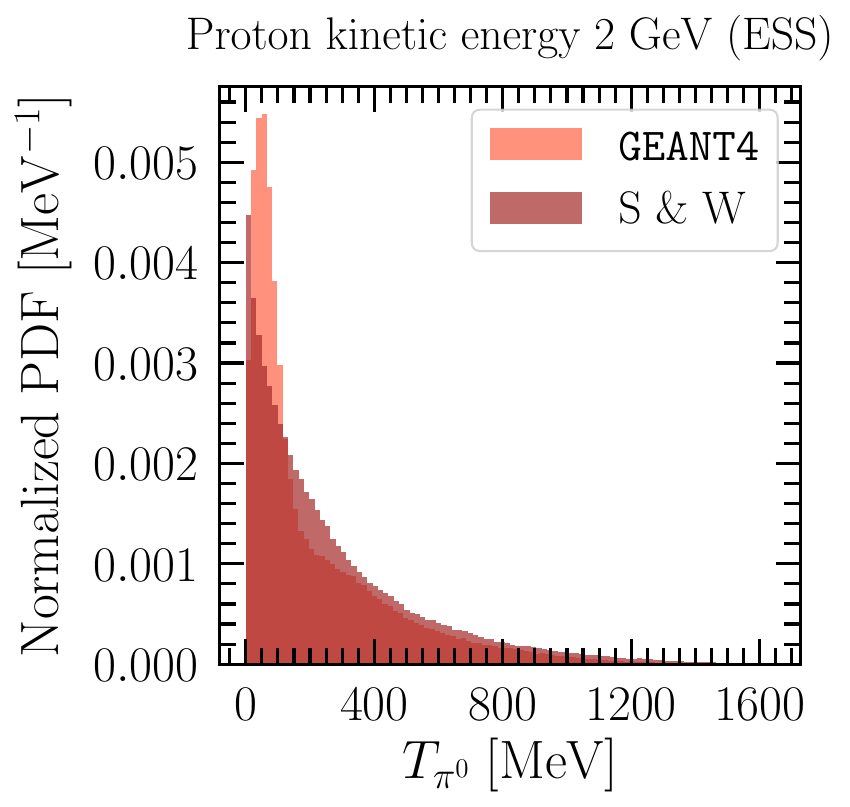}
    \includegraphics[scale=0.6]{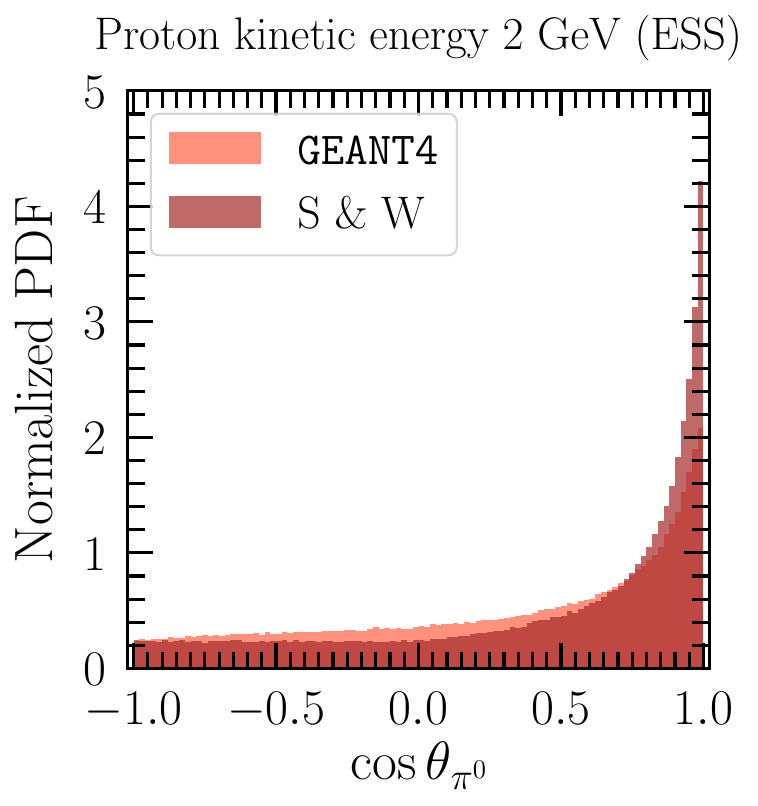}
    \caption{\textbf{Left panel}: One-dimensional, normalized $\pi^0$ kinetic-energy PDFs obtained from the \texttt{GEANT4} simulation and from the Sanford–Wang analytical parametrization. \textbf{Right panel}: Same comparison, but for the $\cos\theta_{\pi^0}$ distribution. In both cases, visible differences arise from subleading physics processes included in \texttt{GEANT4} but not captured by the analytical parametrization.}
    \label{fig:one_dim_PDFs_G4_SW}
\end{figure}

To better quantify these differences, we also construct one-dimensional PDFs for the $\pi^0$ kinetic energy and $\cos\theta_{\pi^0}$. The results are shown in Fig.~\ref{fig:one_dim_PDFs_G4_SW}: the left (right) panel corresponds to the kinetic-energy ($\cos\theta_{\pi^0}$) distribution, from the \texttt{GEANT4} simulation (light red) and from the S \& W parametrization (dark red). We have checked that a good match is found at large binning. However, as the bin width decreases the mismatch gets sizeable. To understand whether this is an \enquote{artifact} we calculated the cumulative distribution function (CDF) and observed that differences persist. Our conclusion is that these differences arise from subleading physics processes underlying neutral pion production.

In summary, the use of the S \& W PDF provides a convenient and
computationally efficient estimate of the neutral pion distributions, reproducing the overall trends observed in the full \texttt{GEANT4} simulation. Small spectral differences are nevertheless present, and most notably the analytical approach yields about 10-20\% more pions.
Assessing the precise impact of these differences on the projected sensitivities would require a careful comparison with the intrinsic uncertainties of the simulation. In a pragmatic spirit, we will run the whole exercise for the ESS  Xe detector using both PDFs, and discuss the overall impact on the final sensitivities in Sec.~\ref{sec:results}.  

%----------------------------------------
\section{Production and detection of DM: \texttt{GEANT4}}
\label{sec:numerical_approach}
%----------------------------------------

\begin{figure}[t]
    \centering
       \includegraphics[width=0.85\textwidth]{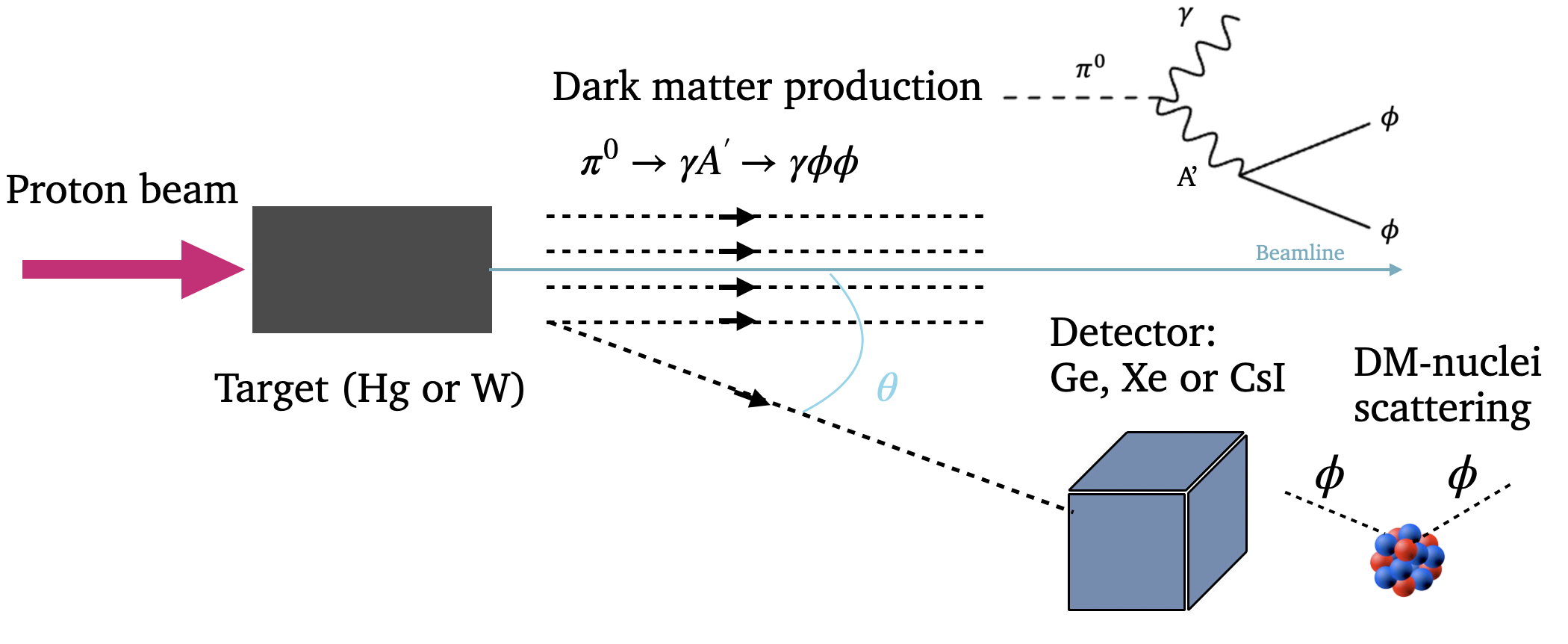}
       \caption{Schematic illustration of the sub-GeV DM production (through a vector portal), and detection pipeline.}
       \label{fig:DM:dist}
\end{figure}

\begin{table}[t]
\centering
\resizebox{\textwidth}{!}{
\renewcommand{\arraystretch}{2}
\begin{tabular}{|c||c|c|c|c|c|c|c|}
\hline
{\bf Facility} & {\bf Target} & $T_\text{p}$ [GeV] & {\bf P [MW]} & {\bf B [m]} & {\bf Detector} & {\bf $a$ [cm]} & {\bf Pulse
Timing}  \\
\hline\hline % First row
ESS~\cite{Baxter:2019mcx} 
& W
& 2 & 5 & 20 
& \makecell[c]{Xe (20 kg), Ge (7 kg) \\ CsI (22.5 kg)}  
& \makecell[c]{50, 11\\ 17} & 14 Hz of 2.8-ms pulses
  \\
\hline % Second row 
J-PARC~\cite{Collar:2025sle} & Hg & 3 & $1 \to 1.3$& 20
& \makecell[c]{Xe (20 kg), Ge (7 kg) \\ CsI (44 kg)} 
& \makecell[c]{50, 11\\ 21.4} & 25 Hz of 0.1-$\mu$s double pulses
 \\
\hline % Third row
CSNS~\cite{CICENNS2025talk}  & W & 1.6 & $0.14 \to 0.5$ & 10.5
& CsI (300 kg) 
& 40.5 & 25 Hz of 0.5-$\mu$s pulses
\\
\hline
\end{tabular}
}
\caption{Parameter specifications for the spallation neutron sources under consideration, including target materials, proton kinetic energies ($T_\text{p}$), power (P), baseline (B), detectors' nuclear target, volume and size, beam pulse repetition rate and width. Detectors are assumed to be cubes with sides given by $a$. For J-PARC and the CSNS, the operation power, P, is expected to increase as shown in the table.}
\label{tab:dim_Geant4}
\end{table}

In this section DM PDFs are fully derived from 
distributions generated with \texttt{GEANT4} simulations. As it has been pointed out, couplings of the new gauge boson to quarks open an additional $\pi^0$ decay channel, sourcing a DM flux. The $\pi^0$ yield is therefore the starting point of the full production–detection chain. The sequence dark photon production, decay into DM  pairs\footnote{Note that although three-body decays are possible, better sensitivities are achieved in the kinematic regime where the narrow-width approximation holds (see discussion below). }, and DM detection, is sketched in Fig.~\ref{fig:DM:dist}. Simulations are prepared for the target materials and proton kinetic energies of the upcoming ESS and the already operating J-PARC and CSNS installations~\cite{Abele:2022iml,Collar:2025sle,CSNS,Su:2023klh}. 

The parameter specifications are listed in Tab. \ref{tab:dim_Geant4}. Spallation-targets are modeled as rectangular cuboids with dimensions: $39.9\times 10.4\times 50\,\text{cm}^3$ for the Spallation Neutron Source at the Oak Ridge National Laboratory (SNS-ORNL)\footnote{We use it to validate our simulation vs the COHERENT results~\cite{COHERENT:2021yvp}.}, $48\times 8\times 15\,\text{cm}^3$ (ESS), $210\times 54\times 19\,\text{cm}^3$ (J-PARC) and $65\times 17\times 7\,\text{cm}^3$ (CSNS)~\cite{AGUILAR201799,JSNS2:2017gzk,WEI20221535}.  Statistical stability of the four-momentum distributions was verified by comparing $10^6$ and $10^7$ POT. Since they remain unchanged, $10^6$ POT is adopted for computational efficiency in the calculation of neutral pion fluxes.

\begin{figure}[t]
    \centering
    \includegraphics[scale=0.56]{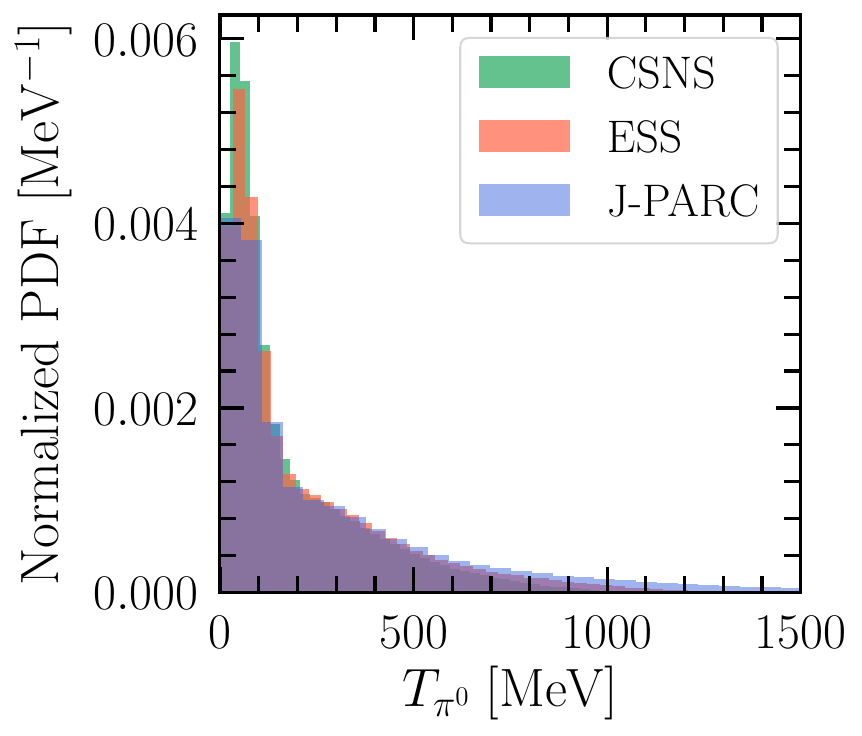}
    \includegraphics[scale=0.56]{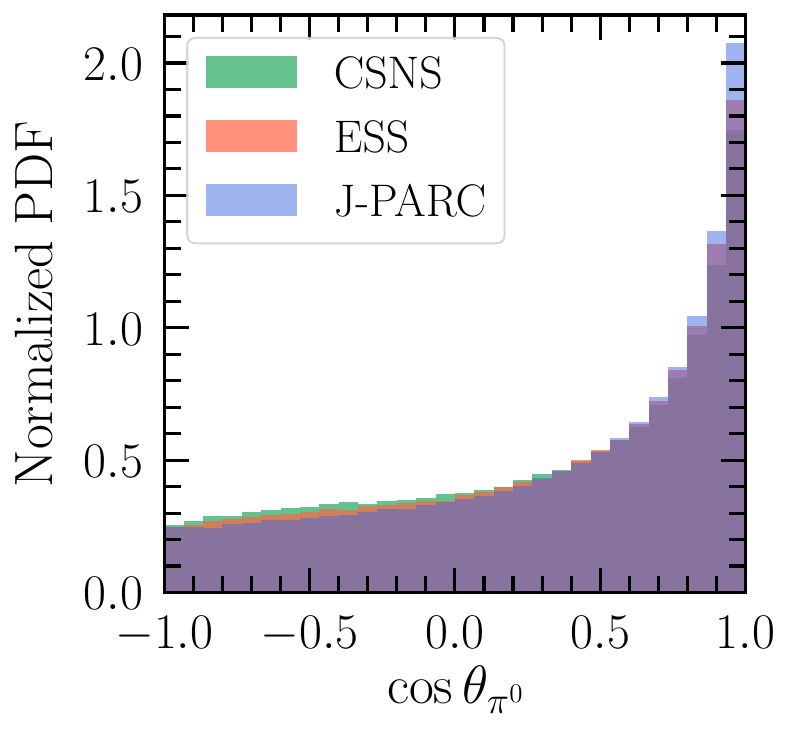}
   \caption{\textbf{Left graph}: Normalized $\pi^0$ kinetic energy PDFs for the CSNS, ESS and J-PARC. \textbf{Right graph}: Same as the left panel, but for the $\pi^0$ polar-angle PDFs, defined with respect to the proton beam direction (see Fig.~\ref{fig:DM:dist}). Results were obtained from \texttt{GEANT4} simulations of proton collisions against a fixed target of tungsten (ESS and CSNS) and mercury (J-PARC) and with proton kinetic energies as given in Tab. \ref{tab:dim_Geant4}. For the simulation the \texttt{FTFP\_BERT} physics list has been used~\cite{Allison:2016lfl}.}
   \label{fig:geant4_pion_PDFs}
\end{figure}

For our simulations we use the \texttt{GEANT4} \texttt{FTFP\_BERT} physics list, a well-validated and extensively used default model which assumes the nucleus as a gas of quasi-free nucleons \cite{Allison:2016lfl}. In the context of neutrino fluxes produced by charged pions decay-at-rest, the COHERENT collaboration has relied on the \texttt{QGSP\_BERT} model instead \cite{Apostolakis:2009egq}. According to their findings, differences below 10 GeV are negligible~\cite{COHERENT:2021yvp}. We validated our simulations by different means. First, we ran a simulation with \texttt{QGSP\_BERT} and found results inline with those reported by COHERENT. Second, a dedicated simulation of SNS-ORNL yields a $\pi^0$ two-dimensional distribution equivalent to that shown in Fig.~\ref{fig:2_dim_PDFs_Geant4_analytical_result}, and matching Ref.~\cite{COHERENT:2021yvp}. Moreover, we computed the $\pi^+$ yield per POT at SNS-ORNL, obtaining 0.078, in good agreement with the reported value 0.08~\cite{COHERENT:2021yvp}. In the case of J-PARC, our results regarding the $\pi^+$ and neutrino yields are consistent with those reported in Ref.~\cite{JSNS2:2017gzk}.
Additional tests with alternative physics lists reveal a small $\eta^0$ component; however, its low yield renders dark photon production through this channel negligible. Our conclusion is that meson activity following from intranuclear cascades is entirely dominated by $\pi^+$ and $\pi^0$ mesons, as expected for sub-GeV to few-GeV proton beams and consistent with previous studies \cite{deNiverville:2016rqh}. The resulting $\pi^0$ kinetic-energy and polar-angle PDFs for our considered experimental setups are shown in Fig.~\ref{fig:geant4_pion_PDFs}.

In Sec. \ref{sec:analytical_assessment} we have provided a brief description of the procedure that follows from a neutral pion four momenta set to obtain the DM flux, we now provide a more detailed discussion. Two kinematic DM regimes can be identified: a three-body decay regime where $2m_\phi<m_{\pi^0}<m_{A^\prime}$, and a regime where the narrow-width approximation is valid and  $2m_\phi < m_{A^\prime} < m_{\pi^0}$. For the latter to be valid, not only the kinematic condition should be satisfied (on-shell dark photon production), but also $\Gamma_\text{Tot}^{A^\prime}\ll m_{A^\prime}$. To see whether this is the case, we recall the dark photon partial decay width
\begin{equation}
    \label{eq:dark_photon_to_gamma_gammaD}
    \Gamma(A^\prime\to \phi + \phi^*)
    =\frac{g_D^2}{48\pi}m_{A^\prime}
    \left(
    1 - 4\frac{m_\phi^2}{m_{A^\prime}^2}
    \right)^{3/2}\, ,
\end{equation}
from which it becomes evident that, provided the kinematic on-shell condition is satisfied, the narrow-width approximation limit holds. Note that this region is actually where better sensitivities are achieved. In the off-shell regime, extra phase space factors further suppress the $\pi^0\to A^\prime+\phi+\phi^*$ mode. Relevant for the DM flux energy distribution is the amount of $\pi^0$ mesons that undergo decays to the \enquote{exotic} mode. The dominant $\pi^0\to\gamma+\gamma$ decay mode has a known partial decay width which reads
\begin{equation}
    \label{eq:pi0_gamma_gamma_decay_width}
    \Gamma(\pi^0\to \gamma + \gamma) = \frac{\alpha^2}{2\times 32\pi^3}\frac{m_{\pi^0}^3}{f_\pi^2} \ ,
\end{equation}
where $f_\pi$ is the neutral pion decay constant and the factor of two comes from the phase space for identical particles. The dark photon counterpart can be inferred from Eq.~\eqref{eq:pi0_gamma_gamma_decay_width} by suppressing the symmetry factor and including a kinematic factor that accounts for the massive dark photon, resulting in
\begin{equation}
    \label{eq:pi0_gamma_gammaD_decay_width}
    \Gamma(\pi^0\to \gamma + A^\prime)
    = 2 \Gamma(\pi^0\to \gamma + \gamma)\epsilon^2
    \left(1 - \frac{m_{A^\prime}^2}{m_{\pi^0}^2}\right)^3\ .
\end{equation}
This result applies to the generic $U(1)^\prime$ dark photon scenario, in which couplings of $A^\prime$ to SM particles arise via kinetic mixing. The corresponding expression for the gauged baryon-number model is given in App.~\ref{app:baryophilic_results}. 
In the narrow-width approximation, the two-body decay $\pi^0 \to \gamma + A^\prime$ is followed by the subsequent two-body decay $A^\prime\to \phi + \phi^*$. Normalizing to the total $\pi^0$ width, the branching ratio for $\pi^0 \to \gamma +\phi+\phi^*$ can be written as \cite{Batell:2009di, DeRomeri:2019kic}
\begin{equation}
    \label{eq:pi0_to_gamma_2DM_BR}
    \mathcal{B}(\pi^0 \to \gamma + \phi + \phi^*) = \mathcal{B}(\pi^0 \to \gamma + \gamma)\times 2 \varepsilon^2  \left(1 - \frac{m_{A'}^2}{m_{\pi^0}^2} \right)^3 \times
\mathcal{B}(A' \to \phi + \phi^*)\ .
\end{equation}
Here $\mathcal{B}(\pi^0 \to \gamma \gamma) = 0.99$ \cite{ParticleDataGroup:2024cfk} and we assume that 
$\mathcal{B}(A^\prime \to \phi \phi^*)=1$. One should emphasize that the latter assumption is never exact. In the $U(1)^\prime$ generic dark photon scenario discussed in Sec. \ref{sec:theory}, the kinetic mixing parameter unavoidably couples $A^\prime$ to $e^+e^-$ pairs. Hence, taking $\mathcal{B}(A' \to \phi + \phi^*)=1$ effectively corresponds to the limit $g_D \gg \varepsilon$ (and similarly in baryon-number models). Even when tree-level lepton couplings are absent, loop-induced decays into $e^+e^-$ are present, implying that a fraction of $A'$ bosons will decay visibly. From Eq.~\eqref{eq:pi0_to_gamma_2DM_BR}, one can see that the fraction of DM particles produced in $\pi^0$ decays is small, even for a moderately large kinetic mixing parameter. This suppression makes high-intensity facilities such as spallation neutron sources particularly well suited for these searches, as their large statistics can compensate for the small branching fraction.

Coming back to our simulation procedure, for every $\pi^0$, we generate dark photons isotropically in its rest frame, with energies fixed by two-body kinematics, and subsequently boost their four-momenta to the laboratory frame. For each facility, the simulated neutral pion four-momenta determine the corresponding Lorentz boost parameters and propagation directions. The same procedure is then applied to the DM particles produced in the decay $A^\prime \to \phi \phi^*$, generating isotropic configurations in the mediator rest frame and boosting them along the $A^\prime$ direction (more details are given in App.~\ref{app:boost}). This MC algorithm is applied to the full $\pi^0$ sample, producing 100 (or 300) DM four-momenta per pion to enhance statistics. From the resulting events, we construct normalized energy and angular PDFs, which are finally rescaled by the number of protons on target, the $\pi^0$ yield per POT (listed in Tab.~\ref{tab:pi0_nu_per_POT_detailed}), and the relevant decay branching ratio of Eq.~\eqref{eq:pi0_to_gamma_2DM_BR} to obtain the DM flux at production.
\begin{table}[t]
\centering
%\resizebox{0.4\textwidth}{!}{
\renewcommand{\arraystretch}{2}
\begin{tabular}{|l||c|c|c|c|c|}
\hline
\bf{Facility} & $N^{\mathrm{POT}}$ & $c_{\pi^0}$ & $c_{\nu_\mu}$ & $c_{\bar\nu_\mu}$  & $c_{\nu_e}$  \\
\hline\hline
ESS   & $2.8 \times 10^{23}$ & 0.31  & 0.15  & 0.15  & 0.15  \\\hline
J-PARC & $3.5 \times 10^{22}$ & 0.54 & 0.24 & 0.24 & 0.24 
\\\hline
CSNS  & $9.8 \times 10^{21}$ & 0.22 & 0.12 & 0.12 & 0.12 \\
\hline
\end{tabular}
%}
\caption{Number of POT per year assuming 208 effective days of operation per year at the ESS and CSNS, and 3600 hours per year at J-PARC. The fractions of particles produced per POT follow from our \texttt{GEANT4} simulations using the FTFP\_BERT physics list.}
\label{tab:pi0_nu_per_POT_detailed}
\end{table}
The differential DM energy ($E_\phi$) and angular ($\theta_\phi$) distributions at production are thus
\begin{align}
    \label{eq:DM_PDF_at_prod}
    \left.
    \dfrac{\d N^\phi}{\d X}\right|^{\rm{prod}} &= N^{\rm{POT}} \times  c_{\pi^0} 
    \times 
    2~\mathcal{B}(\pi^0 \to \gamma + \phi + \phi^*)
    \times
    \left.
    \dfrac{\d N^\phi}{\d X}\right|^{\rm{norm=1}}\; \ .
\end{align}
Here $X = \{E_\phi, \cos\theta_\phi\}$, $N^\text{POT}$ refers to number of POT per year (see Tab.~\ref{tab:pi0_nu_per_POT_detailed}), and a factor of 2 has been included to account for DM pair production. 
\begin{figure}[t!]
    \centering
     \includegraphics[scale=0.56]{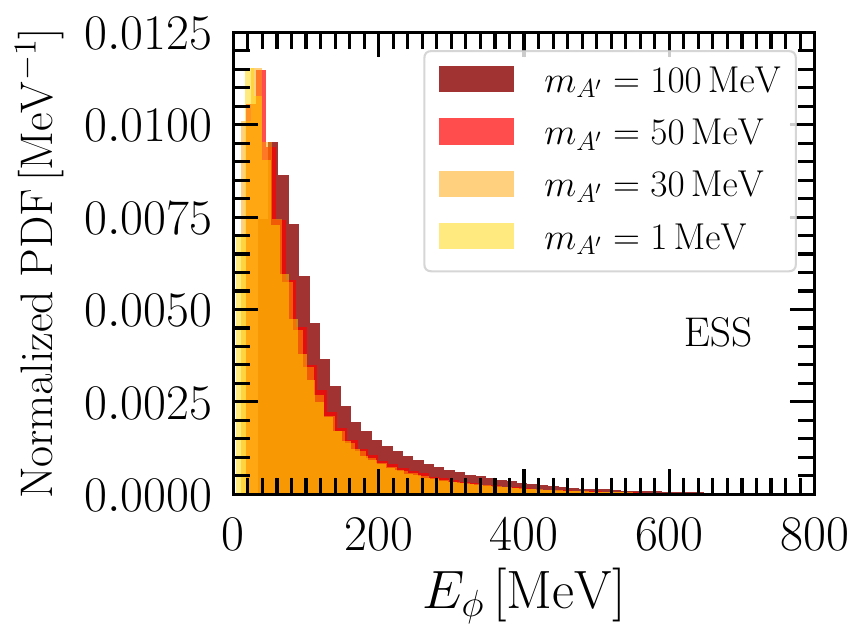}
     \includegraphics[scale=0.56]{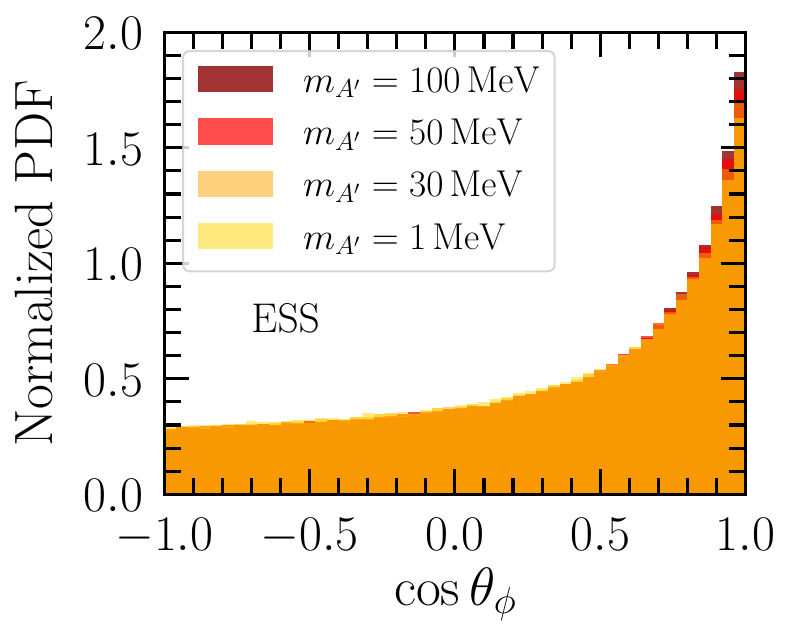}\\
    \includegraphics[scale=0.56]{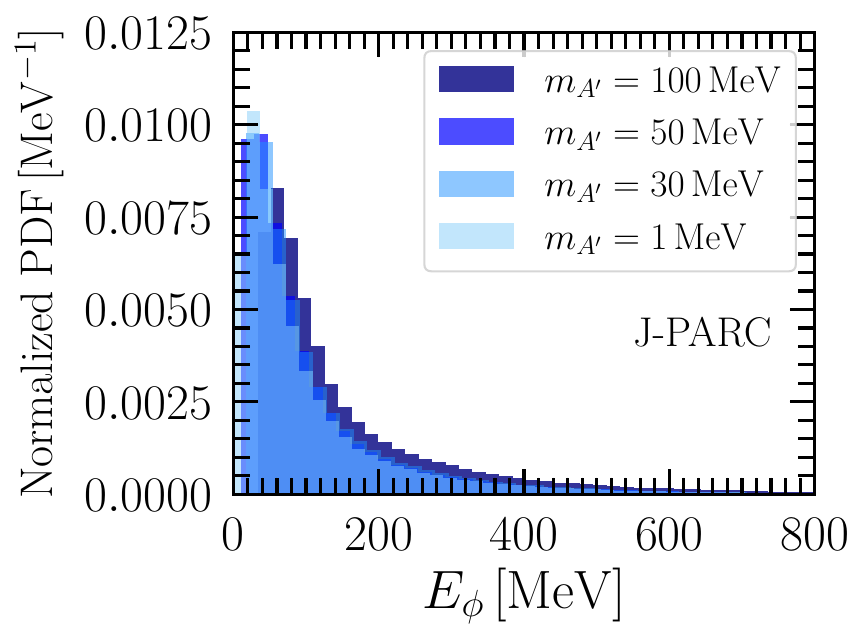}
    \includegraphics[scale=0.56]{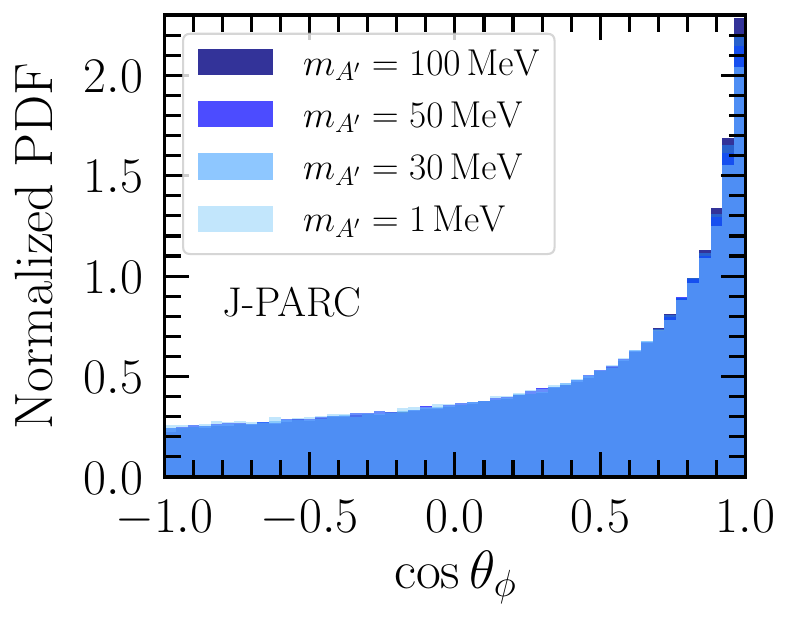}\\
    \includegraphics[scale=0.56]{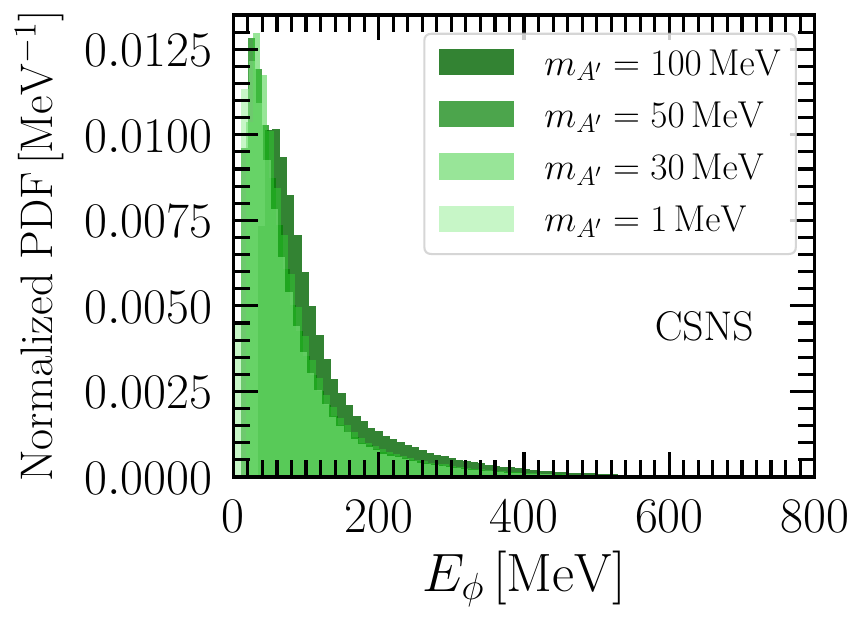}
    \includegraphics[scale=0.56]{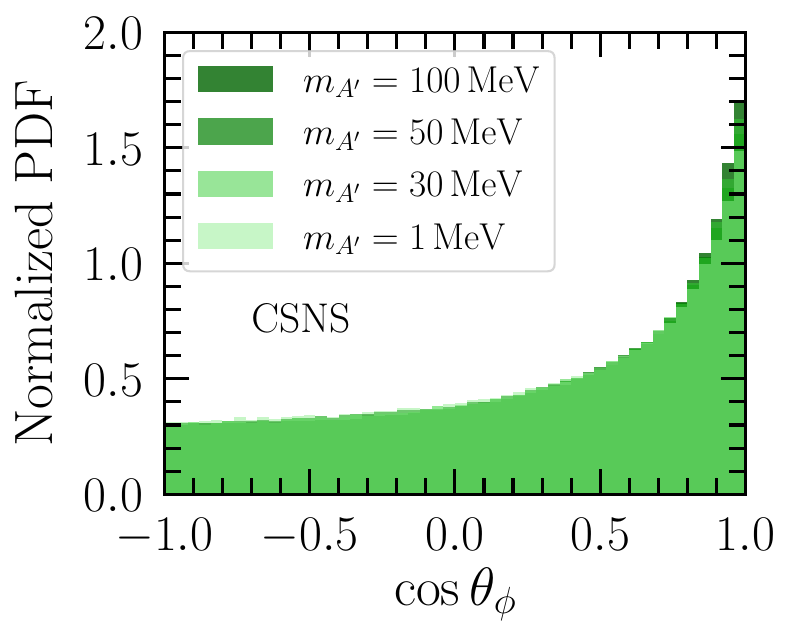} 
    \caption{{\bf Left (right) graphs}: Expected DM total energy (angular) normalized PDFs at production. Results are for ESS (top row), J-PARC (middle row), and CSNS (bottom row). Each PDF is calculated for different dark photon masses, from lighter to heavier (lighter to darker color) assuming $m_{A'} = 3m_{\phi}$. Results follow from \texttt{GEANT4} simulations with $10^6$ POT (see text for further details).}
    \label{fig:ESS:DM:flux}
\end{figure}

Separate distributions are obtained for each choice of DM mass, $m_\phi$, and kinetic mixing parameter.
For illustration, the left (right) panels of Fig.~\ref{fig:ESS:DM:flux} show the DM energy (angular) PDFs at each production site assuming $m_{A'}=3\,m_{\phi}$.
The angular distributions clearly indicate that DM particles are predominantly produced in the forward direction. This feature is expected: the proton beam kinetic energy exceeds the $\pi^0$ mass by roughly one order of magnitude (a factor $\sim 15$-$20$, depending on the facility), so neutral pions are moderately boosted and, as shown in Fig.~\ref{fig:geant4_pion_PDFs}, strongly collimated along
the beam axis. The DM particles inherit this kinematic configuration from their parent $\pi^0$ mesons.
The energy PDFs peak at sub-$100$~MeV values, with a mild shift toward larger $E_\phi$ as the dark photon mass increases. This behavior can be traced back to the intranuclear cascade dynamics, where low-energy secondaries are produced more copiously. As a consequence, lighter DM states are more abundant at low energies, while increasing $m_\phi$ naturally shifts the distribution toward higher total energies.

Once the DM energy spectra at detection are determined, we compute the differential nuclear recoil rate. For a detector with target nuclei of mass number $A$, the convolution of the DM flux with the
DM-nucleus cross section reads
\begin{equation}
    \label{eq:diff:dm:events}
    \dfrac{\d N}{\d E_r} = t_{\mathrm{run}} n_A \langle \ell\rangle  \int_{E_\phi^\mathrm{min}}^{E_\phi^\mathrm{max}}\mathrm{d}E_{\phi} \left.\dfrac{\d N^\phi}{\d E_\phi}\right|^{\rm{det}} \dfrac{\d \sigma_{\phi \mathcal{N}}}{\d E_r}\, ,    
\end{equation}
where $t_{\mathrm{run}}$ refers to data taking time and $n_A = (N_A/m_A^\text{mol})\rho_A$ is the target material number density, with $N_A$ the Avogadros number, $m_A^\text{mol}$ the target material molar mass in mol/kg, and $\rho_A$ the volumetric mass density of the target material. 
We assume cubic detectors of side length $a = \sqrt[3]{\frac{m_\text{det}}{\rho_A}}$, 
with $a$ determined from the detector masses listed in
Tab.~\ref{tab:dim_Geant4}.
The upper integration limit in Eq.~\eqref{eq:diff:dm:events} is set by the kinematic endpoint of the DM spectrum, typically $E_\phi^\text{max}\simeq 600$~MeV. The lower limit follows from kinematics
\begin{equation}
    E_\phi^\text{min}=
\frac{E_r}{2}
+
\frac{\sqrt{m_\mathcal{N}(E_r + 2 m_{\mathcal{N}})\left(m_{\mathcal{N}} E_r + 2 m_\phi^2\right)}}{2m_{\mathcal{N}}} \, ,
\end{equation}
which can be simplified to $E_\phi^\text{min}\simeq \sqrt{m_\mathcal{N} E_r/2}$. The average DM path length inside the fiducial volume is approximated by
\begin{equation}
    \label{eq:DM_FV_path}
    \langle \ell \rangle = a\times \left(1 + \frac{\theta^2}{2}\right)\ ,
\end{equation}
where $\theta=a/(2\times L)$ and $L$ stands for the production-detection baseline (shown in Tab. \ref{tab:dim_Geant4}). The correction factor accounts for the fact that DM particles do not traverse the detector perfectly parallel to its faces when $a/L\ll 1$. Although geometric assumptions introduce uncertainties, they are subdominant compared to those associated with the $\pi^0$ production spectra, which constitute the leading systematic in this class of calculations.

The differential cross section in Eq.~\eqref{eq:diff:dm:events} depends on the energy regime of the incoming DM particles. In the regime relevant for spallation sources, the DM energy distributions fall off rapidly and peak below 100 MeV (see Fig.~\ref{fig:ESS:DM:flux}). This is 
exactly the region where \textit{coherent elastic DM-nucleus scattering} is expected to dominate over electron-DM elastic scattering and other scattering processes. For instance, from 200--300 MeV on, DM-nucleon elastic scattering (the neutral current counterpart of quasi-elastic scattering) will open up. However, at those energies the cross section is still small and the DM population has been almost entirely depleted. Contributions from this process are therefore negligible. The coherent DM-nucleus elastic scattering cross section reads 
\begin{equation}
    \label{eq:cross:scalar:DM}
    \left.\frac{\d \sigma_{\phi \mathcal{N}}}{\d E_r}\right|^{\rm{dark~photon}}
    =
    \frac{g_D^2}{2\pi}
    \frac{ m_{\mathcal{N}} E_\phi^2}
    {\left(E_\phi^2 - m_\phi^2\right)
    \left(2 m_{\mathcal{N}} E_r + m_{A'}^2\right)^2} 
    \left(
1
- \frac{m_\phi^2 E_r}{2 m_{\mathcal{N}} E_\phi^2}
- \frac{E_r}{E_\phi}
\right)
\epsilon^2\,Z^2\,e^2\mathcal{F}_Z^2(E_r)\, .
%\label{eq:scalar_DM_xsec}
\end{equation}
As discussed in Sec. \ref{sec:theory}, the dark photon couples to quarks through electric millicharge. Thus, to nuclei at zero momentum transfer it couples with strength $\epsilon\,Z\,e$ (maximal nuclear isospin-breaking couplings). At non-zero  momentum transfer the electric charge cloud becomes smaller and so the coupling diminishes. Such effect, as in the neutrino case, is accounted for through the introduction of an electric charge nuclear form factor which defines an \enquote{effective} energy-dependent coupling $\epsilon\,Z\,e\,\mathcal{F}_Z(E_r)$. Regarding the charge nuclear form factor $\mathcal{F}_Z$, here we adopt the Helm parametrization~\cite{Helm:1956zz}. In baryophilic scenarios there is no such millicharge coupling. The gauge boson couples to baryon number, thus to nuclei proportionally to total baryon number charge $A$ (see App. \ref{app:baryophilic_results}).

\begin{figure}[t]
    \centering
    \includegraphics[width=0.49\textwidth]{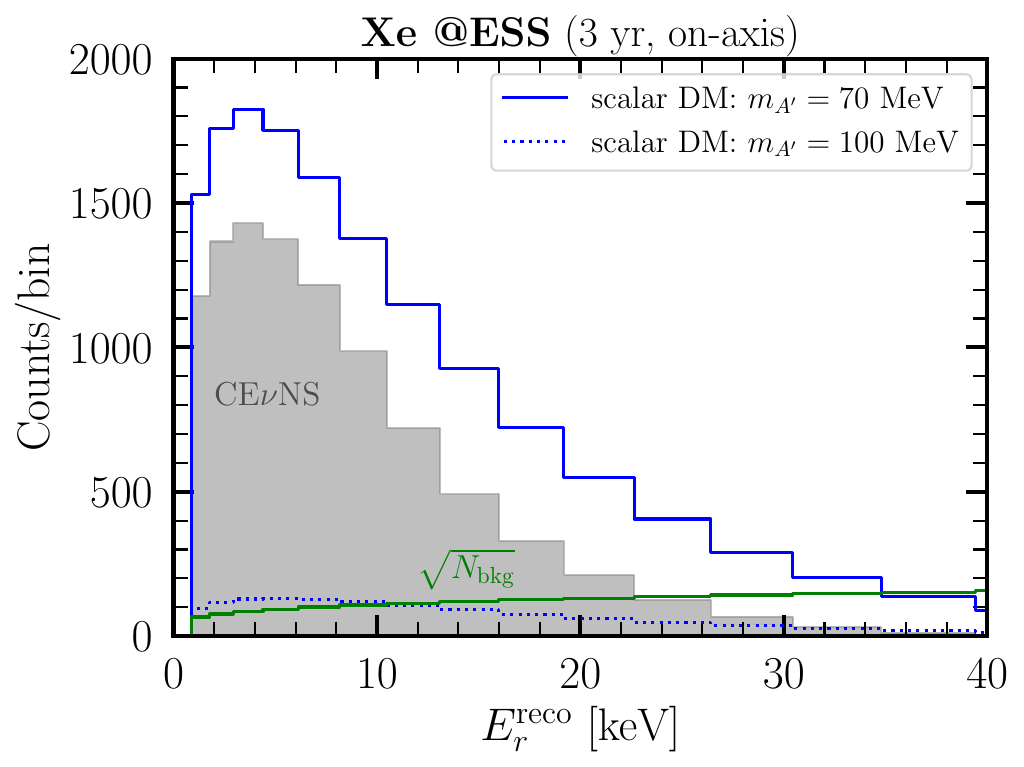}
    \includegraphics[width=0.47\textwidth]{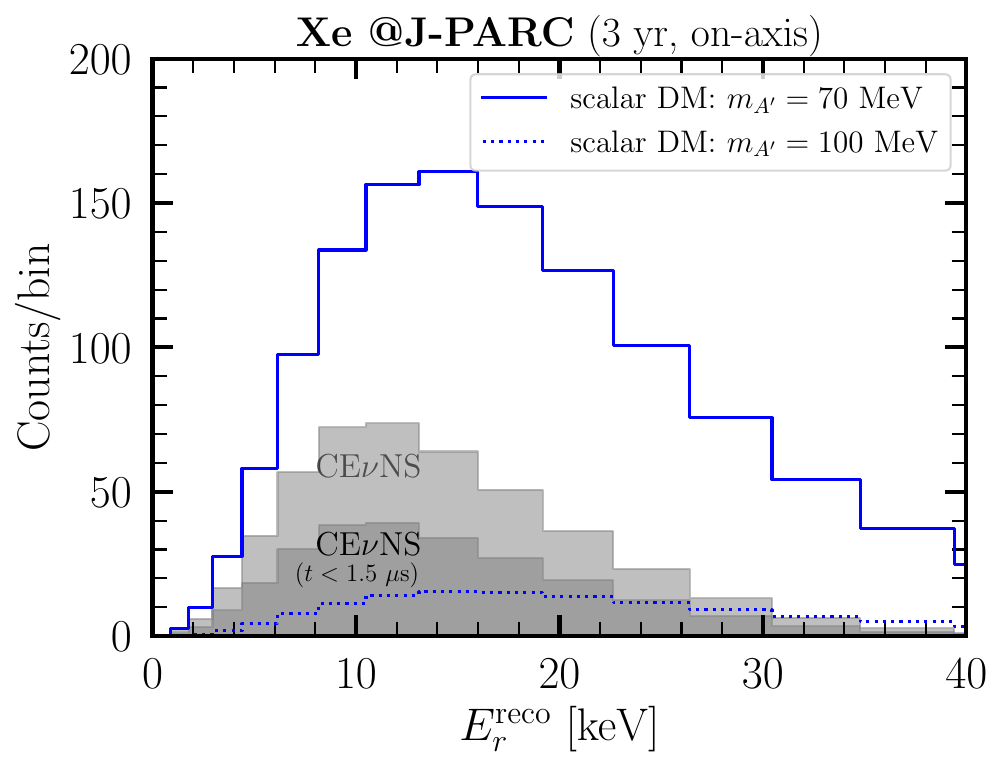}
        \includegraphics[width=0.48\textwidth]{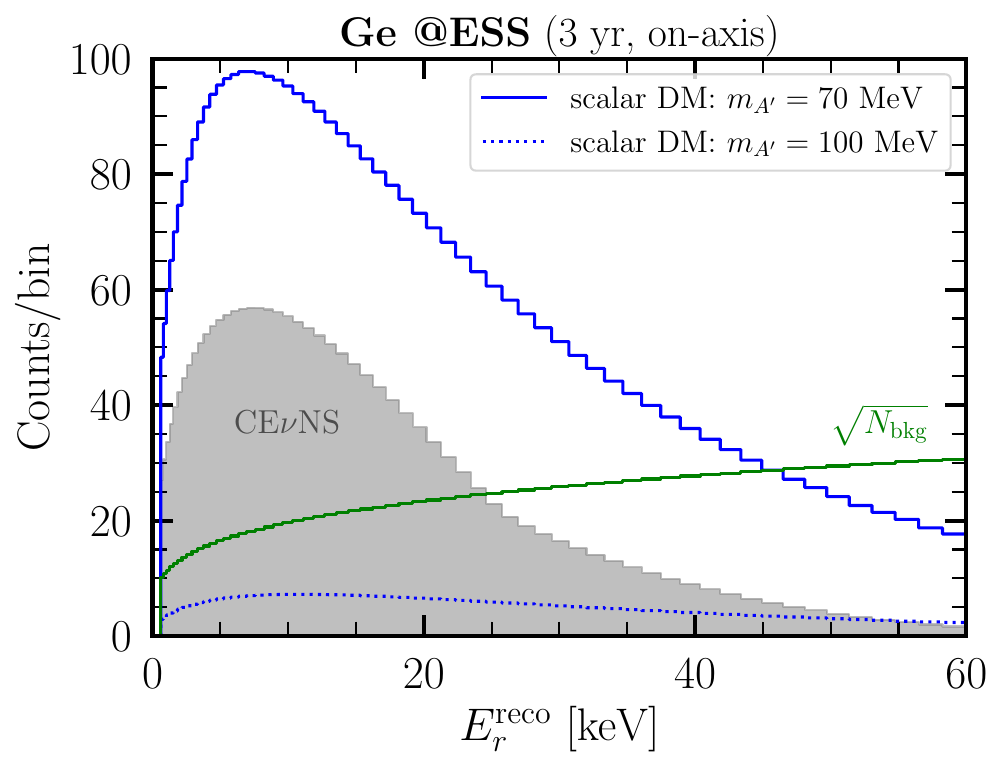}
    \includegraphics[width=0.47\textwidth]{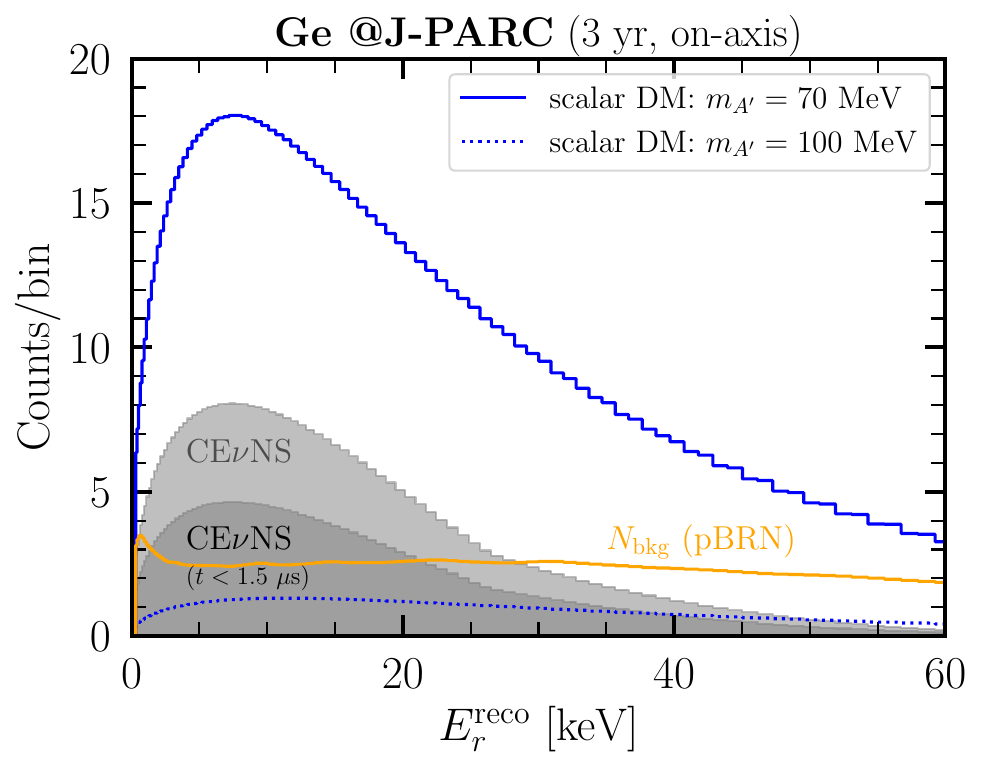}
    \caption{\textbf{Top panels:} Comparison of \cevns~(shaded gray), DM (blue histogram) and steady-state-background (SSB, green histogram)  spectra for Xe targets at the ESS  (left) and J-PARC (right). \textbf{Bottom panels:} Same as top panels but for for Ge targets. For the Ge detector at J-PARC, the relevant pBRN background is shown (orange histogram).  In all cases, the detector is assumed to be on-axis and we present the spectra for two representative DM masses in the simple dark photon scenario. We fix $m_{A^\prime} = 3\,m_{\phi}$, $\alpha_D = 0.5$, and $\epsilon = 2 \times 10^{-4}$.}
       \label{fig:spectra_comparison_Xe_Ge}
\end{figure}
The differential number of DM particles reaching the detector can be obtained by accounting for its finite polar and azimuthal angular acceptance, namely
\begin{equation}
    \label{eq:DM_at_detection}
    \left.\dfrac{\d N^\phi}{\d E_\phi}\right|^{\mathrm{det}} 
    = \left( \dfrac{N^\phi_{\mathrm{cut}}}{N^\phi_{\mathrm{prod}}} \right)
      \left( \dfrac{\Delta\varphi}{2\pi} \right)
      \left.\dfrac{\d N^\phi}{\d E_\phi}\right|^{\mathrm{prod}} \, .
\end{equation}
Here $N_\text{prod}$ follows from Eq.~\eqref{eq:DM_PDF_at_prod}, while $N_\text{cut}$ corresponds to the subset of produced DM
particles whose laboratory polar angle lies within the detector
acceptance for a given baseline.
For example, for a detector located at $90^\circ$ off-axis this means $\theta_\phi^\text{Lab}\subset [\pi-\Delta\theta_\text{det},\pi+\Delta\theta_\text{det}]/2$. The angle $\theta_\text{det}$ is \enquote{measured} from the proton beam direction (along $\widehat{z}$) to the center of the detector. The half-aperture angle is determined
geometrically from the detector size and baseline,
$\Delta\theta_{\mathrm{det}} \simeq a/(2L)$ for $a/L \ll 1$.
The second factor accounts for the finite azimuthal coverage.
Assuming azimuthal symmetry of the DM flux
(inherited from the $\pi^0$ distribution),
the effective azimuthal acceptance is
\begin{equation}
    \label{eq:azimuthal_det_resolution}
    \Delta\varphi =
    \begin{cases}
    \dfrac{a}{L \sin\theta_{\mathrm{det}}} \, , & \text{off-axis} \, , \\[8pt]
    2\pi \, , & \text{on-axis} \, ,
    \end{cases}
\end{equation}
where the off-axis expression follows from the projection of the detector transverse size onto the azimuthal direction.
This factor reduces to full azimuthal coverage in the on-axis configuration. For our analyses, we consider two benchmark detector deployment configurations:
(i) On-axis location, corresponding to the direction of maximal DM particles;
(ii) off-axis location at $\theta_{\rm det} = 90^\circ$, following the ESS proposal~\cite{Baxter:2019mcx}.

\begin{figure}[t]
    \centering
        \includegraphics[width=\textwidth]{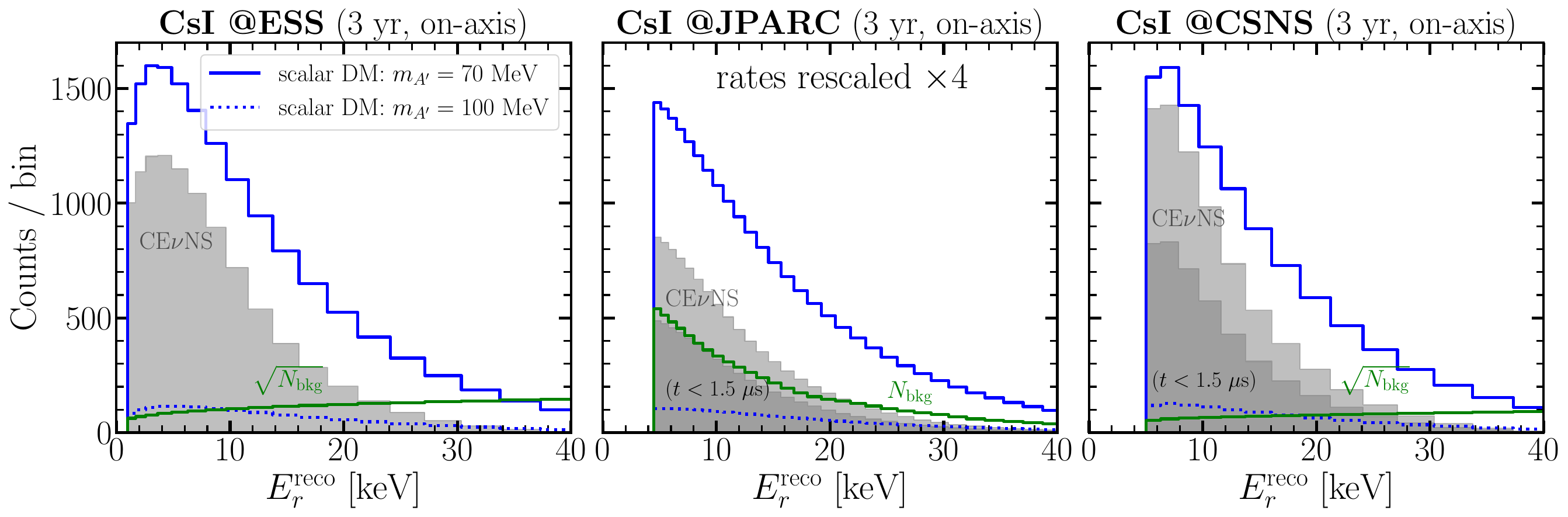}
       \caption{Comparison of \cevns~(shaded gray), DM (blue histogram) and steady-state-background (SSB, green histogram) spectra for the CsI targets at the ESS (\textbf{left panel}), J-PARC  (\textbf{middle panel}, rescaled by $\times 4$), and CSNS (\textbf{right panel}). In all cases the detector is assumed to be on axis and we present the spectra for two representative DM masses in the simple dark photon scenario. We fix $m_{A^\prime} = 3\,m_{\phi}$, $\alpha_D = 0.5$, and $\epsilon = 2 \times 10^{-4}$.}
       \label{fig:spectra_comparison_CsI}
       \end{figure}

The predicted nuclear-recoil DM signals are calculated from Eq.~\eqref{eq:diff:dm:events} assuming a flat 80\% recoil acceptance and accounting for detector-specific resolution effects following the prescriptions of Refs.~\cite{Baxter:2019mcx,Collar:2025sle}. Moreover, for the High-Pressure (HP) gaseous time-projection chamber (TPC) Xe detector proposed at J-PARC, we include an additional efficiency correction to account for the probability of detecting at least two primary scintillation photons, following Ref.~\cite{Collar:2025sle}. More details are given in the next subsection. For the sake of illustration, we show in Fig.~\ref{fig:spectra_comparison_Xe_Ge} the DM event rates for Xe detectors at the ESS (top row, left panel) and at J-PARC (top row, right panel), and for Ge detectors at the same facilities (bottom row). In all cases, we fix the detector to be on axis and we present event rates for two representative DM masses, $m_\phi =70, 100\,\text{MeV}$ (solid and dotted lines, respectively). We focus on the simple dark photon scenario and fix $m_{A^\prime} = 3\,m_{\phi}$, $\alpha_D = 0.5$, and $\epsilon = 2.3 \times 10^{-4}$. Similarly, we present in Fig.~\ref{fig:spectra_comparison_CsI} the DM event rates for CsI detectors at the three facilities.
For comparison, we also show the expected \cevns~event rates (gray-shaded histograms) as well as other background events: mainly steady-state backgrounds (SSB) (green solid histograms), and a beam-related prompt neutron background (orange histogram), dubbed pBRN, which is relevant in the case of Ge target at J-PARC. With increasing dark photon mass, the available phase space decreases and so do the event rates, as clearly visible in our results.
Finally, note that for the J-PARC and CSNS analyses we account for timing information, whose details will be given in the next subsection. We illustrate in Figs.~\ref{fig:spectra_comparison_Xe_Ge} and~\ref{fig:spectra_comparison_CsI} how the inclusion of a cut on the arrival time of neutrino events allows to reduce the \cevns~background rate, displayed as different gray shades in some of the panels.
% -----------------
% Section
% ----------------
\subsection{Statistical Analysis}
\label{sec:analysis}
To assess the sensitivity of \cevns\ detectors to DM signals, once the predicted event rates are obtained, we perform a $\chi^2$ analysis by minimizing the Poissonian least-squares function
\begin{equation}
  \chi^2 = 2\sum_{i}\left [ N^{\textrm{th}}_{i} - N^{\textrm{exp}}_{i} + N^{\textrm{exp}}_{i}\ln\left ( \frac{N^{\textrm{exp}}_{i}}{N^{\textrm{th}}_{i}} \right )\right ] +  \left(\frac{\alpha}{\sigma_{\alpha}}\right)^2 + \left(\frac{\beta_1}{\sigma_{\beta_1}}\right)^2 + \left(\frac{\beta_2}{\sigma_{\beta_2}}\right)^2  ,
    \label{eq:chi}
\end{equation}
where $N_i^{\textrm{th}}$ denotes the expected number of events in the $i$-th energy bin and is given by

\begin{equation}
    N_i^{\textrm{th}} = N_i^{\textrm{DM}}(1+\alpha) + N_i^{\textrm{CEvNS}}(1+\beta_1) + N_i^{\textrm{SSB/pBRN}}(1+\beta_2).
    \label{eq:Nth}
\end{equation}

The integrated DM and \cevns~spectra denoted by $N_i^\text{DM}$ and $N_i^\mathrm{CE\nu NS}$ in Eq.~\eqref{eq:Nth}, are computed in bins of reconstructed recoil energy $E_r^\mathrm{reco}$, and they are calculated as
\begin{equation}
    \frac{\d N}{\d E_r^\mathrm{reco} } = \int_{E_r^\mathrm{min}}^{E_r^\mathrm{max}} \frac{\d N}{\d E_r} \mathcal{G}(E_r^\text{reco}, E_r) \, \d E_r \, ,
    \label{eq:diff_rate_reco}
\end{equation}
where $E_r^\mathrm{min}=0$, while $E_r^\mathrm{max}$ is obtained from the kinematics corresponding to the maximum neutrino or DM energy. Following Refs.~\cite{Baxter:2019mcx,Collar:2025sle}, the binning is chosen such that each bin width is twice the resolution at the bin centre. The adopted recoil-energy thresholds are $E_r^\mathrm{thres}=(0.6, 0.9, 1.0)$~keV for (Ge, Xe and CsI) at ESS~\cite{Baxter:2019mcx}, $E_r^\mathrm{thres}=(0.3, 0.9, 4.5)$~keV for (Ge, Xe and CsI) at  J-PARC~\cite{Collar:2025sle} and $E_r^\mathrm{thres}=5$~keV for CsI at the CSNS~\cite{Su:2023klh}.
Here,  $\mathcal{G}(E_r^\text{reco}, E_r)$ is a normalized Gaussian smearing function that is written as
\begin{equation}
    \mathcal{G}(E_r^\mathrm{reco},E_r) = \frac{1}{\sqrt{2\pi}\sigma}
\exp\left[
-\frac{(E_r^\mathrm{reco}-E_r)^2}{2\sigma^2}
\right] \, ,
\end{equation}
where the resolution power depends on the recoil energy according to $\sigma(E_r)= \sigma_0 \sqrt{E_r/E_r^\mathrm{thres}}$. The values of $\sigma_0$ and $E_r^\mathrm{thres}$ for each nuclear target  can be readily taken from Table~I in Ref.~\cite{Baxter:2019mcx} for the ESS and from Table~II in Ref.~\cite{Collar:2025sle} for J-PARC, noting that the parameters reported in these two references are identical. To our knowledge, this information is not available for the CsI detector at CSNS; therefore, we adopt the same resolution parameters as for the other two facilities in this case.

Eq. \eqref{eq:Nth} also includes the typical backgrounds at a spallation neutron source. The dominant background originates from SSB, denoted by $N_i^{\textrm{SSB}}$, which is primarily induced by cosmic rays. The latter is relevant for all the detectors considered in the present study, with the only exception being the case of a Ge target at J-PARC where the most important background component is instead the pBRN. Furthermore, we neglect the backgrounds for the case of the Xe time-projection chamber at J-PARC~\cite{Collar:2025sle}. We estimate the expected background rates following the prescriptions of Refs.~\cite{Baxter:2019mcx,Collar:2025sle,Su:2023klh} for the ESS, J-PARC and CSNS, respectively, which are shown as green histograms in the panels of Figs.~\ref{fig:spectra_comparison_Xe_Ge} and~\ref{fig:spectra_comparison_CsI}. Given the sizeable magnitude of this contribution, in some cases we display $\sqrt{N_i^{\textrm{SSB}}}$ in the figure for improved readability. Unless otherwise mentioned, we neglect other subdominant backgrounds like pBRN and neutrino-induced neutron, barring the Ge target at J-PARC where, as already mentioned, pBRN is the dominant source of background.

Since our analysis is focused on DM searches, another relevant background in all detectors arises from \cevns\ interactions. 
Their corresponding expected rates can be computed using an expression analogous to Eq.~\eqref{eq:diff:dm:events}, by employing the neutrino flux produced at spallation sources~\cite{COHERENT:2021yvp} together with the SM \cevns\ cross section~\cite{PhysRevD.9.1389}
\begin{equation}
\frac{\d \sigma_{\nu \mathcal{N}}}{\d E_r}
= \frac{G_F^2 m_\mathcal{N}}{\pi}
\left[ Z F_Z(q^2) g_V^p + N F_N(q^2) g_V^n \right]^2
\left(1 - \frac{m_\mathcal{N} E_r}{2 E_\nu^2} \right),
\end{equation}
where the vector couplings are $g_V^p = \tfrac{1}{2} - 2\sin^2\theta_W$ and $g_V^n = -\tfrac{1}{2}$. We have checked that our predicted \cevns~events are in good agreement with the expectations given in Refs.~\cite{Baxter:2019mcx,Chatterjee:2022mmu,Collar:2025sle,Su:2023klh}, when adopting the neutrino yield values reported in those works. However, for consistency, in the actual calculations performed in this study we use the pion and neutrino yields obtained with our \texttt{GEANT4} simulations and summarized in Tab.~\ref{tab:pi0_nu_per_POT_detailed}. 
Going back to Eq.~\eqref{eq:chi}, the quantity $N_i^{\textrm{exp}}$ corresponds to the experimentally measured event rate. Since we are assessing the sensitivity of future facilities, we assume this to be given by the expected SM contribution, namely the sum of the \cevns\ signal and the SSB background (prompt neutrons in the case of Ge at J-PARC). The quantities $\alpha$, $\beta_1$, and $\beta_2$ are nuisance parameters associated with the normalization uncertainties of the DM signal, \cevns, and SSB (or pBRN in the case of Ge at J-PARC) contributions, respectively, with uncertainties $\sigma_\alpha = 10\%$, $\sigma_{\beta_1} = 10\%$, and $\sigma_{\beta_2} = 1\%$ ($\sigma_{\beta_2} = 25\%$ in the case of pBRN).

A combination of energy and timing cuts has been shown to be highly effective in suppressing the SM \cevns~background at COHERENT~\cite{Dutta:2019nbn,Dutta:2020vop,COHERENT:2021pvd}. Motivated by this statement, we include timing cuts in our analysis. At ESS, timing information is not available due to the much longer beam spills of approximately 2.8~ms~\cite{Baxter:2019mcx}. Indeed, a key difference between SNS-ORNL and ESS lies in their proton beam time structure: the former operates at 60~Hz with $\sim$1~$\mu$s-wide POT spills, while the ESS delivers 14~Hz spills lasting 2.8~ms. In contrast, in the cases of J-PARC and CSNS, timing information is available because of the shorter pulses; at J-PARC two 0.1 $\mu$s-wide pulses separated by 0.44 $\mu$s with a repetition rate of 25 Hz \cite{JSNS2:2017gzk}, while at CSNS 0.5 $\mu$s-wide pulses with a rate of 25 Hz \cite{CICENNS2025talk}.
Hence, both at J-PARC and CSNS timing information can be used to enhance signal-to-background discrimination. Although we do not perform a full two-dimensional (energy-time) statistical analysis, we implement a timing cut by rejecting \cevns~events arriving at $t > 1.5\,\mu$s. This choice is rather conservative: our simulations indicate that sub-GeV DM particles are more prompt and typically reach the detector at $t \lesssim 1\,\mu\text{s}$, in agreement with the findings of Ref.~\cite{Dutta:2020vop}. At J-PARC, the double-pulse time structure of the proton beam results in two distinct peaks of DM-induced events within a $\sim1\,\mu\text{s}$ window. We model this feature following the prescriptions of Ref.~\cite{Collar:2025sle}.

In more detail, for both J-PARC and CSNS, we include timing information as follows. The double differential event rate in terms of reconstructed nuclear recoil energy $E_r^\mathrm{reco}$ and reconstructed time $t_\mathrm{reco}$ can be cast in the form
\begin{equation}
       \frac{\d^2 N_i}{\d t_\mathrm{reco} \, \d E_r^\mathrm{reco}} =\sum_{x=\nu_e, \nu_\mu, \bar{\nu}_\mu} \int \frac{\d N}{\d E_r^\mathrm{reco}} \frac{\d \mathcal{P}_x(t)}{\d t} \mathcal{K}(t_\mathrm{reco},t)\, \d t \, ,
\end{equation}
where $\d \mathcal{P}_x / \d t$ are the time distributions of the different incoming neutrinos $x=\nu_e, \nu_\mu, \bar{\nu}_\mu$, normalized to one, which also incorporate information regarding the proton beam time profile of J-PARC and CSNS. The latter read
\begin{equation}
\frac{\d \mathcal{P}_{\nu_\mu}(t)}{\d t}
=
\int_0^{t}
\frac{\d p}{\d t_p}
\,
\frac{e^{-(t-t_p)/\tau_\pi}}{\tau_\pi} \, \d t_p\, ,
\end{equation}
 \begin{equation}
\frac{\d \mathcal{P}_{\nu_e, \bar{\nu}_\mu}(t)}{\d t} 
=
\int_0^{t}
\frac{\d p}{\d t_p}
\,
\frac{
e^{-(t-t_p)/\tau_\mu}
-
e^{-(t-t_p)/\tau_\pi}
}{
\tau_\mu - \tau_\pi
}\, \d t_p\, ,
\end{equation}
where $\tau_\pi = 0.026033~\mathrm{\mu s}$ and $\tau_\mu = 2.2~\mathrm{\mu s}$ denote the pion and muon lifetimes and $t$ denotes the arrival time (in units of $\mathrm{\mu s}$). The proton pulse, for the case of J-PARC reads~\cite{Collar:2025sle}
\begin{equation}
\frac{\d p}{\d t}(t) =
\begin{cases}
0, & t < 0\, , \\
5~\mathrm{\mu s}^{-1},   & 0 < t < 0.1~\mathrm{\mu s}\, , \\
0, & 0.1 < t < 0.54~\mathrm{\mu s}\, , \\
5~\mathrm{\mu s}^{-1},   & 0.54 < t < 0.64~\mathrm{\mu s}\, , \\
0, & \text{otherwise}\, .
\end{cases}
\end{equation}
In the case of CSNS we have instead adopted the time profile of the proton beam reported by the COHERENT Collaboration~\cite{COHERENT:2018imc,COHERENT:2020ybo},
\begin{equation}
 \frac{\d p}{\d t}(t) =
\begin{cases}
0\, , & t < 0\, , \\
\exp \left[-\frac{1}{2} \left(\frac{t}{ \sigma_\mathrm{POT}}\right)^2 \right]\, , &t>0\,  ,
\end{cases}   
\end{equation}
where $\sigma_\mathrm{POT}=\frac{\mathrm{FWHM_{POT}}}{\sqrt{8 \ln(2)}}$, with a full width at half maximum (FWHM) value $\mathrm{FWHM_{POT}}=0.5~\mathrm{\mu s}$.
Finally, the time reconstruction of the event spectra is performed via the smearing function $\mathcal{K}(t_\mathrm{reco},t)$ that is taken to be a normalized Gaussian
\begin{equation}
    \mathcal{K}(t_\mathrm{reco},t) =
\frac{1}{\sqrt{2\pi}\sigma_t}
\exp\left[
-\frac{(t_\mathrm{reco}-t)^2}{2\sigma_t^2}
\right] \, ,
\end{equation}
with $\sigma_t$ being the resolution of the different proposed detectors. Following Ref.~\cite{Collar:2025sle} we take $\sigma_t = 85~\mathrm{n s}$ for Ge and $\sigma_t = 100~\mathrm{n s}$ for Xe at J-PARC. For the CsI detector, we use a lognormal smearing with $\sigma_t = 0.7234~\mathrm{\ln (ns)}$, which is also adopted for the case of CsI detector at CSNS due to the absence of relevant information.

Wherever applicable, timing information has important implications for the SSB background as well. Since SSB is expected to be rather flat in time, the timing cut we applied in our calculations corresponds to a simple rescaling of the overall SSB by a factor of 1.5/6, by noting that the neutrino time profiles are approximately 6 $\mathrm{\mu s}$-wide. Finally, we mention that the whole pBRN component falls well within the $1.5~\mathrm{\mu s}$ time window and hence no timing cut is applicable in this case. 
% -------------
% Section
% -------------
\section{Results}
\label{sec:results}
\begin{figure}[t]
    \centering
        \includegraphics[width=0.49\textwidth]{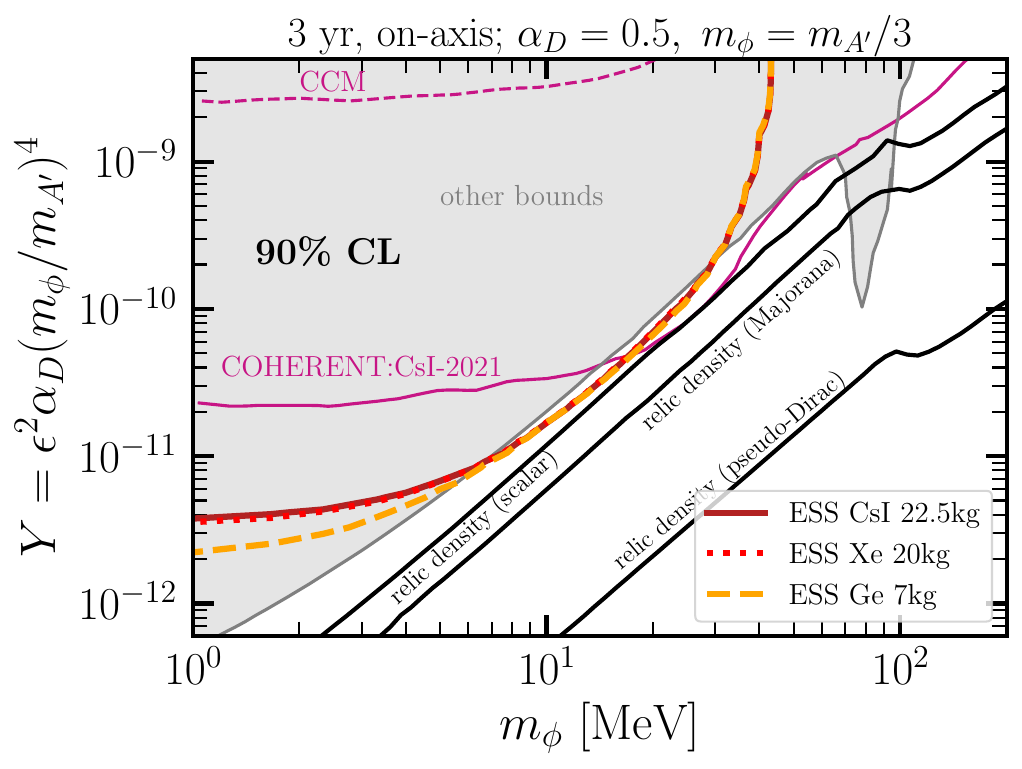}
        \includegraphics[width=0.49\textwidth]{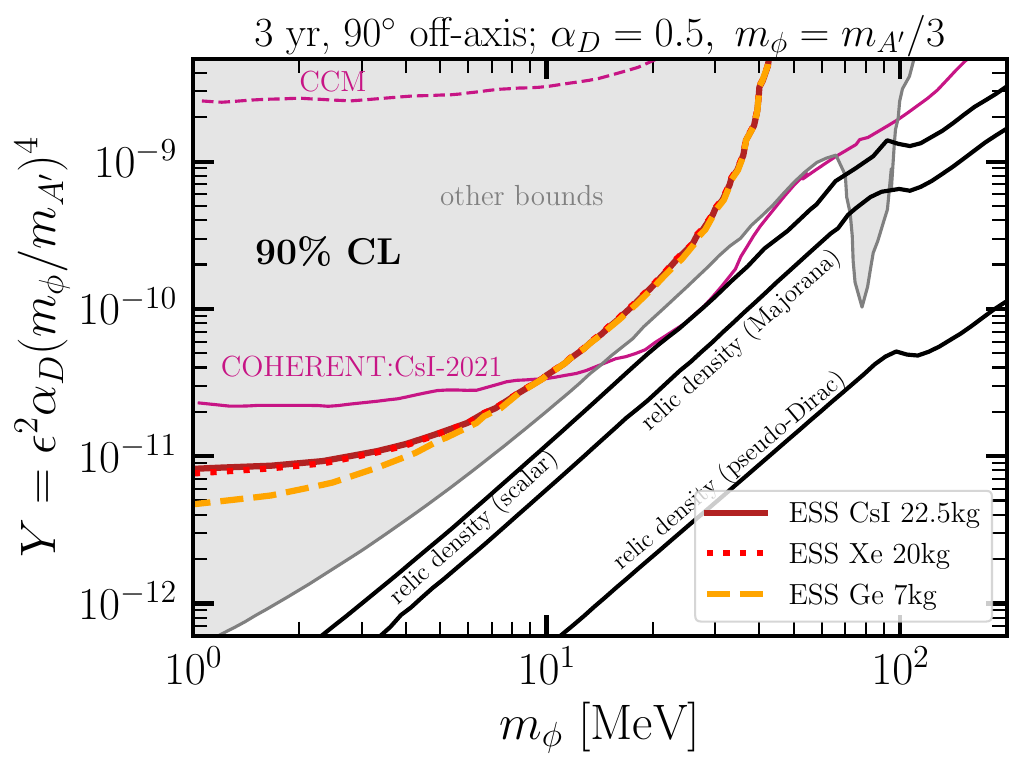}\\
       \includegraphics[width=0.49\textwidth]{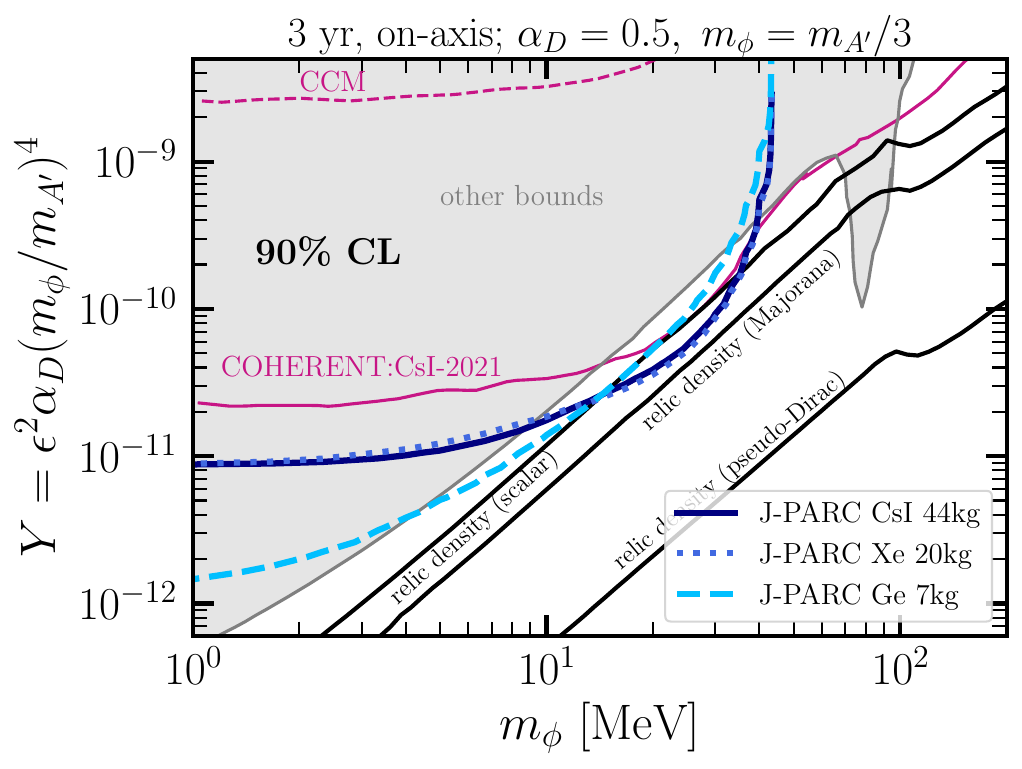}
        \includegraphics[width=0.49\textwidth]{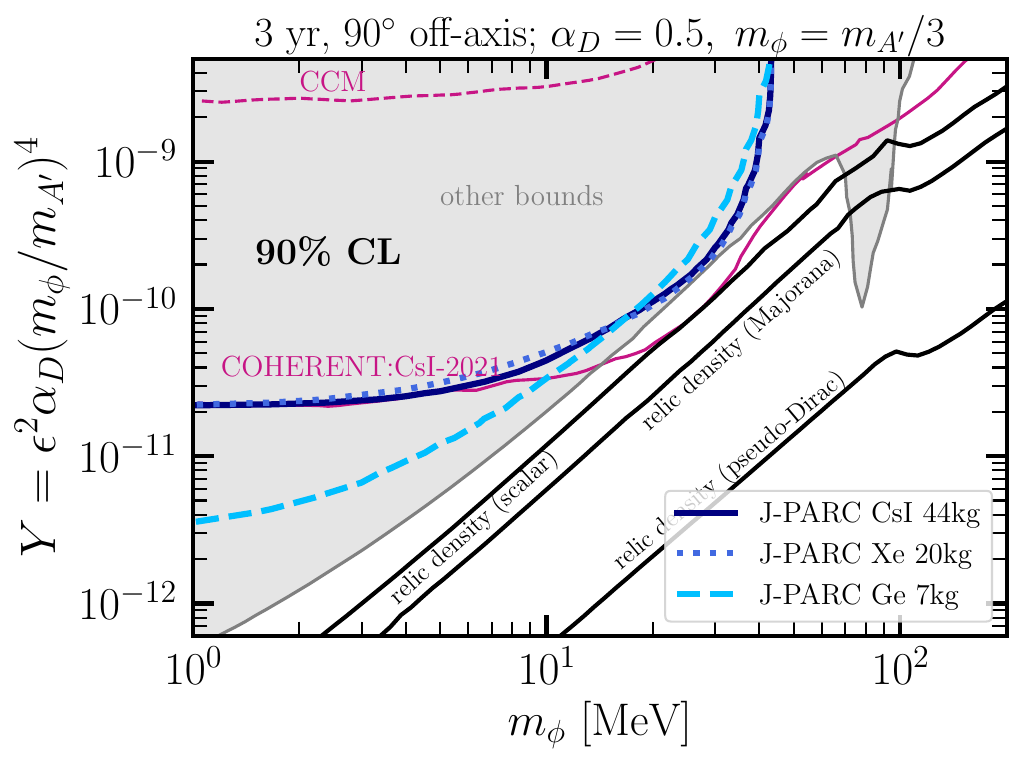}
        \\
       \includegraphics[width=0.49\textwidth]{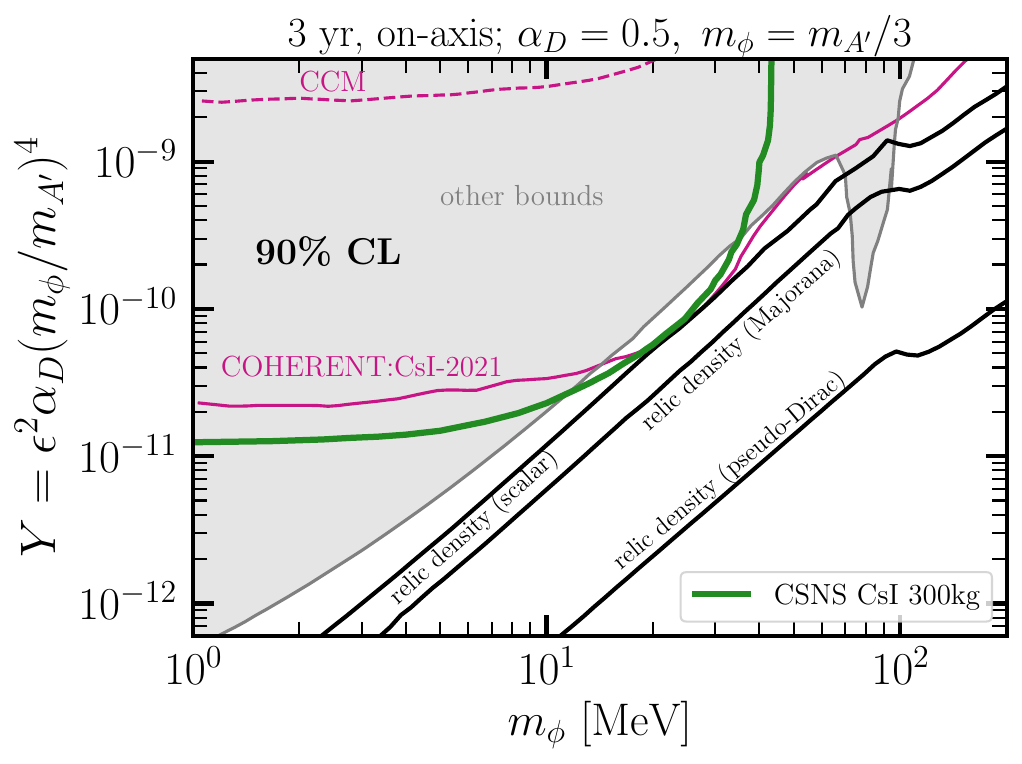}
        \includegraphics[width=0.49\textwidth]{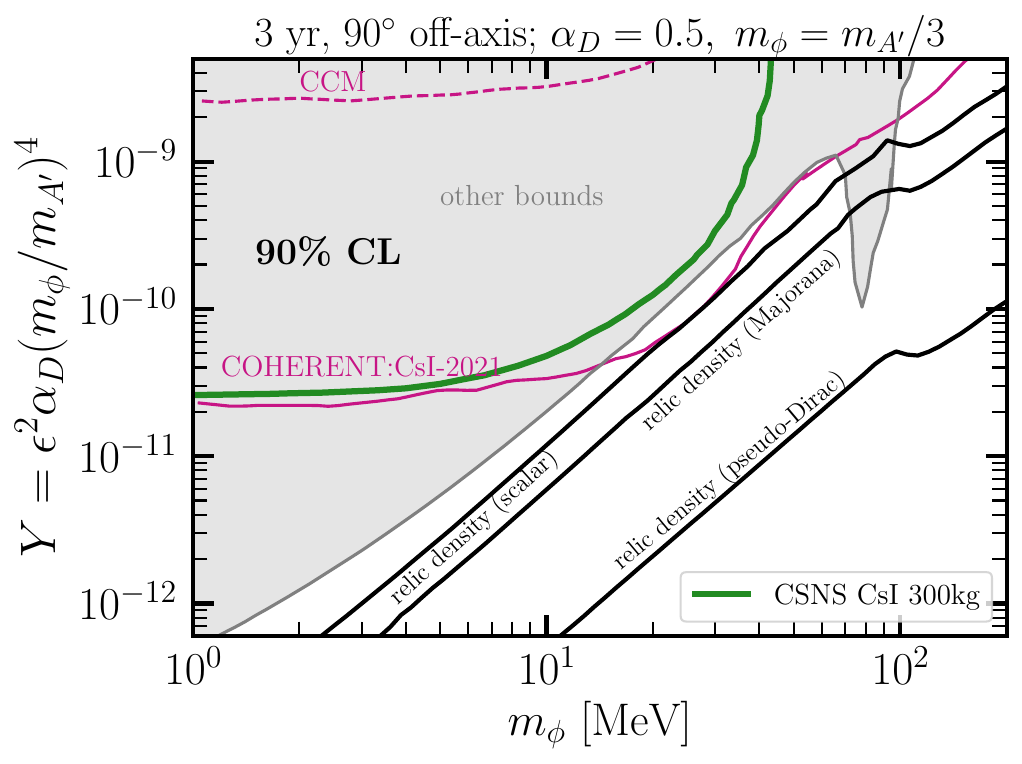}
       \caption{{\bf Projected 90\% CL sensitivities} for scalar DM interacting via a light dark photon at the ESS ({\bf top panels}), J-PARC ({\bf middle}) and CSNS ({\bf bottom}). {\bf Left (right) panels:} Results assume the detector to be on-axis (at $\theta_\mathrm{det} = 90^\circ$ with respect to the proton beam direction). Different target materials and volumes are considered (see main text for more details). The gray shaded regions and pink lines indicate current accelerator-based bounds.}
       \label{fig:scalar_DM_contour}
\end{figure}

In this section, we present the projected sensitivities for sub-GeV DM for the experimental configurations discussed above.
Figure~\ref{fig:scalar_DM_contour} shows the 90\% CL exclusion regions in the $(m_{\phi}, Y)$ parameter space for detectors deployed at the ESS and J-PARC with CsI (solid contours), xenon (dotted) and germanium (dashed contours) targets, as well as at the CSNS with a CsI target (green solid contour). To compare with results in the literature, we stick to the following values: $m_{A'} = 3\,m_{\phi}$ and $\alpha_D = 0.5$ \footnote{Note that theoretical considerations on the running
of the dark fine structure constant imply $\alpha_D \lesssim 0.1-0.5$~\cite{Davoudiasl:2015hxa}.}. The left panels in Fig.~\ref{fig:scalar_DM_contour} show projected sensitivities for detectors placed on-axis, where the DM angular distribution peaks. The right panels, instead, display expected sensitivities for detectors located  $90^\circ$ off-axis with respect to the proton beam direction (and inline with the location of the CsI detector in the COHERENT experiment). In all scenarios, we assume an  exposure time of 3 years, corresponding to 208 days (3600 hours) per calendar year at the ESS and CSNS (J-PARC). In the analysis of the J-PARC and CSNS facilities, we have included a cut on the arrival time of neutrino events  to reduce the \cevns~and SSB backgrounds as discussed in Sec.~\ref{sec:analysis}. We refer the reader to App.~\ref{app:baryophilic_results} where we discuss how our sensitivities are expected to change by varying some of the assumptions on backgrounds' uncertainties in the statistical analysis, and what the impact of not including the timing information on the neutrino events is.
For comparison, we display dark-pink contours corresponding to existing bounds from \cevns-based experiments, namely the COHERENT CsI detector~\cite{COHERENT:2021pvd} (solid) and CCM~\cite{CCM:2021leg} (dashed). We also include additional constraints (light-gray shaded region) obtained from experimental collaborations like MiniBooNE~\cite{MiniBooNEDM:2018cxm},  LSND~\cite{LSND:2001akn,Batell:2009di}, E137~\cite{Batell:2014mga}, BaBar~\cite{BaBar:2017tiz}, and NA64~\cite{NA64:2023wbi}, currently driving the leading bound on this scenario. 
Note that phenomenological limits from KARMEN data might cover some of the regions covered by our analyses \cite{Dutta:2023fij}. All
these limits, however, should be interpreted with care, as they probe different processes and interaction channels. In particular, they are not directly related to \cevns. MiniBooNE and LSND, although neutrino-beam experiments, are based on quasi-elastic scattering and inverse $\beta$ decay, respectively, while E137 and BaBar rely on $e^- - \text{Al}$ fixed-target and $e^+e^-$ collider searches. Consequently, the comparison is indicative rather than strictly equivalent.
We further show the preferred curves in the $Y$ parameter corresponding to the observed relic DM abundance for scalar, pseudo-Dirac, and Majorana dark matter~\cite{Berlin:2018bsc}. As discussed above, these lines represent well-motivated target sensitivities: below them, the DM annihilation cross section is insufficient to reduce the relic abundance to the observed value. Above them, the DM would be underabundant and require an additional candidate to explain the total observed relic density.

As can be seen, future spallation-based neutrino experiments have the potential to improve upon existing constraints, particularly in the DM mass range
$6~\mathrm{MeV} \lesssim m_{\phi} \lesssim 40~\mathrm{MeV}$, provided the detector is located along the proton-beam direction.
In this region, the projected sensitivities can surpass the thermal relic density target for a scalar DM candidate in the case of the CsI and Xe detectors at J-PARC, with optimal performance around
\( m_{\phi} \sim \mathcal{O}(25~\mathrm{MeV}) \).
 In the same mass region, the detectors at ESS and CSNS can probe  new regions of the parameter space up to the relic density target.
Indeed, at the ESS and J-PARC, germanium-based detectors provide the strongest constraints, especially in the low-mass regime, reaching sensitivities down to
\( Y \sim 2 \times 10^{-12} \). This is mostly due to the low recoil-energy threshold achievable in such detectors. In the case of the CsI detector at these facilities, it is clear that doubling its volume (as expected at J-PARC~\cite{Collar:2025sle}), leads to a non-negligible improvement in sensitivity reach, as visible by comparing the J-PARC vs the ESS panels in Fig.~\ref{fig:scalar_DM_contour}. Despite its much larger mass, the CsI detector at the CSNS is slightly less constraining due to its higher operation threshold together with the higher SSB background, but still it allows to probe \( Y \sim 2 \times 10^{-11} \) at tens of MeV dark photon masses, beyond the relic density benchmark.  In the low-mass regime, sensitivities tend to saturate mostly due to detector limitations, namely the loss of efficiency at low recoil energies. 

On the other side of the parameter space instead, the sensitivity of spallation neutron source experiments rapidly reduces for mediator masses above
$m_{A'} \sim 100~\mathrm{MeV}$ (i.e., $m_{\phi} \gtrsim 45~\mathrm{MeV}$), 
reflecting the kinematic constraint for DM production through neutral pion decays. Access to heavier DM masses would require higher energy proton beams capable of producing heavier neutral mesons. Depending on the proton beam collision energy, heavier mesons can be produced, thereby extending the experimental reach to larger values of $m_{\phi}$. In addition, DM production via proton Bremsstrahlung could further expand the accessible parameter space \cite{deNiverville:2016rqh,Blumlein:2013cua}. Extending up to the energy regime of SHIP and FASER, the Drell-Yan production channel is also available. Spallation neutron source facilities provide a unique production channel that facilitates the \enquote{deconstruction} of the physics underlying a DM signal. High statistics combined with high background discrimination power, enabled by timing spectra, make these facilities a unique environment for light secluded sectors. Altogether, these considerations highlight the strengths of spallation neutron source facilities in probing light dark sectors. Information stemming from these measurements are therefore complementary to those at high-energy accelerator facilities.
\begin{figure}[t]
    \centering
        \includegraphics[scale=0.6]{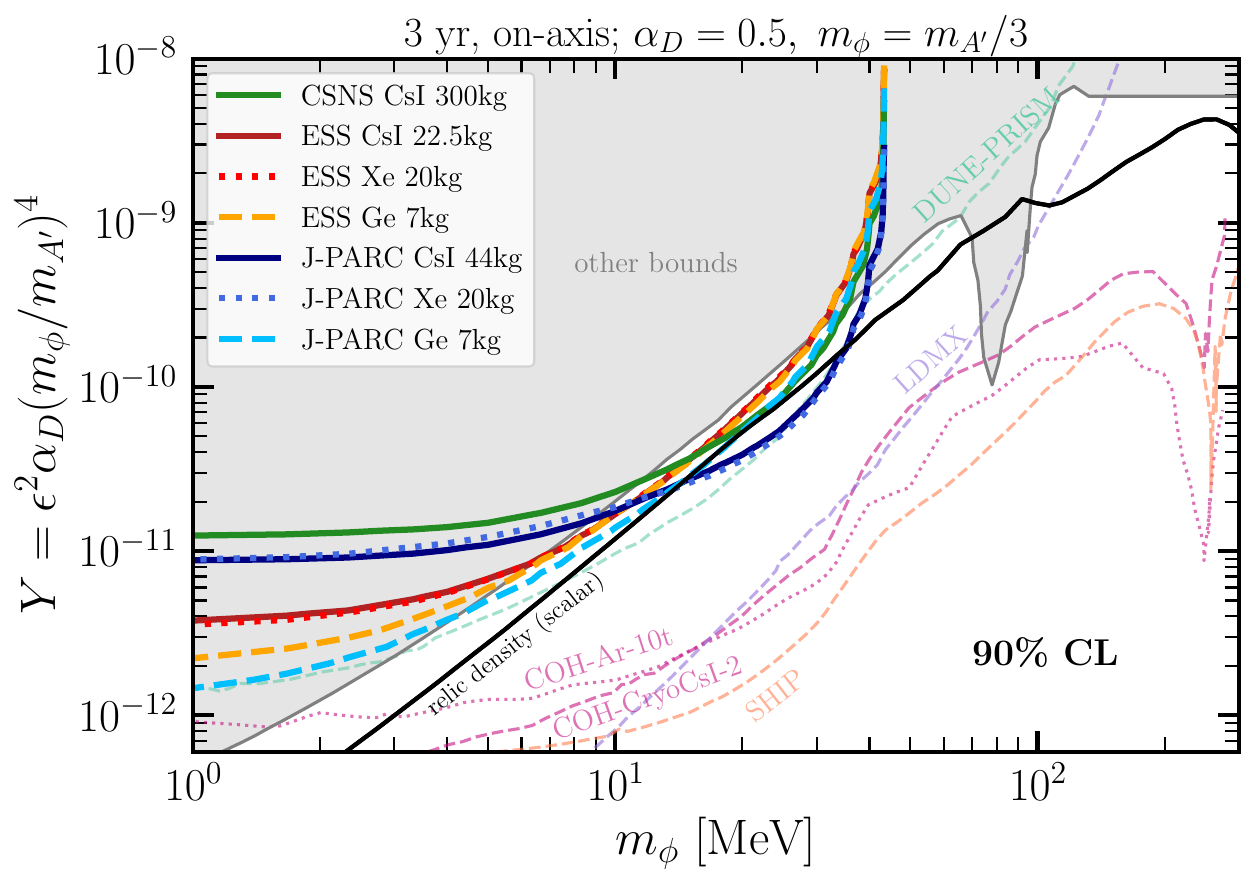}
       \caption{{\bf Projected 90\% CL sensitivities} at the ESS, J-PARC, and CSNS, compared with projections from other future experiments: DUNE-PRISM~\cite{DeRomeri:2019kic} (light green, dashed); the upgraded 700-kg cryogenic CsI detector COH-CryoCsI-2~\cite{COHERENT:2023sol} (pink, dashed); a 10-t liquid-argon detector, COH-Ar-10t~\cite{COHERENT:2023sol} (pink, dotted); SHiP~\cite{SHiP:2020noy,Batell:2022xau} (coral, dashed); and LDMX~\cite{LDMX:2018cma,Batell:2022xau} (light blue, dashed).}
       \label{fig:scalar_DM_contour_future}
\end{figure}

Next, in Fig.~\ref{fig:scalar_DM_contour_future} we compare our projected exclusion reach in the $(m_{\phi},Y)$ plane for the dark photon portal scenario with the sensitivities expected at other future facilities. Our projections assume on-axis detector configurations and three years of data taking.
We include sensitivity estimates for two proposed upgrades of COHERENT detectors optimized for \cevns~detection: a 700 kg undoped cryogenic CsI detector~\cite{COHERENT:2023sol} (pink, dashed) and a 10-tonne LAr detector~\cite{COHERENT:2023sol} (pink, dotted). We also show projections for the first phase of LDMX~\cite{LDMX:2018cma,Batell:2022xau} (light blue, dashed), the movable near detector of DUNE-PRISM with a 75-tonne LAr target~\cite{DeRomeri:2019kic} (light green, dashed), and a five-year run of SHiP~\cite{SHiP:2020noy,Batell:2022xau} (coral, dashed).
These comparisons should also be interpreted with care. LDMX and SHiP are dedicated DM experiments, while DUNE operates at higher beam energies and primarily probes $\text{DM}-e^-$ scattering. Moreover, the projected sensitivities of the future COHERENT upgrades benefit from substantially larger detector masses.
Despite this, our results show that experiments at ESS, J-PARC, and CSNS can competitively explore a broad region of parameter space. In particular, they are sensitive to DM masses in the range $m_{\phi} \simeq 20$–$40$ MeV, touching or  surpassing the thermal relic target for scalar DM. 

\begin{figure}[t!]
    \centering
        \includegraphics[scale=0.6]{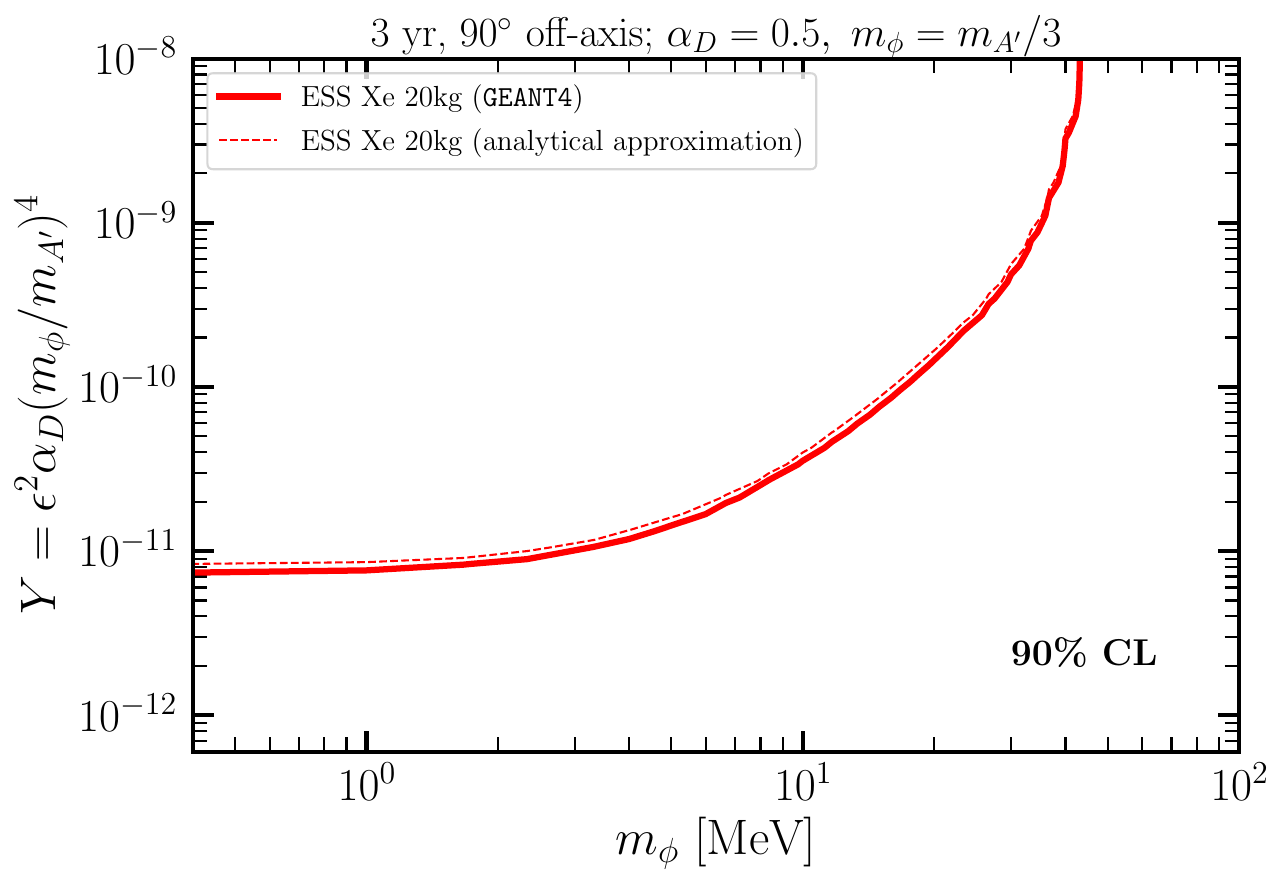}
       \caption{Comparison of the projected 90\% CL sensitivity for a xenon target at ESS, obtained using a \texttt{GEANT4} simulation or a neutral pion PDF derived from the S \& W parametrization. }
    \label{fig:GEANT4_vs_analytical}
\end{figure}

All the sensitivities presented up to this point in this section have been derived from the DM event rate computed using our \texttt{GEANT4}-based simulations, as described in Sec.~\ref{sec:numerical_approach}. As anticipated, for the ESS-Xe detector we repeated the analysis employing instead the neutral pion PDF obtained from the S \& W  parametrization.
Although not shown here, we have verified that their spectral shapes are remarkably consistent between the two approaches, with their normalization differing by less than a factor of 1.3. Overall, the S \& W parametrization systematically predicts fewer events than \texttt{GEANT4}, leading to slightly more conservative sensitivity projections. We present the result of this comparison in Fig.~\ref{fig:GEANT4_vs_analytical} where we compare the projected sensitivities for the Xe detector at the ESS obtained using the two approaches discussed in this work: the neutral pion PDF derived from the S \& W parametrization and that extracted from our dedicated \texttt{GEANT4} simulations. The overall agreement between the two approaches propagates to the final exclusion contours, which are  found to be notably similar over all the mass range probed with our analysis. The differences between the two approaches amount to less than 15\% in the final sensitivities and are barely visible in the resulting plots.  It is worth noting that larger uncertainties  may arise from the simulation framework itself, for instance, when switching between hadronic models in \texttt{GEANT4}.

% --------------
% Section
% --------------
\section{Conclusions}
\label{sec:concl}
Neutrino experiments based on spallation neutron sources have recently experienced significant advances, mainly driven by the first measurements of \cevns. Beyond their primary role in precision neutrino physics, these facilities have demonstrated strong potential as probes of physics beyond the Standard Model, motivating a broad range of additional applications. In light of several new experiments currently under construction or in advanced planning stages, we have investigated the sensitivity of spallation neutron source-based neutrino experiments to sub-GeV DM produced in neutral pion decays.

We have considered representative detector configurations, including planned gaseous xenon (20 kg), high-purity germanium (7 kg), and cryogenic CsI scintillating detectors (22.5 kg and 44 kg) at ESS and J-PARC, as well as a 300 kg CsI detector under construction at CSNS.
A detailed discussion of the neutral pion flux calculation at these facilities has been presented. We compared two approaches: a dedicated \texttt{GEANT4} simulation of proton collisions on mercury and tungsten targets, and the Sanford-Wang parametrization of charged-pion differential cross sections. We find that both methods yield consistent results within the intrinsic uncertainties of the overall calculation. Our analysis focuses on a sub-GeV scalar dark matter candidate in the minimal dark photon portal framework, as well as on a baryophilic scenario based on gauged baryon
number. Since the considered detectors are optimized for nuclear recoil measurements, we have studied coherent elastic dark matter-nucleus scattering signals.

Projected sensitivities have been presented in the $m_{\phi}$-$Y$ plane.
Our results show that spallation-source neutrino experiments can probe interesting regions  of parameter space, particularly in the DM mass range $10~\mathrm{MeV} \lesssim m_{\phi} \lesssim 40~\mathrm{MeV}$. Sensitivities reach or extend beyond the scalar DM thermal relic target .
Even stronger sensitivities are obtained in baryophilic mediator scenarios (gauged baryon number models), as detailed in App.~\ref{app:baryophilic_results}. Our results highlight the strong complementarity between spallation-source facilities, optimized for
\cevns~measurements, and other accelerator-based dark matter searches. By leveraging their nuclear recoil capabilities, these experiments offer a uniquely powerful environment for dark matter searches in the MeV mass range.

% ---------------
% Section
% ---------------
\section*{Acknowledgments}
We thank Francesc Monrabal and Juan Collar for providing us with useful information and input. D.A.S. is supported by ANID grants \enquote{Fondecyt Regular} No. 1221445 and No. 1260595. He thanks IFIC and specially the members of the AHEP group for their kind hospitality during the final stage of this work. VDR acknowledges financial support by the grant CIDEXG/2022/20 (from Generalitat Valenciana) and by the Spanish grants CNS2023-144124 (MCIN/AEI/10.13039/ 501100011033 and “Next Generation EU”/PRTR), PID2023-147306NB-I00, and CEX2023-001292-S (MCIU/AEI/ 10.13039/501100011033).  D.K.P. acknowledges funding from the European Union’s Horizon Europe research and innovation programme under the Marie Skłodowska‑Curie Actions grant agreement No.~101198541 (neutrinoSPHERE).  G.S.G. has been supported by Sistema Nacional de Investigadoras e Investigadores (SNII, México) and by SECIHTI Project No. CBF-2025-I-1589.
%
% ------------
% Appendix
% ------------
\appendix 
\section{Results for other experimental configurations and baryophilic scenarios}
\label{app:baryophilic_results}
In this Appendix, we explore how the sensitivities obtained above change by modifying some aspects of the analysis setup previously assumed.

To begin, we discuss the baryophilic scenario introduced in Sec.~\ref{sec:baryophilic}.  In this model, the DM production follows from the decay $\pi^0 \to \gamma + \phi +\phi$~\cite{CCM:2021leg}
\begin{equation}
\label{eq:pi0_to_gamma_2DM_BR_baryophilic}
    \mathcal{B}(\pi^0 \to \gamma + \phi + \phi^*) \approx \mathcal{B}(\pi^0 \to \gamma + \gamma)\times  2\frac{g_B^2}{e^2} \left(1 - \frac{m_{A'}^2}{m_{\pi^0}^2} \right)^3 \times
\mathcal{B}(A' \to \phi + \phi^*)\, ,
\end{equation}
while the relevant DM-nucleus elastic scattering cross section is modified as follows

\begin{equation}
    \label{eq:cross:scalar:DM_baryoph}
    \left.\frac{\d \sigma_{\phi \mathcal{N}}}{\d E_r}\right|^{\rm{baryophilic}}
    =
    \frac{g_B^4}{2\pi}
    \frac{ m_{\mathcal{N}} E_\phi^2}
    {\left(E_\phi^2 - m_\phi^2\right)
    \left(2 m_{\mathcal{N}} E_r + m_{A'}^2\right)^2} 
    \left(
1
- \frac{m_\phi^2 E_r}{2 m_{\mathcal{N}} E_\phi^2}
- \frac{E_r}{E_\phi}
\right)
A^2\,\mathcal{F}_A^2(E_r)\, ,
%\label{eq:scalar_DM_xsec}
\end{equation}
where by $\mathcal{F}_A$ we denote a Helm-like nuclear baryon-charge form factor.
As one can notice, this DM--nucleus cross section is similar to the one found in the dark photon case given in Eq.~\eqref{eq:cross:scalar:DM}, with the substitutions $Z e \to A$ (atomic mass number) and $\epsilon e \to g_B$, and with $Q_B = 1$.
In principle, one could define a baryonic structure constant \(\alpha_B = \frac{g_B^2 Q_B^2}{4 \pi}\) and fix it to a benchmark value, as done in dark-photon studies. However, unlike previous works that introduce two independent couplings \(g_B\) and \(g_D\) and fix one of them, our framework effectively contains a single gauge coupling. Fixing \(\alpha_B\) in this case would require adjusting the DM baryon charge \(Q_B\) over several orders of magnitude~\cite{Batell:2021snh} to satisfy both relic density and laboratory constraints, since \(g_B\) cannot be varied independently. We therefore present our sensitivities directly in terms of \(\alpha_B\).

\begin{figure}[t!]
    \centering
        \includegraphics[scale=0.6]{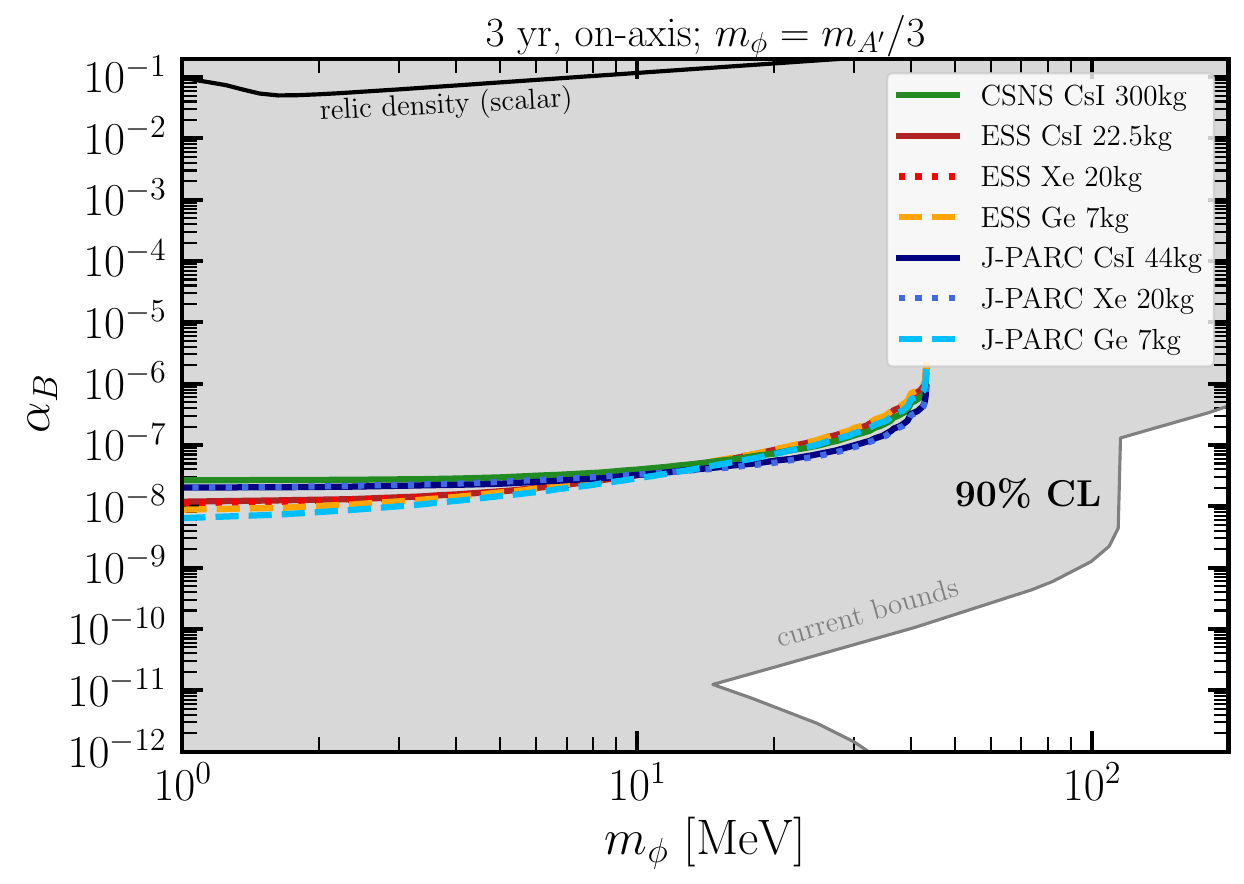}
              \caption{Same as Fig.~\ref{fig:scalar_DM_contour_future} but for the baryophilic model. Other existing bounds are included for comparison in the gray shaded region~\cite{Batell:2021snh}.}
    \label{fig:baryophlic_fig:scalar_DM_contour}
\end{figure}

Our results are shown in Fig.~\ref{fig:baryophlic_fig:scalar_DM_contour}, where we also show the relic density line benchmark, assuming a scalar DM candidate, and other existing limits in shaded gray. The latter are taken from Ref.~\cite{Batell:2021snh} and show that this scenario is already strongly constrained by beam-dump searches in NuCal and by the anomaly-induced rare $K$ and $Z$ decays.
Notice also that a comparison versus other \cevns~based bounds is not straightforward, as both the COHERENT~\cite{COHERENT:2022pli} and CCM~\cite{CCM:2021yzc} Collaborations have presented their results fixing $\alpha_B$. This can be done with very suppressed $g_B$ coupling by assuming very large DM baryon charges. The large baryon structure constant assures the right relic density, while the small $g_B$ coupling instead that laboratory limits are satisfied.
Note that although the theoretical possibility of very large charges can be questioned, such choices remain consistent as long as perturbativity is preserved \cite{Batell:2021snh}.

\begin{figure}[t!]
    \centering
       \includegraphics[width=0.48\textwidth]{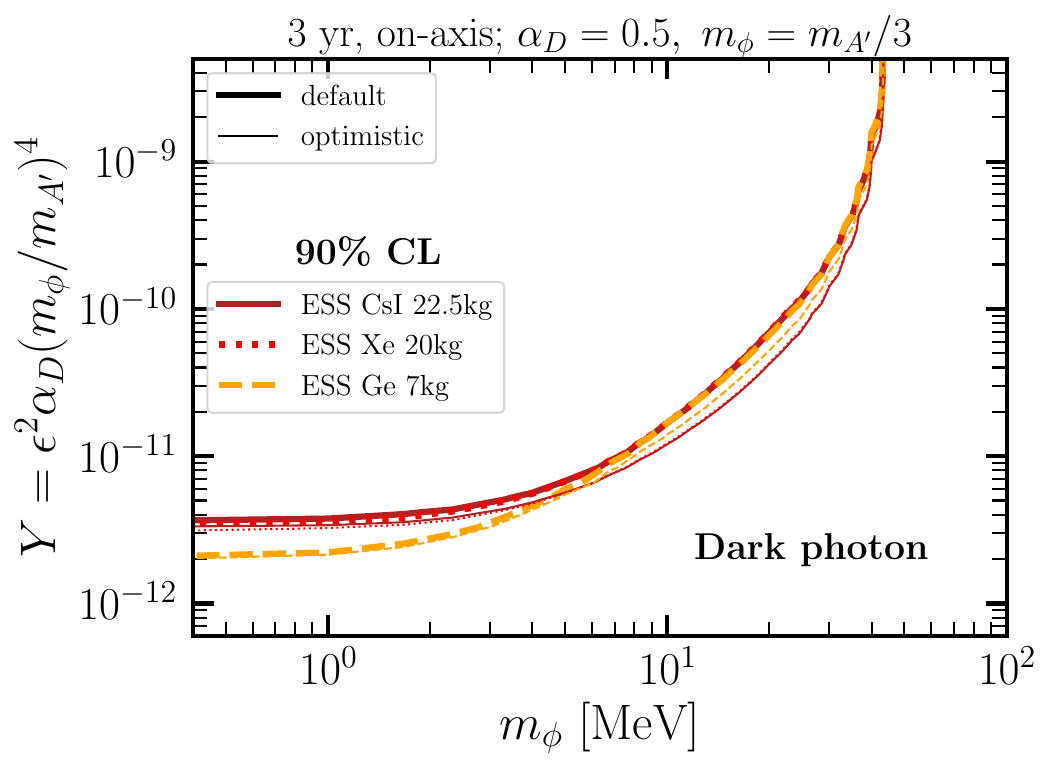}
       \includegraphics[width=0.48\textwidth]{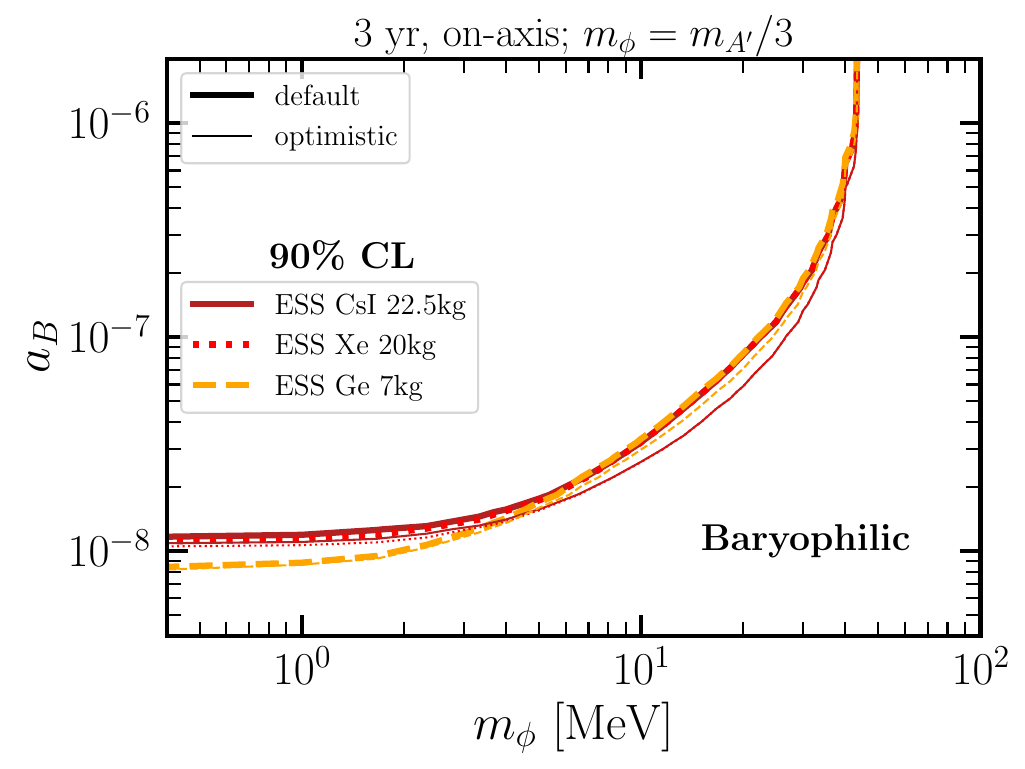}
       \includegraphics[width=0.48\textwidth]{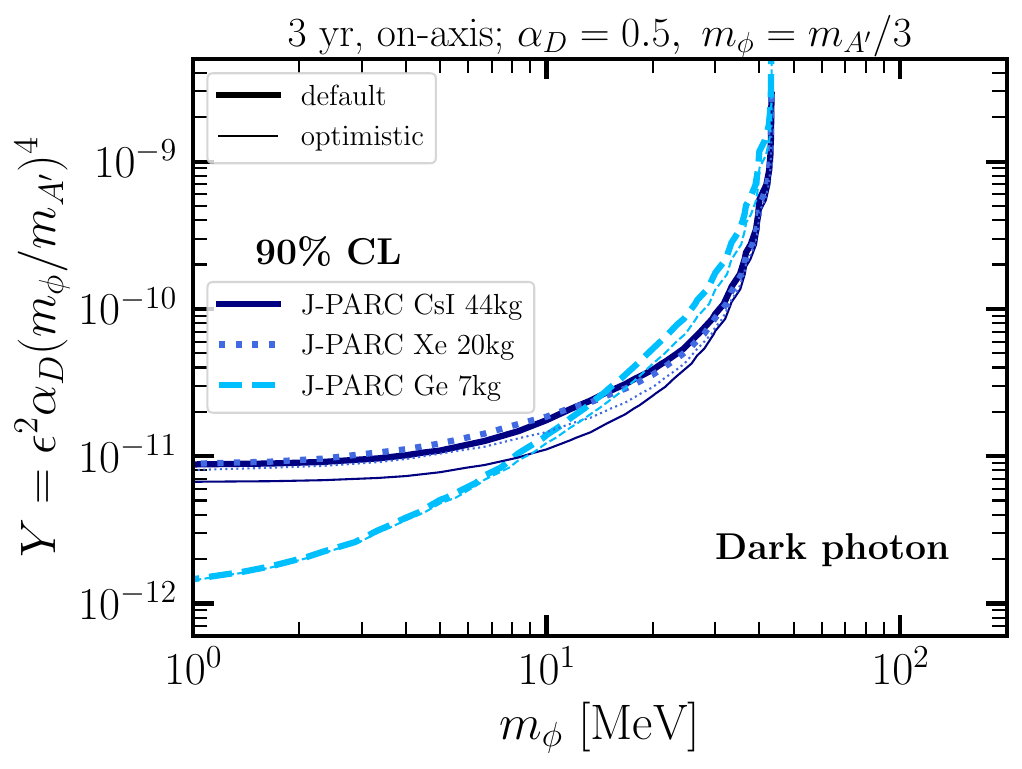}
       \includegraphics[width=0.48\textwidth]{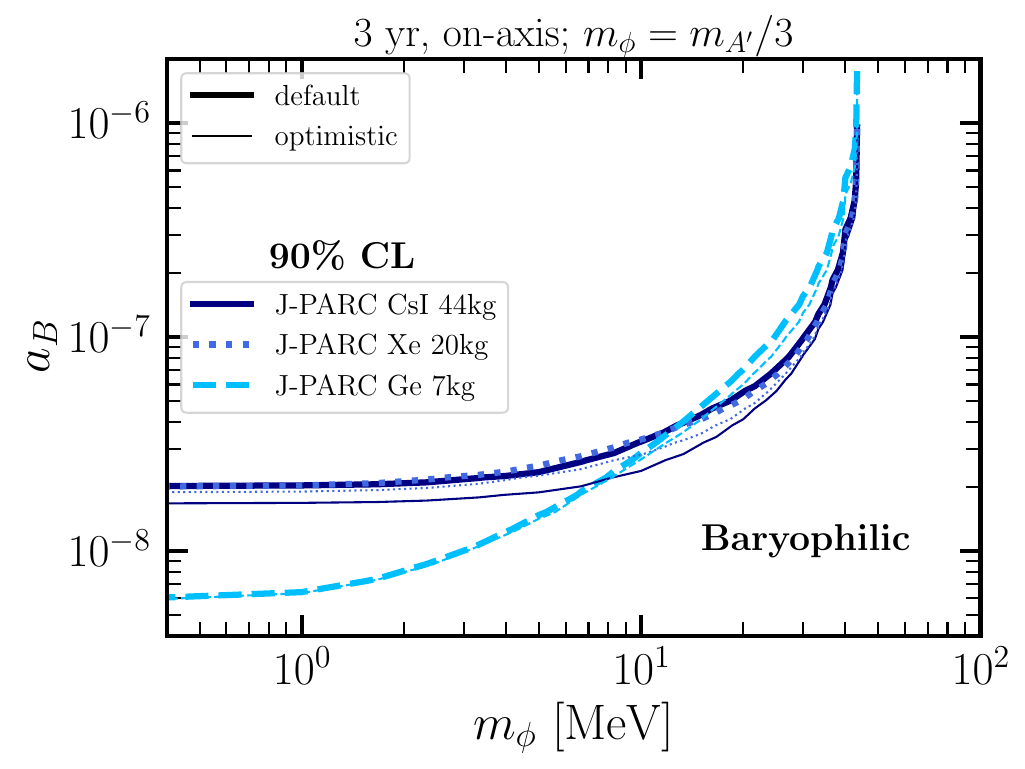}
        \includegraphics[width=0.48\textwidth]{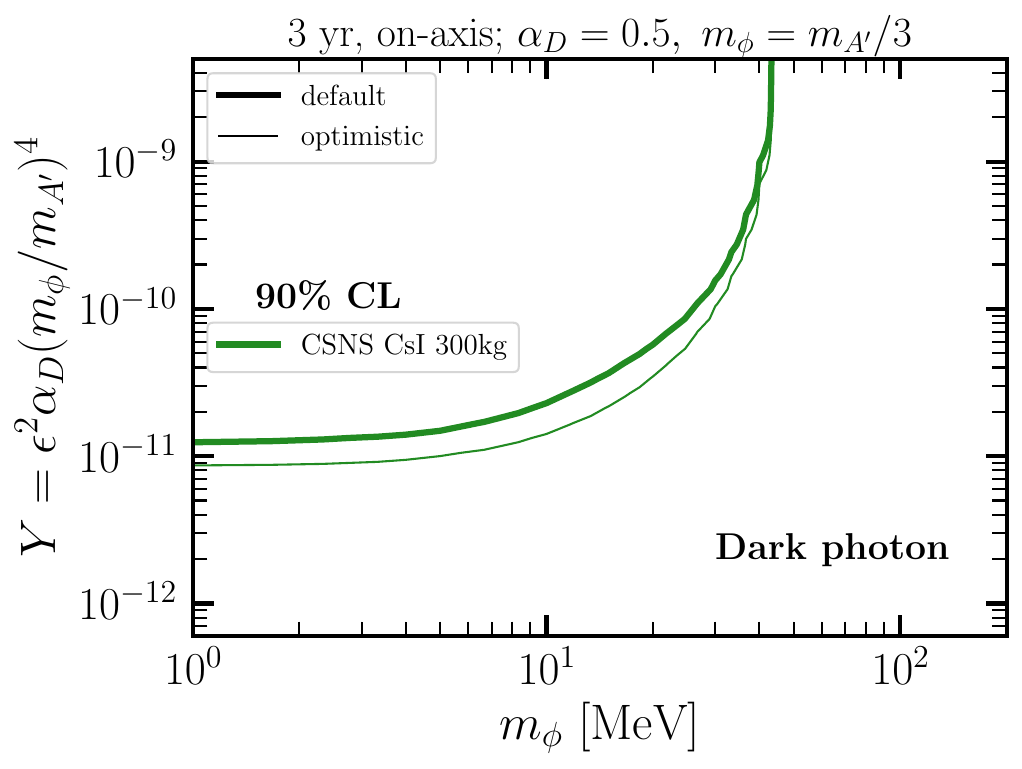}
       \includegraphics[width=0.48\textwidth]{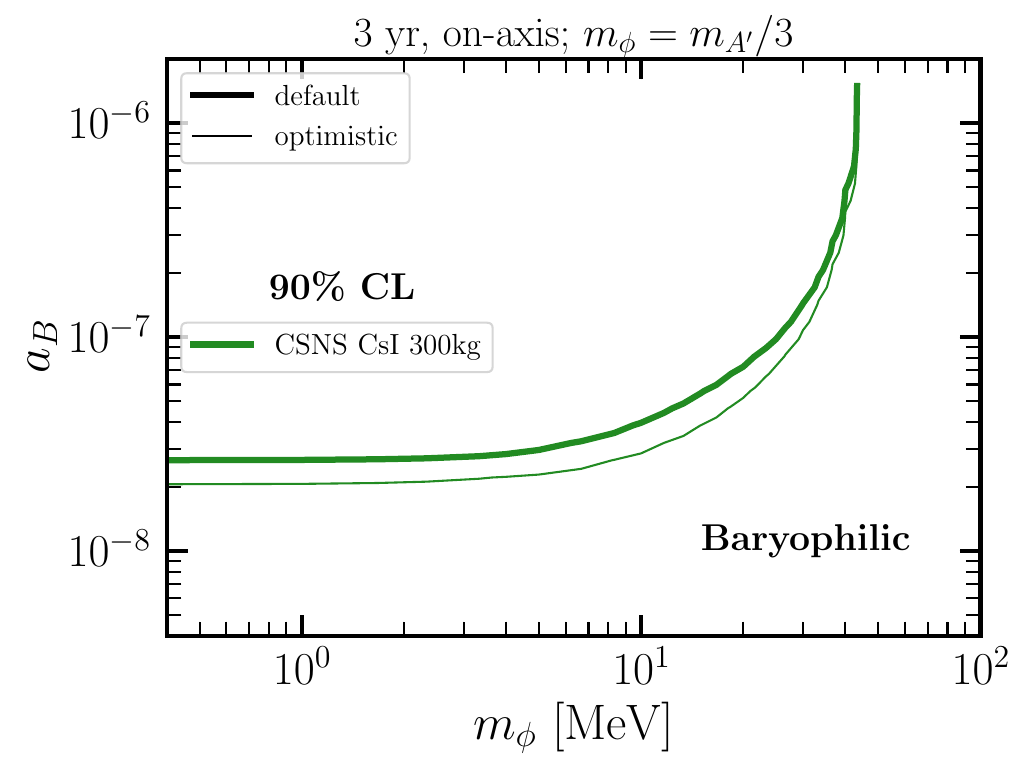}
       \caption{Comparison of 90\% C.L. sensitivities for a Xe, Ge and CsI target at the ESS ({\bf top panels}), Xe, Ge and CsI at the J-PARC ({\bf middle panels}) and CsI at the CSNS ({\bf bottom panels}), assuming an optimistic scenario with improved systematic uncertainties. We consider a scalar DM candidate with a dark photon ({\bf left panels}) or a baryophilic ({\bf right panels}) mediator.}
       \label{fig:ESS_vs_JPARC_scalar_DM_contour_future}
\end{figure}

Next, we discuss how our sensitivities change by varying some of the experimental assumptions. We show in Fig.~\ref{fig:ESS_vs_JPARC_scalar_DM_contour_future} the sensitivities at the different facilities obtained assuming the following uncertainties in the statistical analysis (see Sec.~\ref{sec:analysis}): $\sigma_\alpha = 10\%$ (default)$ \to 5 \%$ (optimistic) for the DM signal, $\sigma_{\beta_1} = 10\%$ (default) $\to 2 \%$ (optimistic)  for the \cevns~background,  respectively while we keep  $\sigma_{\beta_2} = 1\%$ for the SSB background contribution. Moreover, in the case of pBRN we instead use $\sigma_{\beta_2} = 25\%$ (default) $\to 10\%$ (optimistic). The left panels of Fig.~\ref{fig:ESS_vs_JPARC_scalar_DM_contour_future} are obtained for the scalar DM candidate with dark photon mediator (Model 1, see Sec.~\ref{sec:dark_photon_scenario}) while the plots on the right column show the results for a scalar DM candidate in the baryophilic scenario (Model 2, see Sec.~\ref{sec:baryophilic}). From these plots, it becomes evident  that an improvement in the background treatment, while relevant, is not expected to change dramatically the sensitivity reach of these facilities. On the other hand, as already noticed in the discussion of results, larger exposures do have a visible impact leading to stronger bounds. 

\begin{figure}[t]
    \centering
    \includegraphics[width=0.48\textwidth]{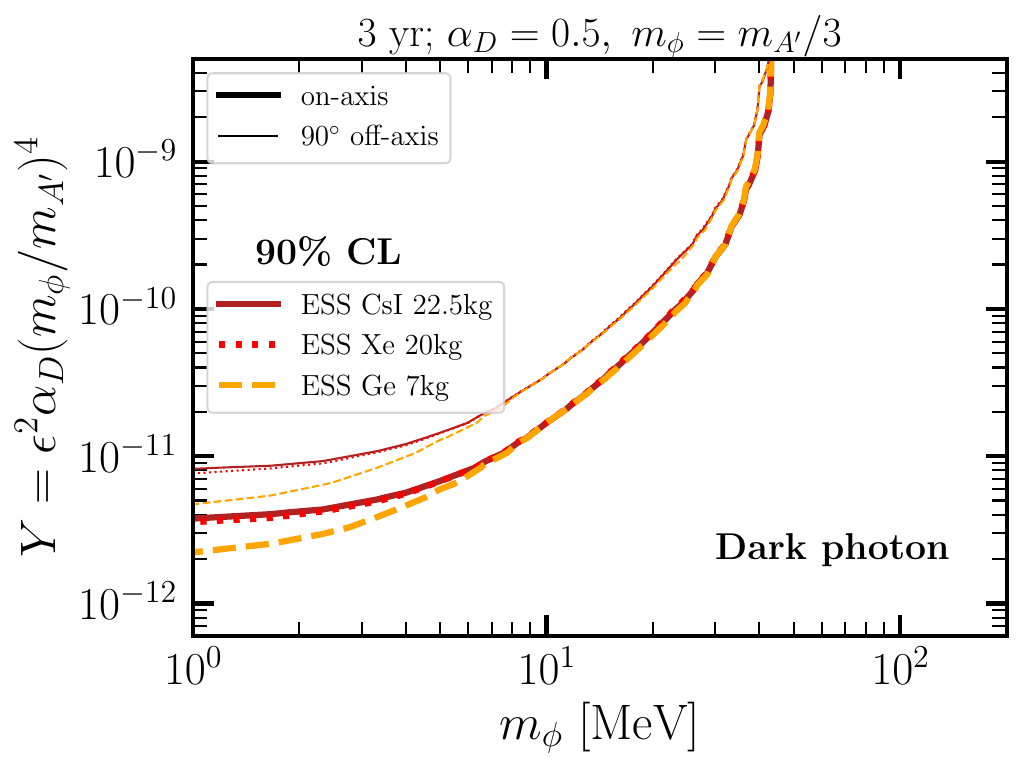}
    \includegraphics[width=0.48\textwidth]{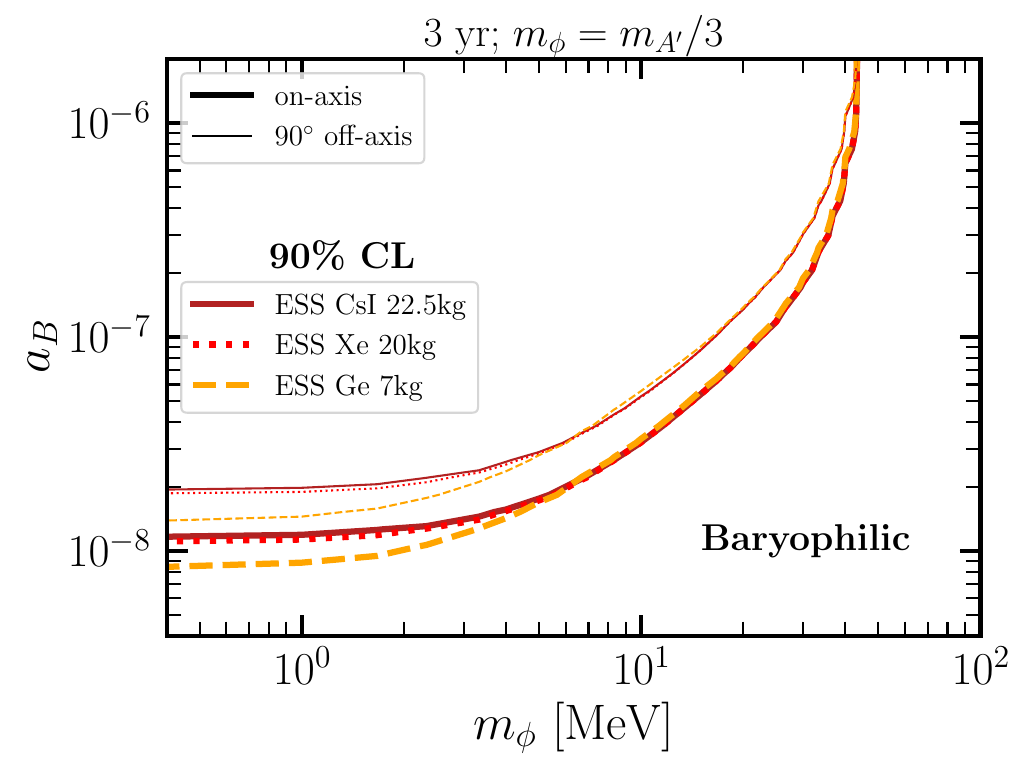}
    \includegraphics[width=0.48\textwidth]{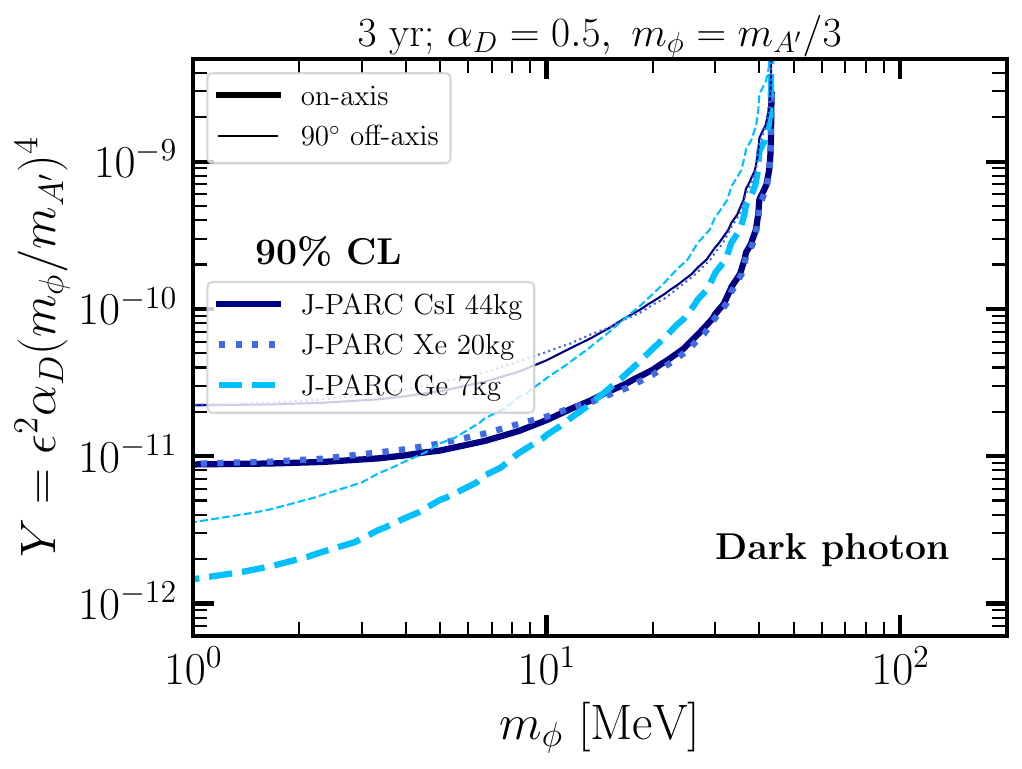}
    \includegraphics[width=0.48\textwidth]{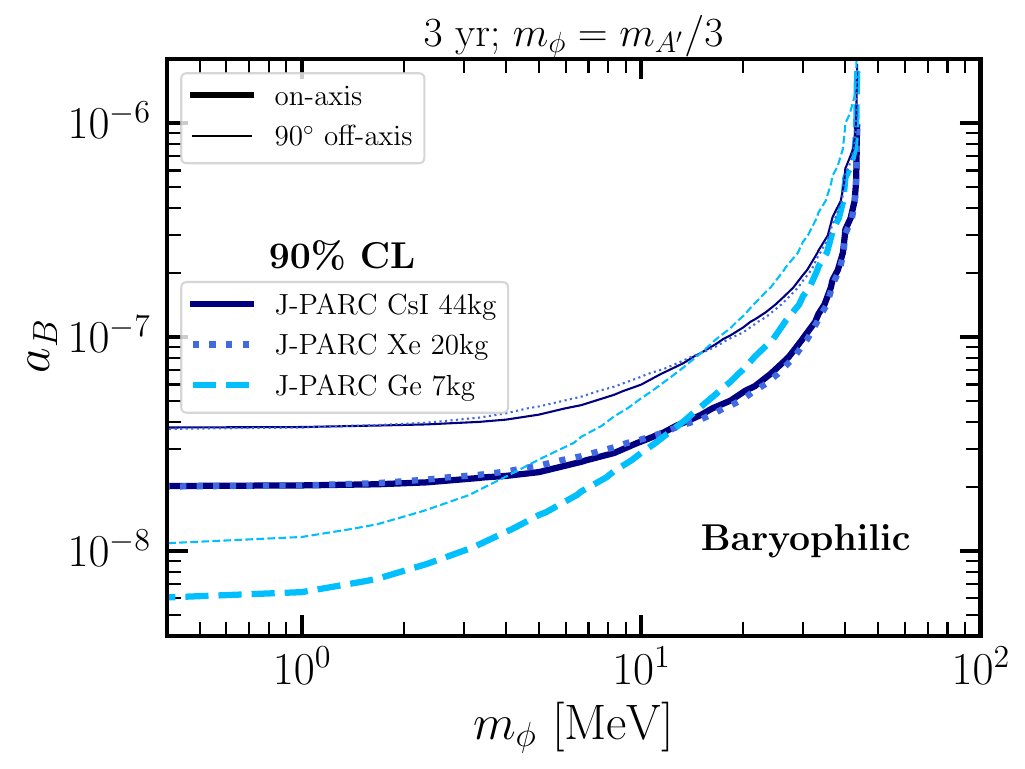}
    \includegraphics[width=0.48\textwidth]{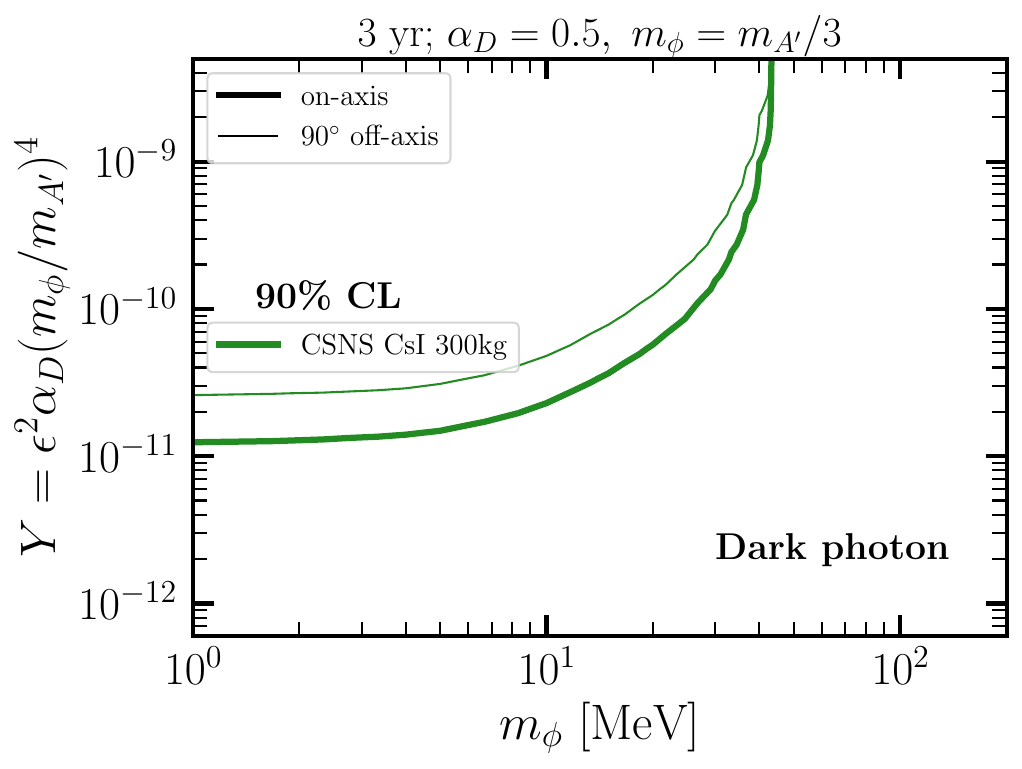}
    \includegraphics[width=0.48\textwidth]{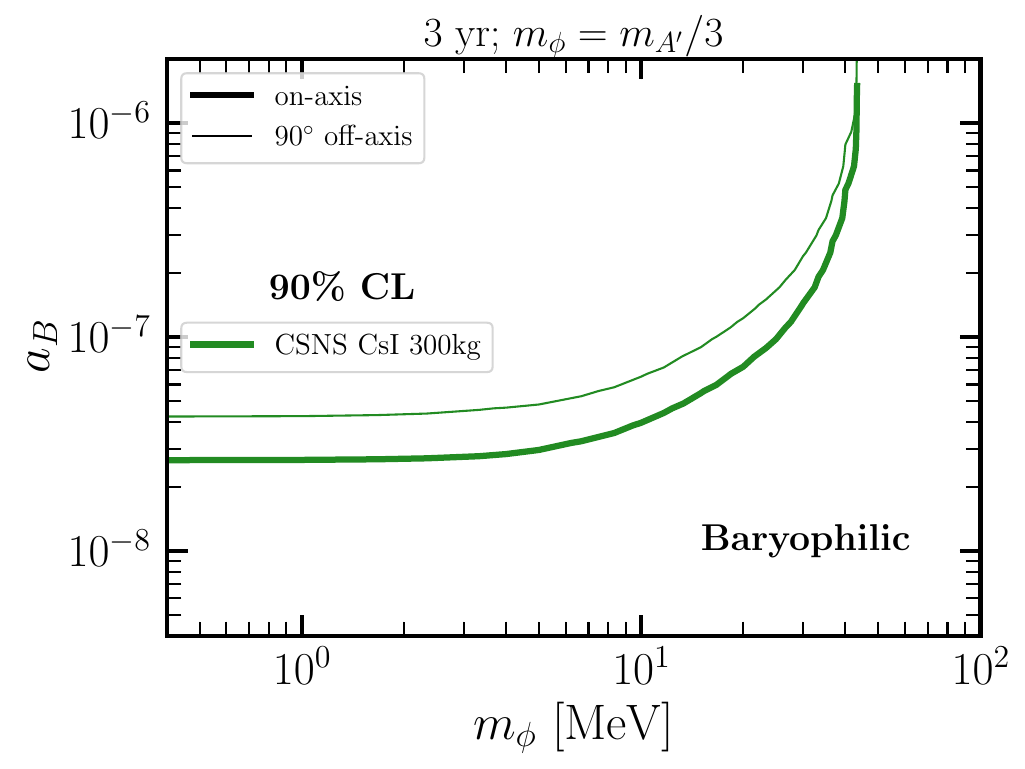}
     \caption{Comparison of the projected 90\% C.L. sensitivities for Xe, Ge, and CsI targets at ESS ({\bf top panels}), Xe, Ge, and CsI at J-PARC ({\bf middle panels}), and CsI at CSNS ({\bf bottom panels}), for detector configurations located either on-axis or at $90^\circ$ off-axis. Results are shown for a scalar DM candidate interacting via a dark photon mediator ({\bf left panels}) and via a baryophilic mediator ({\bf right panels}).}
    \label{fig:ESS_vs_JPARC_scalar_DM_contour_locations}
\end{figure}

In Fig.~\ref{fig:ESS_vs_JPARC_scalar_DM_contour_locations} we instead illustrate how the projected sensitivities at the three facilities vary when relocating the detector from an on-axis configuration to a $90^\circ$ off-axis position with respect to the proton beam direction. As anticipated, placing the detector along the direction of the incoming proton beam yields the strongest sensitivities. This behavior is driven by the forward-peaked nature of neutral pion production and decay: boosted $\pi^0$ mesons preferentially emit dark photons, and consequently DM particles, along their direction of motion, leading to an enhanced DM flux in the forward direction. In contrast, moving the detector off axis reduces both the overall flux and the typical DM energies, resulting in weaker constraints. Quantitatively, we find that the on-axis configuration improves the sensitivities by approximately a factor of $\sim 2.5$ compared to the $90^\circ$ off-axis case, with only mild dependence on the mediator mass and facility considered.
% ---------------
% Section
% ---------------
\section{Lorentz transformations}
\label{app:boost}
For a given facility, the set of neutral pions define a set of boost variables. Consider an arbitrary element of that four-momenta list. Total energy defines the Lorentz factor $\gamma$ and $\beta$
according to
\begin{equation}
    \label{eq:boos_variables}
    \gamma = \frac{E_{\pi^0}}{m_\pi^0}\ ,\qquad \beta = \frac{\sqrt{\gamma^2 - 1}}{\gamma}\ .
\end{equation}
The three-momentum components, instead, define the polar and azimuthal $\pi^0$ propagation angles according to
\begin{align}
    \label{eq:polar_azimuthal_angles}
    \cos\theta_{\pi^0}= \frac{p_{\pi^0}^z}{p_{\pi^0}}\ ,\qquad
    \sin\theta_{\pi^0} = \frac{p_{\pi^0}^{xy}}{p_{\pi^0}}\ ,\qquad
    %\nonumber\\
    \cos\varphi_{\pi^0}= \frac{p_{\pi^0}^x}{p_{\pi^0}^{xy}}\ ,
    \qquad
    \sin\varphi_{\pi^0} = \frac{p_{\pi^0}^y}{p_{\pi^0}^{xy}}\ ,
\end{align}
where $p_{\pi^0}^{xy}=\sqrt{(p_{\pi^0}^x)^2+(p_{\pi^0}^y)^2}$ and $p_{\pi^0}$ is the total three-momentum. These angles, in turn, define the velocity vector
\begin{equation}
    \label{eq:vel_vector}
    \vec{\beta}=\beta\times(
    \sin\theta_{\pi^0}\cos\varphi_{\pi^0},
    \sin\theta_{\pi^0}\sin\varphi_{\pi^0},
    \cos\theta_{\pi^0})\ .
\end{equation}
Once the boost variables are fixed, the dark photon polar and azimuthal angles,
$0\leq\theta_{A^\prime}^*\leq \pi$ and $0\leq\varphi_{A^\prime}^*\leq 2\pi$, are randomly generated in the $\pi^0$ rest frame. For a given mediator mass, the corresponding four-momentum $p_{A^\prime}^*$ follows from two-body kinematics and energy-momentum conservation. In that frame, the dark photon energy is written as
\begin{equation}
    \label{eq:total_energy_dark_photon_pi0_frame}
    E_{A^\prime}^* = \frac{m_{\pi^0}^2 - m_{A^\prime}}{2m_{\pi^0}}\ .
\end{equation}
Thus, with the aid of a generic Lorentz transformation along $\vec\beta$ the dark photon four-momentum in the laboratory frame can be written as $p^\text{Lab}_{A^\prime} = \Lambda\,\cdot p_{A^\prime}^*$, with the entries of the Lorentz transformation matrix given by \cite{Weinberg:1972kfs}
\begin{equation}
    \label{eq:Lambda_Lorentz_transformation_matrix}
    \Lambda_{ij}=
    \delta_{ij} 
    + \beta_i\,\beta_j\,\frac{(\gamma - 1)}{\beta^2}\ .
\end{equation}
The same procedure is applied to the DM particle produced in the decay
$A^\prime \to \phi + \phi^*$. In the dark photon rest frame, the angles
$0\leq\theta_\phi^*\leq \pi$ and $0\leq\varphi_\phi^*\leq 2\pi$ are randomly selected, while the energy is fixed by kinematics, $E_\phi^* = m_{A^\prime}/2$. The four-momentum $p_\phi^*$ is subsequently boosted to the laboratory frame along the $A^\prime$ direction, $p^\text{Lab}_\phi = \Lambda\,\cdot p_\phi^*$,
with boost parameters determined by $p_{A^\prime}^{\rm Lab}$.
\bibliographystyle{utphys}
\bibliography{bibliography}

@article{AGUILAR201799,
title = {Design specification for the European Spallation Source neutron generating target element},
journal = {Nuclear Instruments and Methods in Physics Research Section A: Accelerators, Spectrometers, Detectors and Associated Equipment},
volume = {856},
pages = {99-108},
year = {2017},
issn = {0168-9002},
doi = {https://doi.org/10.1016/j.nima.2017.03.003},
author = {A. Aguilar and F. Sordo and T. Mora and L. Mena and M. Mancisidor and J. Aguilar and G. Bakedano and I. Herranz and P. Luna and M. Magan and R. Vivanco and F. Jimenez-Villacorta and K. Sjogreen and U. Oden and J.M. Perlado and J.L. Martinez and F.J. Bermejo}
}

@article{WEI20221535,
    title = {Overview of CSNS tantalum cladded tungsten solid Target-1 and Target-2},
    journal = {Nuclear Engineering and Technology},
    volume = {54},
    number = {5},
    pages = {1535-1540},
    year = {2022},
    issn = {1738-5733},
    doi = {https://doi.org/10.1016/j.net.2021.10.032},
    author = {Shaohong Wei and Ruiqiang Zhang and Quan Ji and Changfeng Li and Bin Zhou and Youlian Lu and Jun Xu and Ke Zhou and Chongguang Zhao and Ning He and Wen Yin and Tianjiao Liang}
}

@book{Weinberg:1972kfs,
    author = "Weinberg, Steven",
    note = "{{Gravitation and Cosmology: Principles and Applications of the General Theory of Relativity}}",
    isbn = "978-0-471-92567-5, 978-0-471-92567-5",
    publisher = "John Wiley and Sons",
    address = "New York",
    year = "1972"
}

@article{Apostolakis:2009egq,
    author = "Apostolakis, J. and others",
    title = "{Geometry and physics of the Geant4 toolkit for high and medium energy applications}",
    doi = "10.1016/j.radphyschem.2009.04.026",
    journal = "Radiat. Phys. Chem.",
    volume = "78",
    number = "10",
    pages = "859--873",
    year = "2009"
}

@article{Allison:2016lfl,
    author = "Allison, J. and others",
    title = "{Recent developments in Geant4}",
    reportNumber = "FERMILAB-PUB-16-447-CD",
    doi = "10.1016/j.nima.2016.06.125",
    journal = "Nucl. Instrum. Meth. A",
    volume = "835",
    pages = "186--225",
    year = "2016"
}

@article{Ovchynnikov:2023cry,
    author = "Ovchynnikov, Maksym and Tastet, Jean-Loup and Mikulenko, Oleksii and Bondarenko, Kyrylo",
    title = "{Sensitivities to feebly interacting particles: Public and unified calculations}",
    eprint = "2305.13383",
    archivePrefix = "arXiv",
    primaryClass = "hep-ph",
    doi = "10.1103/PhysRevD.108.075028",
    journal = "Phys. Rev. D",
    volume = "108",
    number = "7",
    pages = "075028",
    year = "2023"
}

@article{PandaX:2024muv,
    author = "Bo, Zihao and others",
    collaboration = "PandaX",
    title = "{First Indication of Solar B8 Neutrinos through Coherent Elastic Neutrino-Nucleus Scattering in PandaX-4T}",
    eprint = "2407.10892",
    archivePrefix = "arXiv",
    primaryClass = "hep-ex",
    doi = "10.1103/PhysRevLett.133.191001",
    journal = "Phys. Rev. Lett.",
    volume = "133",
    number = "19",
    pages = "191001",
    year = "2024"
}

@article{XENON:2024ijk,
    author = "Aprile, Elena and others",
    collaboration = "XENON",
    title = "{First Indication of Solar B8 Neutrinos via Coherent Elastic Neutrino-Nucleus Scattering with XENONnT}",
    eprint = "2408.02877",
    archivePrefix = "arXiv",
    primaryClass = "nucl-ex",
    doi = "10.1103/PhysRevLett.133.191002",
    journal = "Phys. Rev. Lett.",
    volume = "133",
    number = "19",
    pages = "191002",
    year = "2024"
}

@article{CMS:2012wdt,
    author = "Chatrchyan, Serguei and others",
    collaboration = "CMS",
    title = "{Combined Search for the Quarks of a Sequential Fourth Generation}",
    eprint = "1209.1062",
    archivePrefix = "arXiv",
    primaryClass = "hep-ex",
    reportNumber = "CMS-EXO-11-098, CERN-PH-EP-2012-251",
    doi = "10.1103/PhysRevD.86.112003",
    journal = "Phys. Rev. D",
    volume = "86",
    pages = "112003",
    year = "2012"
}

@article{Dobrescu:2014fca,
    author = "Dobrescu, Bogdan A. and Frugiuele, Claudia",
    title = "{Hidden GeV-Scale Interactions of Quarks}",
    eprint = "1404.3947",
    archivePrefix = "arXiv",
    primaryClass = "hep-ph",
    reportNumber = "FERMILAB-PUB-14-098-T",
    doi = "10.1103/PhysRevLett.113.061801",
    journal = "Phys. Rev. Lett.",
    volume = "113",
    pages = "061801",
    year = "2014"
}

@article{FileviezPerez:2010gw,
    author = "Fileviez Perez, Pavel and Wise, Mark B.",
    title = "{Baryon and lepton number as local gauge symmetries}",
    eprint = "1002.1754",
    archivePrefix = "arXiv",
    primaryClass = "hep-ph",
    doi = "10.1103/PhysRevD.82.079901",
    journal = "Phys. Rev. D",
    volume = "82",
    pages = "011901",
    year = "2010",
    note = "[Erratum: Phys.Rev.D 82, 079901 (2010)]"
}

@article{Duerr:2013dza,
    author = "Duerr, Michael and Fileviez Perez, Pavel and Wise, Mark B.",
    title = "{Gauge Theory for Baryon and Lepton Numbers with Leptoquarks}",
    eprint = "1304.0576",
    archivePrefix = "arXiv",
    primaryClass = "hep-ph",
    doi = "10.1103/PhysRevLett.110.231801",
    journal = "Phys. Rev. Lett.",
    volume = "110",
    pages = "231801",
    year = "2013"
}

@article{FileviezPerez:2024fzc,
    author = "Fileviez Perez, Pavel",
    title = "{Lepton and baryon numbers as local gauge symmetries}",
    eprint = "2406.06866",
    archivePrefix = "arXiv",
    primaryClass = "hep-ph",
    doi = "10.1103/PhysRevD.110.035018",
    journal = "Phys. Rev. D",
    volume = "110",
    number = "3",
    pages = "035018",
    year = "2024"
}

@article{Butterworth:2024eyr,
    author = "Butterworth, Jon and Debnath, Hridoy and Fileviez Perez, Pavel and Yeh, Yoran",
    title = "{Dark matter from anomaly cancellation at the LHC}",
    eprint = "2405.03749",
    archivePrefix = "arXiv",
    primaryClass = "hep-ph",
    reportNumber = "MCNET-24-15",
    doi = "10.1103/PhysRevD.110.075001",
    journal = "Phys. Rev. D",
    volume = "110",
    number = "7",
    pages = "075001",
    year = "2024"
}

@article{Garcia:2024uwf,
    author = "Garcia, Giovani Dalla Valle and Kahlhoefer, Felix and Ovchynnikov, Maksym and Schwetz, Thomas",
    title = "{Not-so-inelastic Dark Matter}",
    eprint = "2405.08081",
    archivePrefix = "arXiv",
    primaryClass = "hep-ph",
    reportNumber = "P3H-24-028, TTP24-011",
    doi = "10.1007/JHEP02(2025)127",
    journal = "JHEP",
    volume = "02",
    pages = "127",
    year = "2025"
}

@article{Berlin:2018bsc,
    author = "Berlin, Asher and Blinov, Nikita and Krnjaic, Gordan and Schuster, Philip and Toro, Natalia",
    title = "{Dark Matter, Millicharges, Axion and Scalar Particles, Gauge Bosons, and Other New Physics with LDMX}",
    eprint = "1807.01730",
    archivePrefix = "arXiv",
    primaryClass = "hep-ph",
    reportNumber = "FERMILAB-PUB-18-310-A, SLAC-PUB-17297",
    doi = "10.1103/PhysRevD.99.075001",
    journal = "Phys. Rev. D",
    volume = "99",
    number = "7",
    pages = "075001",
    year = "2019"
}

@article{CarrilloGonzalez:2021lxm,
    author = "Carrillo Gonz{\'a}lez, Mariana and Toro, Natalia",
    title = "{Cosmology and signals of light pseudo-Dirac dark matter}",
    eprint = "2108.13422",
    archivePrefix = "arXiv",
    primaryClass = "hep-ph",
    reportNumber = "Imperial/TP/2021/MC/03",
    doi = "10.1007/JHEP04(2022)060",
    journal = "JHEP",
    volume = "04",
    pages = "060",
    year = "2022"
}

@article{Brahma:2023psr,
    author = "Brahma, Nirmalya and Heeba, Saniya and Schutz, Katelin",
    title = "{Resonant pseudo-Dirac dark matter as a sub-GeV thermal target}",
    eprint = "2308.01960",
    archivePrefix = "arXiv",
    primaryClass = "hep-ph",
    doi = "10.1103/PhysRevD.109.035006",
    journal = "Phys. Rev. D",
    volume = "109",
    number = "3",
    pages = "035006",
    year = "2024"
}

@article{Okun:1982xi,
    author = "Okun, L. B.",
    title = "{LIMITS OF ELECTRODYNAMICS: PARAPHOTONS?}",
    reportNumber = "ITEP-48-1982",
    journal = "Sov. Phys. JETP",
    volume = "56",
    pages = "502",
    year = "1982"
}

@article{Holdom:1985ag,
    author = "Holdom, Bob",
    title = "{Two U(1)'s and Epsilon Charge Shifts}",
    reportNumber = "UTPT-85-30",
    doi = "10.1016/0370-2693(86)91377-8",
    journal = "Phys. Lett. B",
    volume = "166",
    pages = "196--198",
    year = "1986"
}

@misc{deniverville_PhD,
    title = {Searching for Hidden Sector Dark Matter with Fixed Target Neutrino Experiments},
    howpublished = {\url{https://dspace.library.uvic.ca/server/api/core/bitstreams/9c3b0420-da5d-4a74-a8e6-4c772f41757b/content}}
}

@article{COHERENT:2021yvp,
    author = "Akimov, D. and others",
    collaboration = "COHERENT",
    title = "{Simulating the neutrino flux from the Spallation Neutron Source for the COHERENT experiment}",
    eprint = "2109.11049",
    archivePrefix = "arXiv",
    primaryClass = "hep-ex",
    doi = "10.1103/PhysRevD.106.032003",
    journal = "Phys. Rev. D",
    volume = "106",
    number = "3",
    pages = "032003",
    year = "2022"
}

@article{Chu:2020ysb,
    author = "Chu, Xiaoyong and Kuo, Jui-Lin and Pradler, Josef",
    title = "{Dark sector-photon interactions in proton-beam experiments}",
    eprint = "2001.06042",
    archivePrefix = "arXiv",
    primaryClass = "hep-ph",
    doi = "10.1103/PhysRevD.101.075035",
    journal = "Phys. Rev. D",
    volume = "101",
    number = "7",
    pages = "075035",
    year = "2020"
}

@article{MiniBooNE:2008hfu,
    author = "Aguilar-Arevalo, A. A. and others",
    collaboration = "MiniBooNE",
    title = "{The Neutrino Flux Prediction at MiniBooNE}",
    eprint = "0806.1449",
    archivePrefix = "arXiv",
    primaryClass = "hep-ex",
    reportNumber = "FERMILAB-PUB-08-161-AD-E",
    doi = "10.1103/PhysRevD.79.072002",
    journal = "Phys. Rev. D",
    volume = "79",
    pages = "072002",
    year = "2009"
}

@article{K2K:2006yov,
    author = "Ahn, M. H. and others",
    collaboration = "K2K",
    title = "{Measurement of Neutrino Oscillation by the K2K Experiment}",
    eprint = "hep-ex/0606032",
    archivePrefix = "arXiv",
    doi = "10.1103/PhysRevD.74.072003",
    journal = "Phys. Rev. D",
    volume = "74",
    pages = "072003",
    year = "2006"
}

@article{Caputo:2025avc,
    author = "Caputo, Andrea and Park, Jaeyoung and Yun, Seokhoon",
    title = "{The Heavy Dark Photon Handbook: Cosmological and Astrophysical Bounds}",
    eprint = "2511.15785",
    archivePrefix = "arXiv",
    primaryClass = "hep-ph",
    reportNumber = "CTPU-PTC-25-42",
    month = "11",
    year = "2025"
}

@article{Bauer:2018onh,
    author = "Bauer, Martin and Foldenauer, Patrick and Jaeckel, Joerg",
    title = "{Hunting All the Hidden Photons}",
    eprint = "1803.05466",
    archivePrefix = "arXiv",
    primaryClass = "hep-ph",
    doi = "10.1007/JHEP07(2018)094",
    journal = "JHEP",
    volume = "07",
    pages = "094",
    year = "2018"
}

@article{COHERENT:2021pvd,
    author = "Akimov, D. and others",
    collaboration = "COHERENT",
    title = "{First Probe of Sub-GeV Dark Matter beyond the Cosmological Expectation with the COHERENT CsI Detector at the SNS}",
    eprint = "2110.11453",
    archivePrefix = "arXiv",
    primaryClass = "hep-ex",
    doi = "10.1103/PhysRevLett.130.051803",
    journal = "Phys. Rev. Lett.",
    volume = "130",
    number = "5",
    pages = "051803",
    year = "2023"
}

@misc{NIST-XCOM,
    title = {National {I}nstitute of {S}tandards and {T}echnology ,
            {XCOM}: {P}hoton cross sections database},
    howpublished = {\url{https://www.nist.gov/pml/xcom-photon-cross-sections-database}}
}

@techreport{SanfordWang1967,
  author      = {Sanford, J. R. and Wang, C. L.},
  title       = {Empirical formulas for particle production in p-Be collisions between 10 and 35 {GeV}/c},
  institution = {Brookhaven National Laboratory},
  number      = {AGS Internal Report},
  year        = {1967},
  note        = {BNL AGS internal report}
}

@misc{Burman:1989ds,
    author = "Burman, R. L. and Smith, E. S.",
    title = {Parametrization of Pion Production and Reaction Cross-sections at {LAMPF} Energies},
    note = "LA-11502-MS",
    year = {1989}
}

@article{CCM:2021leg,
    author = "Aguilar-Arevalo, A. A. and others",
    collaboration = "CCM",
    title = "{First dark matter search results from Coherent CAPTAIN-Mills}",
    eprint = "2105.14020",
    archivePrefix = "arXiv",
    primaryClass = "hep-ex",
    reportNumber = "LA-UR-21-24983",
    doi = "10.1103/PhysRevD.106.012001",
    journal = "Phys. Rev. D",
    volume = "106",
    number = "1",
    pages = "012001",
    year = "2022"
}

@article{PhysRevD.9.1389,
  title = {Coherent effects of a weak neutral current},
  author = {Freedman, Daniel Z.},
  journal = {Phys. Rev. D},
  volume = {9},
  issue = {5},
  pages = {1389--1392},
  numpages = {0},
  year = {1974},
  month = {Mar},
  publisher = {American Physical Society},
  doi = {10.1103/PhysRevD.9.1389},
  url = {https://link.aps.org/doi/10.1103/PhysRevD.9.1389}
}

@article{Chatterjee:2022mmu,
    author = "Chatterjee, Sabya Sachi and Lavignac, St{\'e}phane and Miranda, O. G. and Sanchez Garcia, G.",
    title = "{Constraining nonstandard interactions with coherent elastic neutrino-nucleus scattering at the European Spallation Source}",
    eprint = "2208.11771",
    archivePrefix = "arXiv",
    primaryClass = "hep-ph",
    doi = "10.1103/PhysRevD.107.055019",
    journal = "Phys. Rev. D",
    volume = "107",
    number = "5",
    pages = "055019",
    year = "2023"
}

@article{Helm:1956zz,
    author = "Helm, Richard H.",
    title = "{Inelastic and Elastic Scattering of 187-Mev Electrons from Selected Even-Even Nuclei}",
    doi = "10.1103/PhysRev.104.1466",
    journal = "Phys. Rev.",
    volume = "104",
    pages = "1466--1475",
    year = "1956"
}

@article{deNiverville:2015mwa,
    author = "deNiverville, Patrick and Pospelov, Maxim and Ritz, Adam",
    title = "{Light new physics in coherent neutrino-nucleus scattering experiments}",
    eprint = "1505.07805",
    archivePrefix = "arXiv",
    primaryClass = "hep-ph",
    doi = "10.1103/PhysRevD.92.095005",
    journal = "Phys. Rev. D",
    volume = "92",
    number = "9",
    pages = "095005",
    year = "2015"
}

@article{Su:2023klh,
    author = "Su, Chenguang and Liu, Qian and Liang, Tianjiao",
    collaboration = "CLOVERS, CE{\ensuremath{\nu}}NS@CSNS",
    title = "{CE{\ensuremath{\nu}}NS Experiment Proposal at CSNS {\textdagger}}",
    eprint = "2303.13423",
    archivePrefix = "arXiv",
    primaryClass = "physics.ins-det",
    doi = "10.3390/psf2023008019",
    journal = "Phys. Sci. Forum",
    volume = "8",
    number = "1",
    pages = "19",
    year = "2023"
}

@article{Baxter:2019mcx,
    author = "Baxter, D. and others",
    title = "{Coherent Elastic Neutrino-Nucleus Scattering at the European Spallation Source}",
    eprint = "1911.00762",
    archivePrefix = "arXiv",
    primaryClass = "physics.ins-det",
    reportNumber = "IFIC/19-45, YITP-SB-19-37, FERMILAB-PUB-19-612-V",
    doi = "10.1007/JHEP02(2020)123",
    journal = "JHEP",
    volume = "02",
    pages = "123",
    year = "2020"
}

@article{ParticleDataGroup:2024cfk,
    author = "Navas, S. and others",
    collaboration = "Particle Data Group",
    title = "{Review of particle physics}",
    doi = "10.1103/PhysRevD.110.030001",
    journal = "Phys. Rev. D",
    volume = "110",
    number = "3",
    pages = "030001",
    year = "2024"
}

@article{Batell:2009di,
    author = "Batell, Brian and Pospelov, Maxim and Ritz, Adam",
    title = "{Exploring Portals to a Hidden Sector Through Fixed Targets}",
    eprint = "0906.5614",
    archivePrefix = "arXiv",
    primaryClass = "hep-ph",
    doi = "10.1103/PhysRevD.80.095024",
    journal = "Phys. Rev. D",
    volume = "80",
    pages = "095024",
    year = "2009"
}

@article{Abele:2022iml,
    author = "Abele, H. and others",
    title = "{Particle Physics at the European Spallation Source}",
    eprint = "2211.10396",
    archivePrefix = "arXiv",
    primaryClass = "physics.ins-det",
    doi = "10.1016/j.physrep.2023.06.001",
    journal = "Phys. Rept.",
    volume = "1023",
    pages = "1--84",
    year = "2023"
}

@article{GEANT4:2002zbu,
    author = "Agostinelli, S. and others",
    collaboration = "GEANT4",
    title = "{GEANT4 - A Simulation Toolkit}",
    reportNumber = "SLAC-PUB-9350, FERMILAB-PUB-03-339, CERN-IT-2002-003",
    doi = "10.1016/S0168-9002(03)01368-8",
    journal = "Nucl. Instrum. Meth. A",
    volume = "506",
    pages = "250--303",
    year = "2003"
}

@article{Dutta:2020vop,
    author = "Dutta, Bhaskar and Kim, Doojin and Liao, Shu and Park, Jong-Chul and Shin, Seodong and Strigari, Louis E. and Thompson, Adrian",
    title = "{Searching for dark matter signals in timing spectra at neutrino experiments}",
    eprint = "2006.09386",
    archivePrefix = "arXiv",
    primaryClass = "hep-ph",
    reportNumber = "MI-TH-2014",
    doi = "10.1007/JHEP01(2022)144",
    journal = "JHEP",
    volume = "01",
    pages = "144",
    year = "2022"
}

@article{Dutta:2019nbn,
    author = "Dutta, Bhaskar and Kim, Doojin and Liao, Shu and Park, Jong-Chul and Shin, Seodong and Strigari, Louis E.",
    title = "{Dark matter signals from timing spectra at neutrino experiments}",
    eprint = "1906.10745",
    archivePrefix = "arXiv",
    primaryClass = "hep-ph",
    reportNumber = "MI-TH-1925",
    doi = "10.1103/PhysRevLett.124.121802",
    journal = "Phys. Rev. Lett.",
    volume = "124",
    number = "12",
    pages = "121802",
    year = "2020"
}

@article{Berlin:2020uwy,
    author = "Berlin, Asher and deNiverville, Patrick and Ritz, Adam and Schuster, Philip and Toro, Natalia",
    title = "{Sub-GeV dark matter production at fixed-target experiments}",
    eprint = "2003.03379",
    archivePrefix = "arXiv",
    primaryClass = "hep-ph",
    doi = "10.1103/PhysRevD.102.095011",
    journal = "Phys. Rev. D",
    volume = "102",
    number = "9",
    pages = "095011",
    year = "2020"
}

@article{deNiverville:2011it,
    author = "deNiverville, Patrick and Pospelov, Maxim and Ritz, Adam",
    title = "{Observing a light dark matter beam with neutrino experiments}",
    eprint = "1107.4580",
    archivePrefix = "arXiv",
    primaryClass = "hep-ph",
    doi = "10.1103/PhysRevD.84.075020",
    journal = "Phys. Rev. D",
    volume = "84",
    pages = "075020",
    year = "2011"
}

@article{DeRomeri:2019kic,
    author = "De Romeri, Valentina and Kelly, Kevin J. and Machado, Pedro A. N.",
    title = "{DUNE-PRISM Sensitivity to Light Dark Matter}",
    eprint = "1903.10505",
    archivePrefix = "arXiv",
    primaryClass = "hep-ph",
    reportNumber = "FERMILAB-PUB-19-116-T",
    doi = "10.1103/PhysRevD.100.095010",
    journal = "Phys. Rev. D",
    volume = "100",
    number = "9",
    pages = "095010",
    year = "2019"
}

@article{NA64:2023wbi,
    author = "Andreev, Yu. M. and others",
    collaboration = "NA64",
    title = "{Search for Light Dark Matter with NA64 at CERN}",
    eprint = "2307.02404",
    archivePrefix = "arXiv",
    primaryClass = "hep-ex",
    reportNumber = "CERN-EP-2023-130",
    doi = "10.1103/PhysRevLett.131.161801",
    journal = "Phys. Rev. Lett.",
    volume = "131",
    number = "16",
    pages = "161801",
    year = "2023"
}

@article{Cirelli:2024ssz,
    author = "Cirelli, Marco and Strumia, Alessandro and Zupan, Jure",
    title = "{Dark Matter}",
    eprint = "2406.01705",
    archivePrefix = "arXiv",
    primaryClass = "hep-ph",
    month = "6",
    year = "2024"
}

@article{XENON:2024hup,
    author = "Aprile, E. and others",
    collaboration = "XENON",
    title = "{First Search for Light Dark Matter in the Neutrino Fog with XENONnT}",
    eprint = "2409.17868",
    archivePrefix = "arXiv",
    primaryClass = "hep-ex",
    doi = "10.1103/PhysRevLett.134.111802",
    journal = "Phys. Rev. Lett.",
    volume = "134",
    number = "11",
    pages = "111802",
    year = "2025"
}

@article{XENON:2025vwd,
    author = "Aprile, E. and others",
    collaboration = "XENON",
    title = "{WIMP Dark Matter Search using a 3.1 tonne $\times$ year Exposure of the XENONnT Experiment}",
    eprint = "2502.18005",
    archivePrefix = "arXiv",
    primaryClass = "hep-ex",
    month = "2",
    year = "2025"
}

@article{Zhang:2025ajc,
    author = "Zhang, Minzhen and others",
    title = "{Search for Light Dark Matter with 259-day data in PandaX-4T}",
    eprint = "2507.11930",
    archivePrefix = "arXiv",
    primaryClass = "hep-ex",
    month = "7",
    year = "2025"
}

@article{RevModPhys.90.045002,
  title = {History of dark matter},
  author = {Bertone, Gianfranco and Hooper, Dan},
  journal = {Rev. Mod. Phys.},
  volume = {90},
  issue = {4},
  pages = {045002},
  numpages = {32},
  year = {2018},
  month = {Oct},
  publisher = {American Physical Society},
  doi = {10.1103/RevModPhys.90.045002},
  url = {https://link.aps.org/doi/10.1103/RevModPhys.90.045002}
}

@article{Billard:2021uyg,
    author = "Billard, Julien and others",
    title = "{Direct detection of dark matter{\textemdash}APPEC committee report*}",
    eprint = "2104.07634",
    archivePrefix = "arXiv",
    primaryClass = "hep-ex",
    doi = "10.1088/1361-6633/ac5754",
    journal = "Rept. Prog. Phys.",
    volume = "85",
    number = "5",
    pages = "056201",
    year = "2022"
}

@article{Schumann:2019eaa,
    author = "Schumann, Marc",
    title = "{Direct Detection of WIMP Dark Matter: Concepts and Status}",
    eprint = "1903.03026",
    archivePrefix = "arXiv",
    primaryClass = "astro-ph.CO",
    doi = "10.1088/1361-6471/ab2ea5",
    journal = "J. Phys. G",
    volume = "46",
    number = "10",
    pages = "103003",
    year = "2019"
}

@article{LZ:2024zvo,
    author = "Aalbers, J. and others",
    collaboration = "LZ",
    title = "{Dark Matter Search Results from 4.2{\,}{\,}Tonne-Years of Exposure of the LUX-ZEPLIN (LZ) Experiment}",
    eprint = "2410.17036",
    archivePrefix = "arXiv",
    primaryClass = "hep-ex",
    reportNumber = "FERMILAB-PUB-24-0796-V",
    doi = "10.1103/4dyc-z8zf",
    journal = "Phys. Rev. Lett.",
    volume = "135",
    number = "1",
    pages = "011802",
    year = "2025"
}

@article{PandaX:2024qfu,
    author = "Bo, Zihao and others",
    collaboration = "PandaX",
    title = "{Dark Matter Search Results from 1.54{\,}{\,}Tonne{\textperiodcentered}Year Exposure of PandaX-4T}",
    eprint = "2408.00664",
    archivePrefix = "arXiv",
    primaryClass = "hep-ex",
    doi = "10.1103/PhysRevLett.134.011805",
    journal = "Phys. Rev. Lett.",
    volume = "134",
    number = "1",
    pages = "011805",
    year = "2025"
}

@article{DarkSide-50:2022qzh,
    author = "Agnes, P. and others",
    collaboration = "DarkSide-50",
    title = "{Search for low-mass dark matter WIMPs with 12~ton-day exposure of DarkSide-50}",
    eprint = "2207.11966",
    archivePrefix = "arXiv",
    primaryClass = "hep-ex",
    reportNumber = "FERMILAB-PUB-22-589-ND-PPD-SCD",
    doi = "10.1103/PhysRevD.107.063001",
    journal = "Phys. Rev. D",
    volume = "107",
    number = "6",
    pages = "063001",
    year = "2023"
}

@article{deNiverville:2012ij,
    author = "deNiverville, Patrick and McKeen, David and Ritz, Adam",
    title = "{Signatures of sub-GeV dark matter beams at neutrino experiments}",
    eprint = "1205.3499",
    archivePrefix = "arXiv",
    primaryClass = "hep-ph",
    doi = "10.1103/PhysRevD.86.035022",
    journal = "Phys. Rev. D",
    volume = "86",
    pages = "035022",
    year = "2012"
}

@article{Izaguirre:2013uxa,
    author = "Izaguirre, Eder and Krnjaic, Gordan and Schuster, Philip and Toro, Natalia",
    title = "{New Electron Beam-Dump Experiments to Search for MeV to few-GeV Dark Matter}",
    eprint = "1307.6554",
    archivePrefix = "arXiv",
    primaryClass = "hep-ph",
    doi = "10.1103/PhysRevD.88.114015",
    journal = "Phys. Rev. D",
    volume = "88",
    pages = "114015",
    year = "2013"
}

@article{Izaguirre:2014bca,
    author = "Izaguirre, Eder and Krnjaic, Gordan and Schuster, Philip and Toro, Natalia",
    title = "{Testing GeV-Scale Dark Matter with Fixed-Target Missing Momentum Experiments}",
    eprint = "1411.1404",
    archivePrefix = "arXiv",
    primaryClass = "hep-ph",
    doi = "10.1103/PhysRevD.91.094026",
    journal = "Phys. Rev. D",
    volume = "91",
    number = "9",
    pages = "094026",
    year = "2015"
}

@article{Batell:2014yra,
    author = "Batell, Brian and deNiverville, Patrick and McKeen, David and Pospelov, Maxim and Ritz, Adam",
    title = "{Leptophobic Dark Matter at Neutrino Factories}",
    eprint = "1405.7049",
    archivePrefix = "arXiv",
    primaryClass = "hep-ph",
    reportNumber = "EFI-14-13",
    doi = "10.1103/PhysRevD.90.115014",
    journal = "Phys. Rev. D",
    volume = "90",
    number = "11",
    pages = "115014",
    year = "2014"
}

@article{Pospelov:2007mp,
    author = "Pospelov, Maxim and Ritz, Adam and Voloshin, Mikhail B.",
    title = "{Secluded WIMP Dark Matter}",
    eprint = "0711.4866",
    archivePrefix = "arXiv",
    primaryClass = "hep-ph",
    doi = "10.1016/j.physletb.2008.02.052",
    journal = "Phys. Lett. B",
    volume = "662",
    pages = "53--61",
    year = "2008"
}

@article{Arkani-Hamed:2008hhe,
    author = "Arkani-Hamed, Nima and Finkbeiner, Douglas P. and Slatyer, Tracy R. and Weiner, Neal",
    title = "{A Theory of Dark Matter}",
    eprint = "0810.0713",
    archivePrefix = "arXiv",
    primaryClass = "hep-ph",
    doi = "10.1103/PhysRevD.79.015014",
    journal = "Phys. Rev. D",
    volume = "79",
    pages = "015014",
    year = "2009"
}

@article{Ge:2017mcq,
    author = "Ge, Shao-Feng and Shoemaker, Ian M.",
    title = "{Constraining Photon Portal Dark Matter with Texono and Coherent Data}",
    eprint = "1710.10889",
    archivePrefix = "arXiv",
    primaryClass = "hep-ph",
    reportNumber = "IPMU17-0149",
    doi = "10.1007/JHEP11(2018)066",
    journal = "JHEP",
    volume = "11",
    pages = "066",
    year = "2018"
}

@article{COHERENT:2019kwz,
    author = "Akimov, D. and others",
    collaboration = "COHERENT",
    title = "{Sensitivity of the COHERENT Experiment to Accelerator-Produced Dark Matter}",
    eprint = "1911.06422",
    archivePrefix = "arXiv",
    primaryClass = "hep-ex",
    doi = "10.1103/PhysRevD.102.052007",
    journal = "Phys. Rev. D",
    volume = "102",
    number = "5",
    pages = "052007",
    year = "2020"
}

@article{COHERENT:2023sol,
    author = "Barbeau, P. S. and others",
    collaboration = "COHERENT",
    title = "{Accessing new physics with an undoped, cryogenic CsI CEvNS detector for COHERENT at the SNS}",
    eprint = "2311.13032",
    archivePrefix = "arXiv",
    primaryClass = "hep-ex",
    doi = "10.1103/PhysRevD.109.092005",
    journal = "Phys. Rev. D",
    volume = "109",
    number = "9",
    pages = "092005",
    year = "2024"
}

@article{CCM:2021yzc,
    author = "Aguilar-Arevalo, A. A. and others",
    collaboration = "CCM",
    title = "{First Leptophobic Dark Matter Search from the Coherent{\textendash}CAPTAIN-Mills Liquid Argon Detector}",
    eprint = "2109.14146",
    archivePrefix = "arXiv",
    primaryClass = "hep-ex",
    reportNumber = "Report-no: LA-UR-21-28552, LA-UR-21-28552",
    doi = "10.1103/PhysRevLett.129.021801",
    journal = "Phys. Rev. Lett.",
    volume = "129",
    number = "2",
    pages = "021801",
    year = "2022"
}

@inproceedings{Batell:2022xau,
    author = "Batell, Brian and others",
    title = "{Dark Sector Studies with Neutrino Beams}",
    booktitle = "{Snowmass 2021}",
    eprint = "2207.06898",
    archivePrefix = "arXiv",
    primaryClass = "hep-ph",
    reportNumber = "FERMILAB-FN-1180-ND-T",
    doi = "10.2172/1882578",
    month = "7",
    year = "2022"
}

@article{deNiverville:2016rqh,
    author = "deNiverville, Patrick and Chen, Chien-Yi and Pospelov, Maxim and Ritz, Adam",
    title = "{Light dark matter in neutrino beams: production modelling and scattering signatures at MiniBooNE, T2K and SHiP}",
    eprint = "1609.01770",
    archivePrefix = "arXiv",
    primaryClass = "hep-ph",
    doi = "10.1103/PhysRevD.95.035006",
    journal = "Phys. Rev. D",
    volume = "95",
    number = "3",
    pages = "035006",
    year = "2017"
}

@article{Buonocore:2019esg,
    author = "Buonocore, Luca and Frugiuele, Claudia and deNiverville, Patrick",
    title = "{Hunt for sub-GeV dark matter at neutrino facilities: A survey of past and present experiments}",
    eprint = "1912.09346",
    archivePrefix = "arXiv",
    primaryClass = "hep-ph",
    doi = "10.1103/PhysRevD.102.035006",
    journal = "Phys. Rev. D",
    volume = "102",
    number = "3",
    pages = "035006",
    year = "2020"
}

@article{Abdullah:2022zue,
    author = "Abdullah, M. and others",
    title = "{Coherent elastic neutrino-nucleus scattering: Terrestrial and astrophysical applications}",
    eprint = "2203.07361",
    archivePrefix = "arXiv",
    primaryClass = "hep-ph",
    month = "3",
    year = "2022"
}

@article{Coloma:2015pih,
    author = "Coloma, Pilar and Dobrescu, Bogdan A. and Frugiuele, Claudia and Harnik, Roni",
    title = "{Dark matter beams at LBNF}",
    eprint = "1512.03852",
    archivePrefix = "arXiv",
    primaryClass = "hep-ph",
    reportNumber = "FERMILAB-PUB-15-535-T",
    doi = "10.1007/JHEP04(2016)047",
    journal = "JHEP",
    volume = "04",
    pages = "047",
    year = "2016"
}

@article{Dror:2017ehi,
    author = "Dror, Jeff A. and Lasenby, Robert and Pospelov, Maxim",
    title = "{New constraints on light vectors coupled to anomalous currents}",
    eprint = "1705.06726",
    archivePrefix = "arXiv",
    primaryClass = "hep-ph",
    doi = "10.1103/PhysRevLett.119.141803",
    journal = "Phys. Rev. Lett.",
    volume = "119",
    number = "14",
    pages = "141803",
    year = "2017"
}

@article{Boehm:2003hm,
    author = "Boehm, C. and Fayet, Pierre",
    title = "{Scalar dark matter candidates}",
    eprint = "hep-ph/0305261",
    archivePrefix = "arXiv",
    doi = "10.1016/j.nuclphysb.2004.01.015",
    journal = "Nucl. Phys. B",
    volume = "683",
    pages = "219--263",
    year = "2004"
}

@article{Fayet:2004bw,
    author = "Fayet, Pierre",
    title = "{Light spin 1/2 or spin 0 dark matter particles}",
    eprint = "hep-ph/0403226",
    archivePrefix = "arXiv",
    reportNumber = "LPTENS-04-15",
    doi = "10.1103/PhysRevD.70.023514",
    journal = "Phys. Rev. D",
    volume = "70",
    pages = "023514",
    year = "2004"
}

@article{Lee:1977ua,
    author = "Lee, Benjamin W. and Weinberg, Steven",
    editor = "Srednicki, M. A.",
    title = "{Cosmological Lower Bound on Heavy Neutrino Masses}",
    reportNumber = "FERMILAB-PUB-77-041-T",
    doi = "10.1103/PhysRevLett.39.165",
    journal = "Phys. Rev. Lett.",
    volume = "39",
    pages = "165--168",
    year = "1977"
}

@article{Jordan:2018gcd,
    author = "Jordan, Johnathon R. and Kahn, Yonatan and Krnjaic, Gordan and Moschella, Matthew and Spitz, Joshua",
    title = "{Signatures of Pseudo-Dirac Dark Matter at High-Intensity Neutrino Experiments}",
    eprint = "1806.05185",
    archivePrefix = "arXiv",
    primaryClass = "hep-ph",
    reportNumber = "FERMILAB-PUB-18-148-A, PUPT 2563, PUPT-2563",
    doi = "10.1103/PhysRevD.98.075020",
    journal = "Phys. Rev. D",
    volume = "98",
    number = "7",
    pages = "075020",
    year = "2018"
}

@article{Planck:2018vyg,
    author = "Aghanim, N. and others",
    collaboration = "Planck",
    title = "{Planck 2018 results. VI. Cosmological parameters}",
    eprint = "1807.06209",
    archivePrefix = "arXiv",
    primaryClass = "astro-ph.CO",
    doi = "10.1051/0004-6361/201833910",
    journal = "Astron. Astrophys.",
    volume = "641",
    pages = "A6",
    year = "2020",
    note = "[Erratum: Astron.Astrophys. 652, C4 (2021)]"
}

@article{Gondolo:2011eq,
    author = "Gondolo, P. and Ko, P. and Omura, Y.",
    title = "{Light dark matter in leptophobic Z' models}",
    eprint = "1106.0885",
    archivePrefix = "arXiv",
    primaryClass = "hep-ph",
    doi = "10.1103/PhysRevD.85.035022",
    journal = "Phys. Rev. D",
    volume = "85",
    pages = "035022",
    year = "2012"
}

@article{Aranda:1998fr,
    author = "Aranda, Alfredo and Carone, Christopher D.",
    title = "{Limits on a light leptophobic gauge boson}",
    eprint = "hep-ph/9809522",
    archivePrefix = "arXiv",
    reportNumber = "WM-98-113",
    doi = "10.1016/S0370-2693(98)01309-4",
    journal = "Phys. Lett. B",
    volume = "443",
    pages = "352--358",
    year = "1998"
}

@article{CSNS,
  author={S. W. Xu and Others},
  title={China Spallation Neutron Source - an overview of application and development},
  journal={Chinese Physics C},
  volume={33},
  number={11},
  pages={1--5},
  year={2009},
  publisher={IOP Publishing},
  doi={10.1088/1674-1137/33/11/021},
url = {https://doi.org/10.1088/1674-1137/33/11/021}}

@article{LSND:2001akn,
    author = "Auerbach, L. B. and others",
    collaboration = "LSND",
    title = "{Measurement of electron - neutrino - electron elastic scattering}",
    eprint = "hep-ex/0101039",
    archivePrefix = "arXiv",
    doi = "10.1103/PhysRevD.63.112001",
    journal = "Phys. Rev. D",
    volume = "63",
    pages = "112001",
    year = "2001"
}

@article{Batell:2014mga,
    author = "Batell, Brian and Essig, Rouven and Surujon, Ze'ev",
    title = "{Strong Constraints on Sub-GeV Dark Sectors from SLAC Beam Dump E137}",
    eprint = "1406.2698",
    archivePrefix = "arXiv",
    primaryClass = "hep-ph",
    doi = "10.1103/PhysRevLett.113.171802",
    journal = "Phys. Rev. Lett.",
    volume = "113",
    number = "17",
    pages = "171802",
    year = "2014"
}

@article{MiniBooNEDM:2018cxm,
    author = "Aguilar-Arevalo, A. A. and others",
    collaboration = "MiniBooNE DM",
    title = "{Dark Matter Search in Nucleon, Pion, and Electron Channels from a Proton Beam Dump with MiniBooNE}",
    eprint = "1807.06137",
    archivePrefix = "arXiv",
    primaryClass = "hep-ex",
    reportNumber = "LA-UR-18-26421, FERMILAB-PUB-18-334-ND",
    doi = "10.1103/PhysRevD.98.112004",
    journal = "Phys. Rev. D",
    volume = "98",
    number = "11",
    pages = "112004",
    year = "2018"
}

@article{BaBar:2017tiz,
    author = "Lees, J. P. and others",
    collaboration = "BaBar",
    title = "{Search for Invisible Decays of a Dark Photon Produced in ${e}^{+}{e}^{-}$ Collisions at BaBar}",
    eprint = "1702.03327",
    archivePrefix = "arXiv",
    primaryClass = "hep-ex",
    reportNumber = "BABAR-PUB-17-001, SLAC-PUB-16923",
    doi = "10.1103/PhysRevLett.119.131804",
    journal = "Phys. Rev. Lett.",
    volume = "119",
    number = "13",
    pages = "131804",
    year = "2017"
}

@article{Davoudiasl:2015hxa,
    author = "Davoudiasl, Hooman and Marciano, William J.",
    title = "{Running of the U(1) coupling in the dark sector}",
    eprint = "1502.07383",
    archivePrefix = "arXiv",
    primaryClass = "hep-ph",
    doi = "10.1103/PhysRevD.92.035008",
    journal = "Phys. Rev. D",
    volume = "92",
    number = "3",
    pages = "035008",
    year = "2015"
}

@article{Collar:2025sle,
    author = "Collar, J. I. and others",
    title = "{Coherent Elastic Neutrino-Nucleus Scattering at the Japan Proton Accelerator Research Complex}",
    eprint = "2512.19788",
    archivePrefix = "arXiv",
    primaryClass = "hep-ph",
    month = "12",
    year = "2025"
}

@article{COHERENT:2020ybo,
    author = "Akimov, D. and others",
    collaboration = "COHERENT",
    title = "{COHERENT Collaboration data release from the first detection of coherent elastic neutrino-nucleus scattering on argon}",
    eprint = "2006.12659",
    archivePrefix = "arXiv",
    primaryClass = "nucl-ex",
    doi = "10.5281/zenodo.3903810",
    month = "6",
    year = "2020"
}

@article{COHERENT:2018imc,
    author = "Akimov, D. and others",
    collaboration = "COHERENT",
    title = "{COHERENT Collaboration data release from the first observation of coherent elastic neutrino-nucleus scattering}",
    eprint = "1804.09459",
    archivePrefix = "arXiv",
    primaryClass = "nucl-ex",
    doi = "10.5281/zenodo.1228631",
    month = "4",
    year = "2018"
}

@article{JSNS2:2017gzk,
    author = "Ajimura, S. and others",
    collaboration = "JSNS2",
    title = "{Technical Design Report (TDR): Searching for a Sterile Neutrino at J-PARC MLF (E56, JSNS2)}",
    eprint = "1705.08629",
    archivePrefix = "arXiv",
    primaryClass = "physics.ins-det",
    month = "5",
    year = "2017"
}

@article{Blumlein:2013cua,
    author = {Bl{\"u}mlein, Johannes and Brunner, J{\"u}rgen},
    title = "{New Exclusion Limits on Dark Gauge Forces from Proton Bremsstrahlung in Beam-Dump Data}",
    eprint = "1311.3870",
    archivePrefix = "arXiv",
    primaryClass = "hep-ph",
    reportNumber = "DESY-13-202, DO-TH-13-29, SFB-CPP-13-87, LPN-13-087",
    doi = "10.1016/j.physletb.2014.02.029",
    journal = "Phys. Lett. B",
    volume = "731",
    pages = "320--326",
    year = "2014"
}

@misc{CICENNS2025talk,
  author       = {Xiang Xiao},
  title        = "{Status of the CICENNS Experiment}",
  year         = {2025},
  url          = {https://indico.cern.ch/event/1458241/contributions/6522284/attachments/3082978/5457435/CICENNS_XiangXiao.pdf},
  note         = {Talk at Magnificent CE$\nu$NS 2025}
}

@article{LDMX:2018cma,
    author = "{\r{A}}kesson, Torsten and others",
    collaboration = "LDMX",
    title = "{Light Dark Matter eXperiment (LDMX)}",
    eprint = "1808.05219",
    archivePrefix = "arXiv",
    primaryClass = "hep-ex",
    reportNumber = "FERMILAB-PUB-18-324-A, SLAC-PUB-17303",
    month = "8",
    year = "2018"
}

@article{SHiP:2020noy,
    author = "Ahdida, C. and others",
    collaboration = "SHiP",
    title = "{Sensitivity of the SHiP experiment to light dark matter}",
    eprint = "2010.11057",
    archivePrefix = "arXiv",
    primaryClass = "hep-ex",
    doi = "10.1007/JHEP04(2021)199",
    journal = "JHEP",
    volume = "04",
    pages = "199",
    year = "2021"
}

@article{Batell:2021snh,
    author = "Batell, Brian and Feng, Jonathan L. and Fieg, Max and Ismail, Ahmed and Kling, Felix and Abraham, Roshan Mammen and Trojanowski, Sebastian",
    title = "{Hadrophilic dark sectors at the Forward Physics Facility}",
    eprint = "2111.10343",
    archivePrefix = "arXiv",
    primaryClass = "hep-ph",
    reportNumber = "PITT-PACC-2122, UCI-TR-2021-16, DESY-21-191",
    doi = "10.1103/PhysRevD.105.075001",
    journal = "Phys. Rev. D",
    volume = "105",
    number = "7",
    pages = "075001",
    year = "2022"
}

@article{COHERENT:2022pli,
    author = "Akimov, D. Yu. and others",
    collaboration = "COHERENT",
    title = "{COHERENT constraint on leptophobic dark matter using CsI data}",
    eprint = "2205.12414",
    archivePrefix = "arXiv",
    primaryClass = "hep-ex",
    reportNumber = "FERMILAB-CONF-22-693-ND",
    doi = "10.1103/PhysRevD.106.052004",
    journal = "Phys. Rev. D",
    volume = "106",
    number = "5",
    pages = "052004",
    year = "2022"
}

@article{Cheek:2025nul,
    author = "Cheek, Andrew and Figueroa, Pablo and Herrera, Gonzalo and Shoemaker, Ian M.",
    title = "{Sub-GeV Dark Matter Under Pressure from Direct Detection}",
    eprint = "2507.15956",
    archivePrefix = "arXiv",
    primaryClass = "hep-ph",
    month = "7",
    year = "2025"
}

@article{Krnjaic:2025noj,
    author = "Krnjaic, Gordan",
    title = "{Testing Thermal-Relic Dark Matter with a Dark Photon Mediator}",
    eprint = "2505.04626",
    archivePrefix = "arXiv",
    primaryClass = "hep-ph",
    reportNumber = "FERMILAB-PUB-25-0234-T",
    month = "5",
    year = "2025"
}

@article{Han:2026ozp,
    author = "Han, Xu and Krnjaic, Gordan",
    title = "{New Thermal-Relic Targets for sub-GeV Dark Matter Direct Detection}",
    eprint = "2603.03444",
    archivePrefix = "arXiv",
    primaryClass = "hep-ph",
    reportNumber = "FERMILAB-PUB-26-0113-T",
    month = "3",
    year = "2026"
}

@article{Dutta:2023fij,
    author = "Dutta, Bhaskar and Huang, Wei-Chih and Newstead, Jayden L.",
    title = "{Probing the Dark Sector with Nuclear Transition Photons}",
    eprint = "2302.10250",
    archivePrefix = "arXiv",
    primaryClass = "hep-ph",
    doi = "10.1103/PhysRevLett.131.111801",
    journal = "Phys. Rev. Lett.",
    volume = "131",
    number = "11",
    pages = "111801",
    year = "2023"
}
\end{document}